\documentclass[a4paper,11pt]{article}
\pdfoutput=1

\usepackage{jheppub}

\usepackage{graphicx}
\usepackage{amsmath}
\usepackage{amssymb}
\usepackage{xspace}
\usepackage[small]{subfigure}
\usepackage{slashed}
\usepackage{mdwlist}
\usepackage[normalem]{ulem}
\usepackage[usenames,dvipsnames,svgnames,table]{xcolor}

\newcommand{\eq}[1]{eq.~\eqref{eq:#1}}
\newcommand{\eqs}[2]{eqs.~\eqref{eq:#1} and \eqref{eq:#2}}
\renewcommand{\sec}[1]{sec.~\ref{sec:#1}}
\newcommand{\secs}[2]{secs.~\ref{sec:#1} and \ref{sec:#2}}

\newcommand{\app}[1]{app.~\ref{app:#1}}
\newcommand{\fig}[1]{fig.~\ref{fig:#1}}
\newcommand{\figs}[2]{figs.~\ref{fig:#1} and \ref{fig:#2}}

\newcommand{\abs}[1]{\lvert#1\rvert}

\newcommand{\ord}[1]{\mathcal{O}(#1)}

\newcommand{\MAe}[3]{\Bigl\langle#1\Bigr\rvert#2\Bigr\rvert#3\Bigr\rangle}

\newcommand{\df}{\mathrm{d}}

\newcommand{\MS}{\overline{\rm MS}}

\newcommand{\Li}{\textrm{Li}}

\newcommand{\bfT}{{\bf T}}

\newcommand{\lra}{\leftrightarrow}

\newcommand{\de}{\delta}
\newcommand{\eps}{\epsilon}

\newcommand{\bT}{\mathbf{T}}

\newcommand{\cL}{{\mathcal L}}

\newcommand{\Tau}{{\mathcal T}}

\newcommand{\nn}{\nonumber}

\newcommand{\Ra}{R}
\newcommand{\Rb}{\mathcal{R}}

\newcommand{\one}{{(1)}}

\newcommand{\SCET}{\ensuremath{{\rm SCET}}\xspace}
\newcommand{\SCETa}{\ensuremath{{\rm SCET}_{\rm I}}\xspace}
\newcommand{\SCETb}{\ensuremath{{\rm SCET}_{\rm II}}\xspace}
\newcommand{\SCETp}{\ensuremath{{\rm SCET}_+}\xspace}

\newcommand{\Taum}[1]{\Tau^{(#1)}}

\renewcommand{\b}{\beta}
\newcommand{\e}{\epsilon}

\newcommand\numberthis{\addtocounter{equation}{1}\tag{\theequation}}

\allowdisplaybreaks[2]

  \newcount\hour \newcount\minute
  \hour=\time \divide \hour by 60 \minute=\time
  \newcommand{\todaytime}{\today \ -- \number\hour :\ifnum \minute<10 0\fi\number\minute}
\title{Soft Functions for Generic Jet Algorithms and Observables at Hadron Colliders}

\author[a,b]{Daniele Bertolini,}
\author[c]{Daniel Kolodrubetz,}
\author[c,d]{Duff Neill,}
\author[e]{Piotr Pietrulewicz,}
\author[c]{\hspace{4cm} Iain W.~Stewart,}
\author[e]{Frank J.~Tackmann,}
\author[f,g]{and Wouter J.~Waalewijn}

\affiliation[a]{Berkeley Center for Theoretical Physics, University of California, Berkeley, CA~94270, U.S.A.}
\affiliation[b]{Theoretical Physics Group, Lawrence Berkeley National Laboratory, Berkeley, CA~94720, U.S.A.}
\affiliation[c]{Center for Theoretical Physics, Massachusetts Institute of Technology, Cambridge, MA~02139, U.S.A.}
\affiliation[d]{Theoretical Division, Los Alamos National Laboratory, Los Alamos, NM 87545, USA}
\affiliation[e]{Theory Group, Deutsches Elektronen-Synchrotron (DESY), D-22607 Hamburg, Germany}
\affiliation[f]{Institute for Theoretical Physics Amsterdam and Delta Institute for Theoretical Physics, University of Amsterdam, Science Park 904, 1098 XH Amsterdam, The Netherlands}
\affiliation[g]{Nikhef, Theory Group, Science Park 105, 1098 XG, Amsterdam, The Netherlands}

\emailAdd{dbertolini@lbl.gov}
\emailAdd{dkolodru@mit.edu}
\emailAdd{duff.neill@gmail.com}
\emailAdd{piotr.pietrulewicz@desy.de}
\emailAdd{iains@mit.edu}
\emailAdd{frank.tackmann@desy.de}
\emailAdd{w.j.waalewijn@uva.nl}

\abstract{
We introduce a method to compute one-loop soft functions for exclusive $N$-jet processes at hadron colliders, allowing for different definitions of the algorithm that determines the jet regions and of the measurements in those regions. In particular, we generalize the $N$-jettiness hemisphere decomposition of ref.~\cite{Jouttenus:2011wh} in a manner that separates the dependence on the jet boundary from the observables measured inside the jet and beam regions. Results are given for several factorizable jet definitions, including  anti-$k_T$, XCone, and other geometric partitionings. We calculate explicitly the soft functions for angularity measurements, including jet mass and jet broadening, in $pp \to L + 1$ jet and explore the differences for various jet vetoes and algorithms. This includes a consistent treatment of rapidity divergences when applicable. We also compute analytic results for these soft functions in an expansion for a small jet radius $R$. We find that the small-$R$ results, including corrections up to $\mathcal{O}(R^2)$, accurately capture the full behavior over a large range of $R$.
}
\preprint{\vbox{
\hbox{DESY 16-137}
\hbox{MIT--CTP 4823} 
\hbox{NIKHEF 2016-035}
\hbox{\todaytime}}}

\begin{document}

\maketitle

%%%%%%%%%%%%%%%%%%%%%%%%%%%%%%%%%%%%%%%%%%%%%%%%%%%%%%%%%%%%%%%%%%%%%%%%%%%%%%%%
\section{Introduction}
\label{sec:intro}
%%%%%%%%%%%%%%%%%%%%%%%%%%%%%%%%%%%%%%%%%%%%%%%%%%%%%%%%%%%%%%%%%%%%%%%%%%%%%%%%

Exclusive jet processes, i.e.~those with a fixed number of hard signal jets in the final state, play a crucial role in the Large Hadron Collider (LHC) physics program. Many important processes, such as Higgs or $W/Z$ boson production or diboson production, are measured in different exclusive jet bins. Furthermore, jet substructure techniques have become increasingly important both in Standard Model and in new physics analyses, and the associated observables often exploit the properties of a fixed number of subjets. Theoretical predictions at increasingly high precision are needed to match the increasing precision of the data. Compared to color-singlet final states, the presence of jets makes perturbative QCD calculations more challenging and the singularity structure more complicated. Furthermore, a fixed number of jets is imposed through a jet veto, which restricts the phase space for additional collinear and soft emissions, and generates large logarithms that often need to be resummed to obtain predictions with the best possible precision.

Soft Collinear Effective Theory (SCET)~\cite{Bauer:2000ew, Bauer:2000yr, Bauer:2001ct, Bauer:2001yt} provides a framework to systematically carry out the resummation of logarithms to higher orders by factorizing the cross section into hard, collinear, and soft functions, and then exploiting their renormalization group evolution. Schematically, the cross section for $pp\to N\text{ jets}$ factorizes for many observables in the singular limit as
\begin{align} \label{eq:simpleFact}
\sigma_N=H_N\times\left[B_aB_b\prod_{i=1}^NJ_i\right]\otimes S_N,
\end{align}
where the hard function $H_N$ contains the virtual corrections to the partonic hard scattering process, the beam functions $B_{a,b}$ contain parton distribution functions and describe collinear initial-state radiation. The jet functions $J_i$ describe final-state radiation collinear to the direction of the hard partons, and the soft function $S_N$ describes wide-angle soft radiation. The resummation of large logarithms is achieved by evaluating each component at its natural scale and then renormalization-group evolving all components to a common scale. For an interesting class of observables, the jet and beam functions are of the inclusive type and do not depend on the precise definition of the jet regions. They are known for a variety of jet and beam measurements, typically at one loop or
beyond~\cite{Bauer:2003pi,Fleming:2003gt,Becher:2006qw,Becher:2009th,Becher:2010pd,Larkoski:2014uqa,Stewart:2010qs,Berger:2010xi,Becher:2010tm,Becher:2012qa,Chiu:2012ir,Gehrmann:2014yya,Gaunt:2014xga,Gaunt:2014cfa,Luebbert:2016itl}. Hard functions are also known for many processes at one loop or beyond (see e.g.\ ref.~\cite{Moult:2015aoa} and references therein). In this paper, we focus on determining the soft functions that appear for a wide class of jet algorithms and jet measurements. The resummation at NLL$^\prime$ and NNLL requires the soft function at one loop. Compared to the beam and jet functions, the perturbative calculation of the soft function generally requires a more sophisticated setup, since it depends not only on the measurements made in the jet and beam regions, but also on the angles between all jet and beam directions and the precise definition of the jet boundaries.

$N$-jettiness \cite{Stewart:2010tn} is a global event shape that allows one to define exclusive $N$-jet cross sections in a manner that is particularly suitable for higher-order analytic resummation. The calculation of the one-loop soft function for exclusive $N$-jet processes using $N$-jettiness has been carried out for arbitrary $N$ in ref.~\cite{Jouttenus:2011wh}. There, $N$-jettiness is used both as the algorithm to partition the phase space into jet and beam regions and as the measurement performed on those regions. To simplify the calculation, the version of $N$-jettiness used in ref.~\cite{Jouttenus:2011wh} was taken to be linear in the constituent four-momenta $p^\mu_i$,
%%%
\begin{align}\label{eq:Nj}
\text{thrust-like $N$-jettiness:} \qquad
\Tau_N &=\sum_i\min_m\left\{\frac{2q_m\cdot p_i}{Q_m}\right\}
 = \sum_i \min_m \left\{\frac{n_m\cdot p_i}{\rho_m}\right\}
\,.\end{align}
%%%
This is essentially a generalization of beam thrust~\cite{Stewart:2009yx} to the case of $N$ jets. In \eq{Nj} the sum runs over the four-momenta $p^\mu_i$ of all particles that are part of the hadronic final state, and the minimization over $m$ runs over the beams and $N$ jets identified by the reference momenta $q^\mu_m=E_m n_m^\mu$ or lightlike vectors $n_m^\mu= (1,\hat n_m)$, where $E_m$ is the jet energy. The directions $\hat n_m$ for the beams are fixed along the beam axis and for the jets are predetermined by a suitable procedure.  Finally, the $Q_m$ or $\rho_m=Q_m/(2E_m)$ are dimension-one or dimension-zero measure factors. The minimization in \eq{Nj} assigns each particle to one of the axes, thus partitioning the phase space into $N$ jet regions and $2$ beam regions.
This definition of $N$-jettiness depends only on the choices of jet directions $\hat n_m$ and measure factors $\rho_m$, which determine the precise partitioning and in particular the size of the jet and beam regions.
For the cross section with a measurement of $\Tau_N$, the $\Tau_N\to 0$ singular region is fully described by a factorization formula of the form in \eq{simpleFact} with inclusive jet and beam functions~\cite{Stewart:2009yx, Stewart:2010tn}. As $\Tau_N\to 0$, different choices of jet axes often differ only by power-suppressed effects in the cross section.

$N$-jettiness can also be used more generally as a means of defining an exclusive jet algorithm, which partitions the particles in an event into a beam region and a fixed number of $N$ jet regions~\cite{Stewart:2010tn, Thaler:2011gf}.  Here particle $i$ is assigned to region $m$ for which some generic distance measure $d_m(p_i)$ is minimal.  These regions are defined by
%%%
\begin{equation} \label{eq:partitioning}
  \text{region}\ m =  \bigl\{ \text{particles}\
   i:\ \text{where}\ d_m(p_i) < d_j(p_i)\ \text{for all}\ j\ne m \bigr\}
\,.\end{equation}
%%%
This partitioning can be obtained from a generalized version of $N$-jettiness defined by
%%%
\begin{equation} \label{eq:Nj2}
  \Tau_N(\{\hat n_m\}) =\sum_i p_{Ti} \min \bigl\{ d_1(p_i) , \ldots, d_N(p_i) , d_a(p_i), d_b(p_i) \bigr\}
\,.\end{equation}
%%%
Here the $d_m$ jet measures depend on pre-defined jet axis $\hat n_m$, while the beam measures $d_a$ and $d_b$
are defined with fixed beam axes along $\pm \hat z$.
Infrared safety requires that all particles in the vicinity of the axis $n_m^\mu =(1,\hat n_m)$ are assigned to the respective $m$th region. More precisely the measures have to satisfy $d_m(p_i) <d_j(p_i)$ for all $j \neq m$ in the limit $p^\mu_i \to E_i n^\mu_m$.
Different choices of the $d_m$ correspond to different $N$-jettiness partitionings, and include for example the Geometric, Conical, and XCone measures~\cite{Stewart:2010tn, Jouttenus:2011wh, Jouttenus:2013hs, Thaler:2010tr, Stewart:2015waa}. The measure in \eq{Nj} corresponds to taking $p_{Ti} d_m(p_i) = (n_m\cdot p_i)/\rho_m$.  The two beam regions can be combined into a single one by defining the common beam measure
%%%
\begin{equation}
  d_0(p_i) = \min \{ d_a(p_i), d_b(p_i) \}
\,.\end{equation}
%%%
Given a common beam region with a single beam measure $d_0(p_i)$, we can always divide it into two separate beam regions for $\eta>0$ and $\eta<0$ by taking for example $d_a(p_i)=[1+\theta(-\eta_i)] d_0(p_i)$ and $d_b(p_i)=[1+\theta(\eta_i)] d_0(p_i)$.

Constructing a full jet algorithm requires in addition to the partitioning an infrared-safe method to determine the jet axes $\hat n_m$. This could be done by simply taking the directions of the $N$ hardest jets obtained from a different (inclusive) jet algorithm. For a standalone $N$-jettiness based jet algorithm, the axes can be obtained by minimizing $N$-jettiness itself over all possible axes,
%%%
\begin{equation} \label{eq:minNj2}
  \Tau_N = \min_{\hat n_1, \ldots, \hat n_N} \Tau_N(\{\hat n_m\})
\,,\end{equation}
%%%
as in refs.~\cite{Thaler:2011gf, Stewart:2015waa}.

For the calculations in this paper, we consider a very general set of distance measures for determining the partitioning into jet and beam regions as in \eq{Nj2}, and a \emph{different} set of fairly general infrared safe observables measured on these regions. We explore and compare properties of different jet partitionings in \sec{jetregions}.  For the measured observables we consider the generic version of $N$-jettiness variables, $\Taum{m}$, given by
%%%
\begin{equation} \label{eq:Tauidef0}
 \Taum{m} = \sum_{i \in \, {\rm region} \,m}f_m(\eta_i,\phi_i) \,p_{Ti}
\,.\end{equation}
%%%
Here, $\eta_i$, $\phi_i$, and $p_{Ti}$ denote the pseudorapidity, azimuthal angle, and transverse momentum
of particle $i$ in region $m$. The dimensionless functions $f_m$ encode the angular dependence of the observable and in the collinear limit behave like an angularity, see \sec{jetmeasure}. When considering a single beam region we have a  common beam measurement $\Taum{0} = \Taum{a} + \Taum{b}$. Earlier analytic calculations of $N$-jettiness cross sections have all been done for the case where the observable and partitioning measure coincide, $f_m = d_m$, in which case the total $N$-jettiness used for the partitioning is equal to the sum over the individual measurements $\Tau_N= \sum_m \Taum{m}$.

The exact definition of the axes $\hat n_m$ is irrelevant for the calculation of the soft function. For our purposes we can therefore separate the jet-axes finding from the partitioning and measurement, and we will assume predetermined axes obtained from a suitable algorithm. However, one should make sure to use recoil-free axes~\cite{Larkoski:2014uqa} for angularities to avoid \SCETb-type perpendicular momentum convolutions between soft and jet functions. This is ensured if one defines the axes through a global minimization as in \eq{minNj2}.

In this paper, we determine factorization theorems, which describe the singular perturbative contributions in the $\Tau_N\to 0$ limit for these generic versions of $N$-jettiness.  We then establish a generalized hemisphere decomposition for computing the corresponding one-loop soft function. We carry out the computations explicitly for a number of interesting cases.
As underlying hard process we consider color-singlet plus jet production, and we discuss results for generic angularities as jet measurements. For the beam measurement we discuss different types of jet vetoes, including beam thrust, beam $C$ parameter, and a jet-$p_T$ veto. We also discuss different partitionings, including anti-$k_T$~\cite{Cacciari:2008gp} and XCone~\cite{Stewart:2015waa,Thaler:2015xaa}.
We find that the one-loop soft function can be written in terms of universal analytic contributions and a set of numerical integrals, which explicitly depend on the partitioning and observable (i.e.\ the specific definitions of the $d_m$ and $f_m$). We show that fully analytical results can be obtained in the limit of small jet radius $R$. Furthermore, we show that the small-$R$ expansion works remarkably well for the soft function even for moderate values of $R$, if one includes corrections up to $\ord{R^2}$.

The rest of the paper is organized as follows. In \sec{genTau}, we discuss in more detail the generalized definition of $N$-jettiness, jet algorithms, and relevant factorization theorems. In \sec{GenHemiDecomp}, we discuss the generalized hemisphere decomposition to calculate the one-loop soft function. In \sec{OneJetCase}, we discuss the explicit results for the case of single-jet production. We conclude in \sec{conclusions}. Details of the calculations are given in \app{anal_soft_pieces} and \app{num_evaluation}, and results for dijet production are discussed in \app{dijets}.

%%%%%%%%%%%%%%%%%%%%%%%%%%%%%%%%%%%%%%%%%%%%%%%%%%%%%%%%%%%%%%%%%%%%%%%%%%%%%%%%
\section{Jet measurements and jet algorithms}
\label{sec:genTau}
%%%%%%%%%%%%%%%%%%%%%%%%%%%%%%%%%%%%%%%%%%%%%%%%%%%%%%%%%%%%%%%%%%%%%%%%%%%%%%%%

In this section, we discuss the general properties we assume for the jet measurements and for the jet algorithms (partitioning). We consider the cross section for events with at least $N$ hard jets in the final state with transverse momenta $p^J_{T,m\geq1} \sim p_T^J \sim Q$, where $Q$ denotes the center-of-mass energy of the hard process.
In \sec{jetmeasure} we define the generalized form of $N$-jettiness measurements, in \sec{jetregions} we discuss and compare different jet algorithms, and in \sec{factorization} we present the form of the factorization theorems for different choices of jet and beam measurements.

%===============================================================================
\subsection{Generalized $N$-jettiness measurements}
\label{sec:jetmeasure}
%===============================================================================

Assuming a partitioning of the phase space into $N$ jet regions ($m=1,\dots,N$) and two beam regions ($m=a,b$), the observable that we will study is defined in each region $m$ by the sum over all particle momenta (but excluding the color-singlet final state),\footnote{We consider only cases without unconstrained phase space domains, i.e.~no regions with nonzero area in $(\eta,\phi)$ coordinates where $f_m=0$.}
%%%
\begin{equation} \label{eq:Tauidef}
\Taum{m} = \sum_{i \in \, {\rm region} \,m}\Taum{m}(p_i)
\qquad\text{with}\qquad
\Taum{m}(p_i) = f_m(\eta_i,\phi_i) \,p_{Ti}
\,.\end{equation}
%%%
Here $\eta_i$ and $\phi_i$ denote the pseudorapidity and azimuthal angle of the particle $i$.
The associated jet and beam axes are normalized lightlike directions, and are given in terms of these coordinates by
%%%
\begin{equation}
n^\mu_{m\geq1} = \frac{1}{\cosh \eta_{m}} \Bigl(\cosh \eta_{m}, \cos \phi_{m}, \sin \phi_{m}, \sinh \eta_{m}\Bigr)
\,,\qquad
n^\mu_{a,b} = (1,0,0,\pm1)
\, .\end{equation}
%%%
The $f_m$ in \eq{Tauidef} are dimensionless functions encoding the angular dependence of the observable. To satisfy infrared safety, we require that $\Taum{m}\to 0$ for soft and $n_m$-collinear emissions, implying in particular that
%%%
\begin{equation}
\lim_{\eta_i \to \infty }  f_a(\eta_i,\phi_i) e^{-\eta_i} = 0
\,,\quad
\lim_{\eta_i \to -\infty }  f_b(\eta_i,\phi_i) e^{\eta_i} = 0
\,,\quad
\lim_{\eta_i\to \eta_m, \phi_i \to \phi_m} f_{m\geq1}(\eta_i,\phi_i) =0
\,.\end{equation}
%%%
For definiteness we will consider the case that the asymptotic behavior of $\Taum{m}$ in the vicinity of its axis is given by an angularity measurement, which holds for all common single-differential observables, i.e.,
%%%
\begin{equation}\label{eq:Taum_coll}
\Taum{m}(p_i) \stackrel{p^\mu_i \to E_i n^\mu_m}{\longrightarrow} c_m\, \bigl(n_m \cdot p_i\bigr)^{\frac{\beta_m}{2}} \bigl(\bar{n}_m \cdot p_i\bigr)^{1-\frac{\beta_m}{2}}
\,,\end{equation}
%%%
with $\beta_m>0$ and some normalization factors $c_m$.
Defining $\gamma \equiv \beta_a = \beta_b$, this is equivalent to 
%%%
\begin{align}\label{eq:Tau0_coll}
f_a(\eta_i,\phi_i) &\stackrel{\eta_i \to \infty}{\longrightarrow} c_a  \,e^{(1-\gamma)\eta_i}
\,,\qquad
f_b(\eta_i,\phi_i) \stackrel{\eta_i \to -\infty}{\longrightarrow} c_b  \,e^{-(1-\gamma)\eta_i}
\, , \nn \\
f_{m\geq1}(\eta_i,\phi_i) &\stackrel{(\eta_i,\phi_i)\to (\eta_m, \phi_m)}{\longrightarrow} c_m\,(2\cosh \eta_m)^{1-\beta_m} \Bigl[(\eta_i-\eta_{m})^2 + (\phi_i-\phi_{m})^2\Bigr]^{\frac{\beta_m}{2}}
\,.\end{align}
%%%
We will discuss several examples in \secs{GenHemiDecomp}{OneJetCase}. The behavior of $f_m$ determines whether the associated collinear and soft sectors are described by a \SCETa-type or \SCETb-type theory. The case $\gamma=\beta_{m} =2$ corresponds to the standard \SCETa situation with a thrust-like measurement $\Taum{m}(p_i) \sim n_m \cdot p_i$.

%===============================================================================
\subsection{Jet algorithms}
\label{sec:jetregions}
%===============================================================================

Given a set of jet and beam axes $\{n_m\}$, the partitioning of the phase space into jet and beam regions
is determined by the distance measures $d_m(p_i)$. As shown in \eq{partitioning},
particle $i$ is assigned to region $m$ if $d_m(p_i) < d_j(p_i)$ for all $j\neq m$, i.e., when it is
closest to the $m$th axis.

For $m\geq1$, the distance measures $d_m(p_i)\equiv d_m(R,n_{m},p_{T,m}^J,\eta_i,\phi_i)$ can depend on the jet size parameter $R$ and the jet transverse momentum $p_{T,m}^J$. In \sec{factorization}, we will show that for $\Tau_N \ll p_T^J$ and for well-separated jets and beams and sufficiently large jet radii, the differential cross section in the $\Taum{m}$ can be factorized into hard, collinear, and soft contributions. This requires a jet algorithm which exhibits soft-collinear factorization, such that $m$-collinear emissions are sufficiently collimated to not be affected by different distance measures $d_{j\neq m}$ and do not play a role for the partitioning of the event. Furthermore, the recoil on the location of the jet axes due to soft emissions is power suppressed for the description of the soft dynamics.\footnote{Note that for angularities with $\beta_m  \leq 1$ the recoil due to soft radiation does matter for the description of the collinear dynamics~\cite{Larkoski:2014uqa}.} Thus the partitioning of soft radiation in the event can be obtained by comparing the distance measures $d_m$ for soft emissions with respect to $N+2$ fixed collinear directions independently of the axes finding and the jet and beam measurements.

\begin{figure}
\centering
\includegraphics[height=5cm]{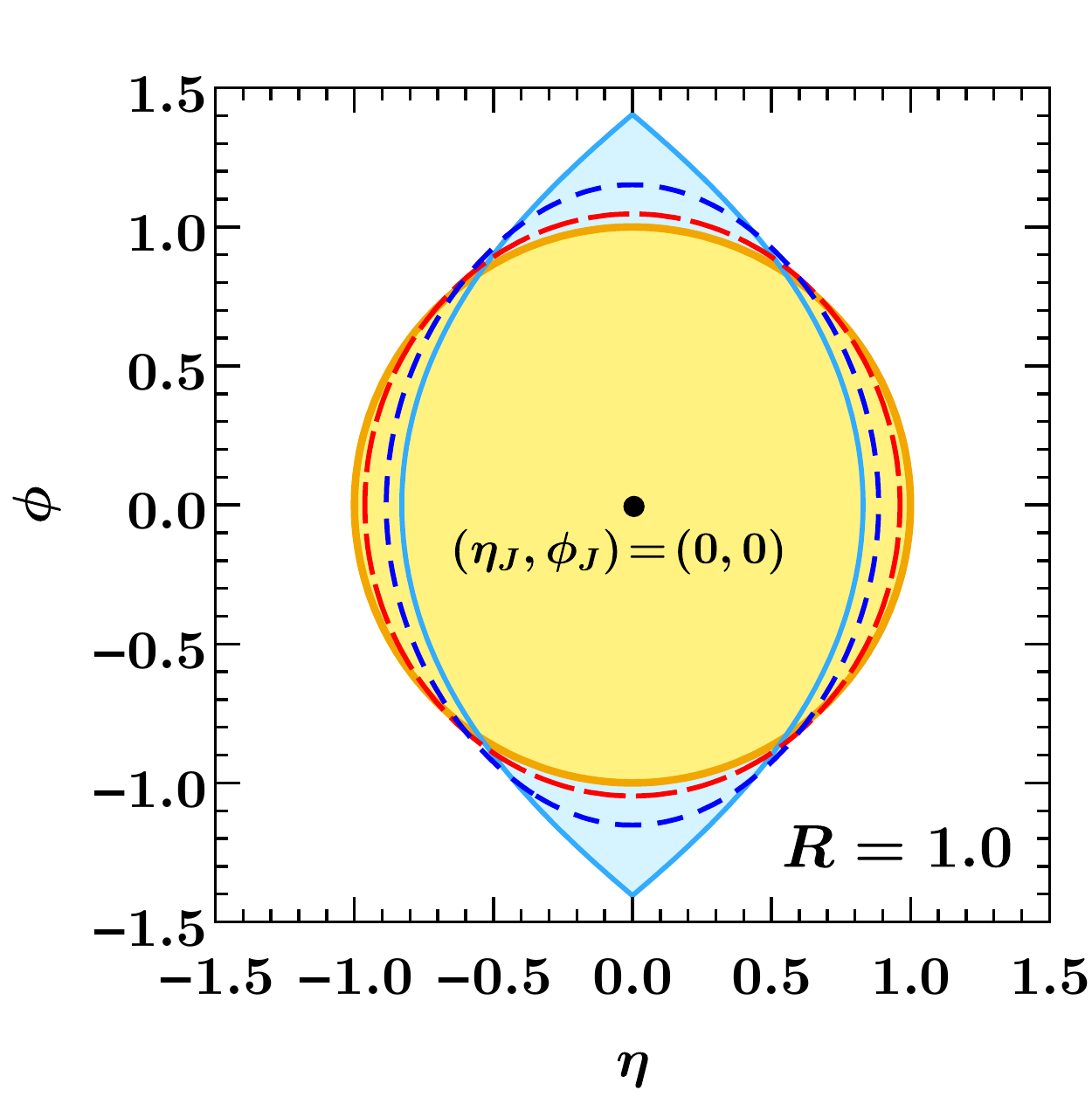}%
\hfill%
\includegraphics[height=5cm]{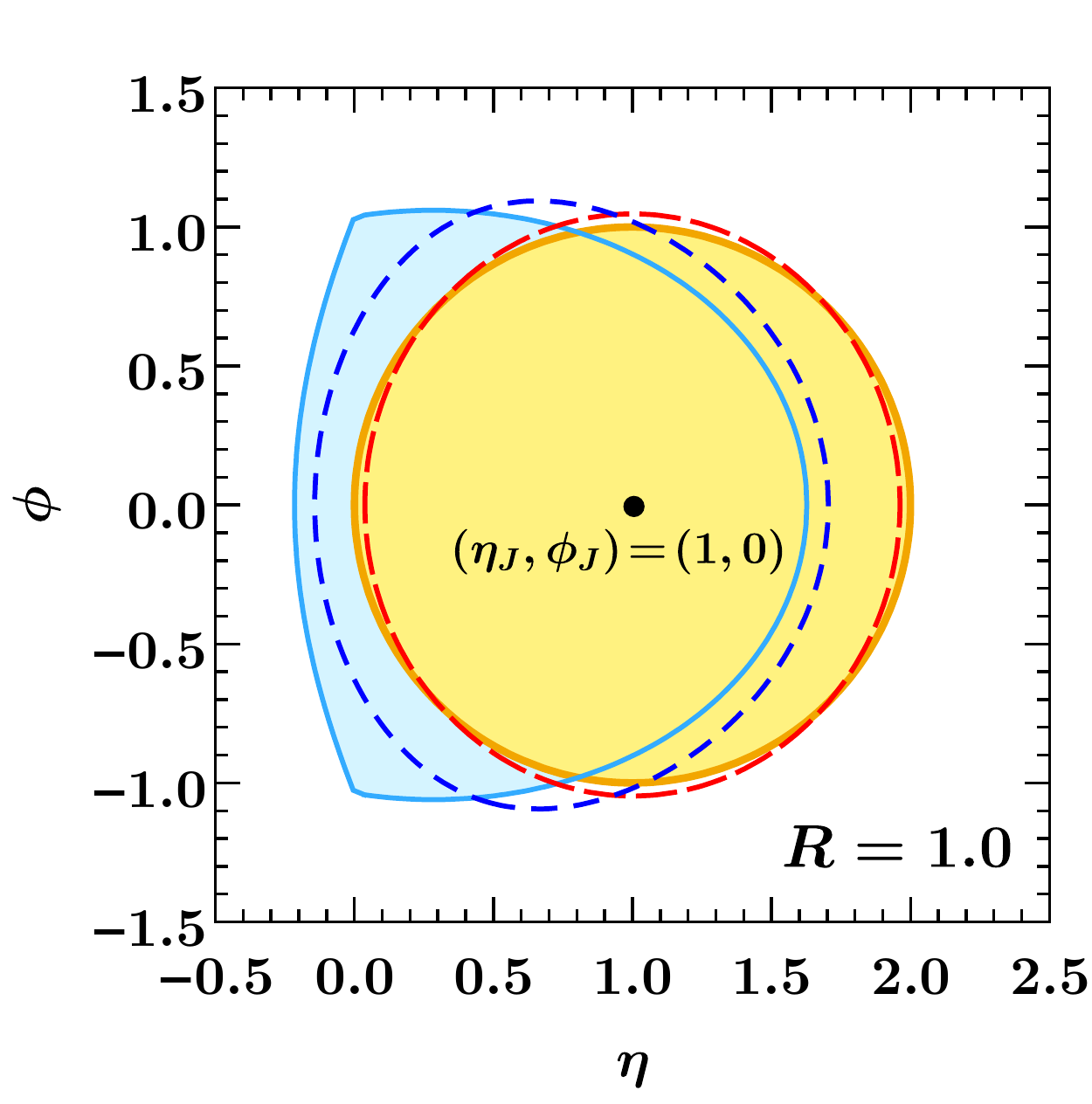}%
\hfill%
\includegraphics[height=5cm]{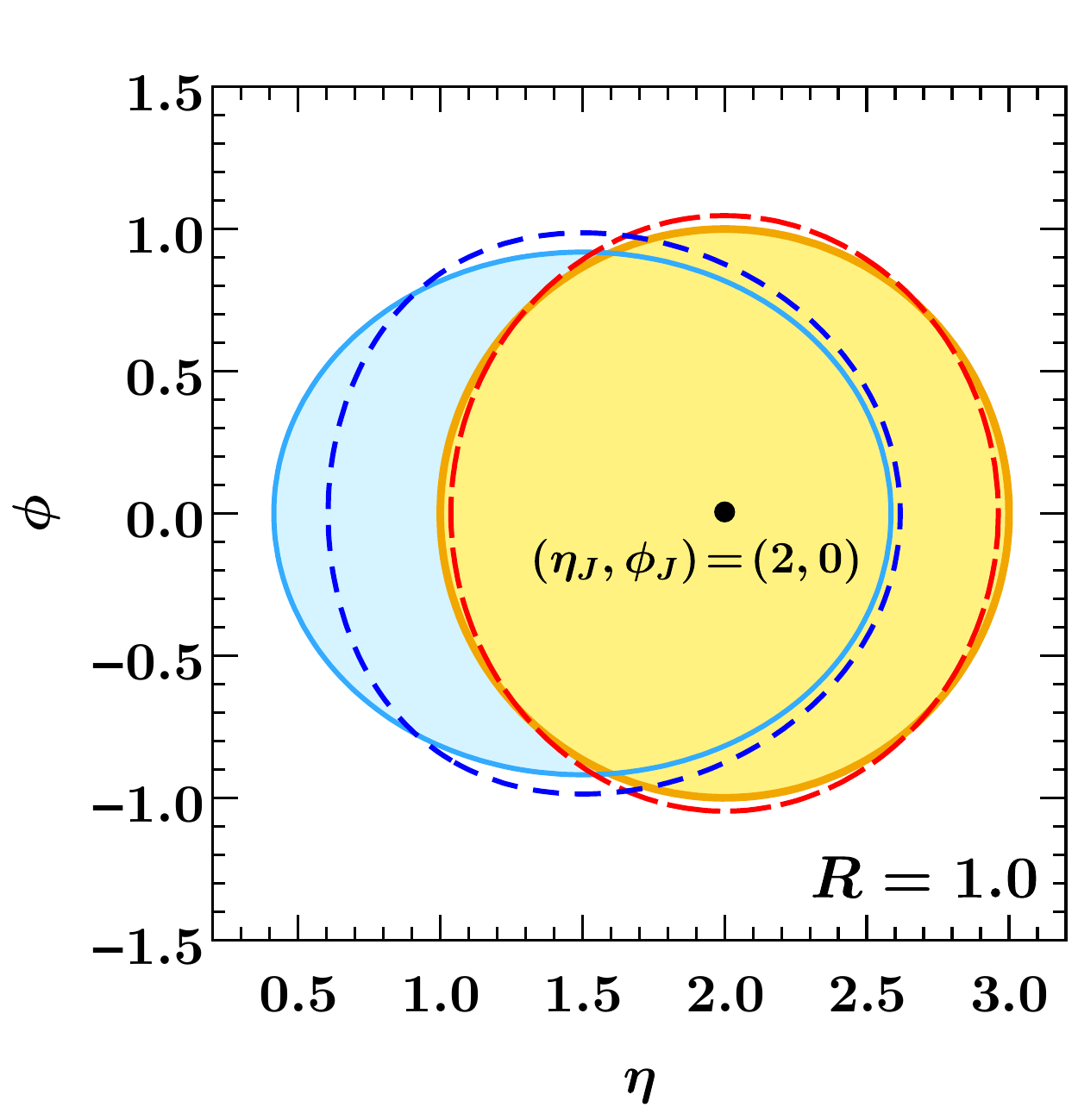}%
\\[0.5ex]
\hspace{0.25cm}\includegraphics[height=5cm]{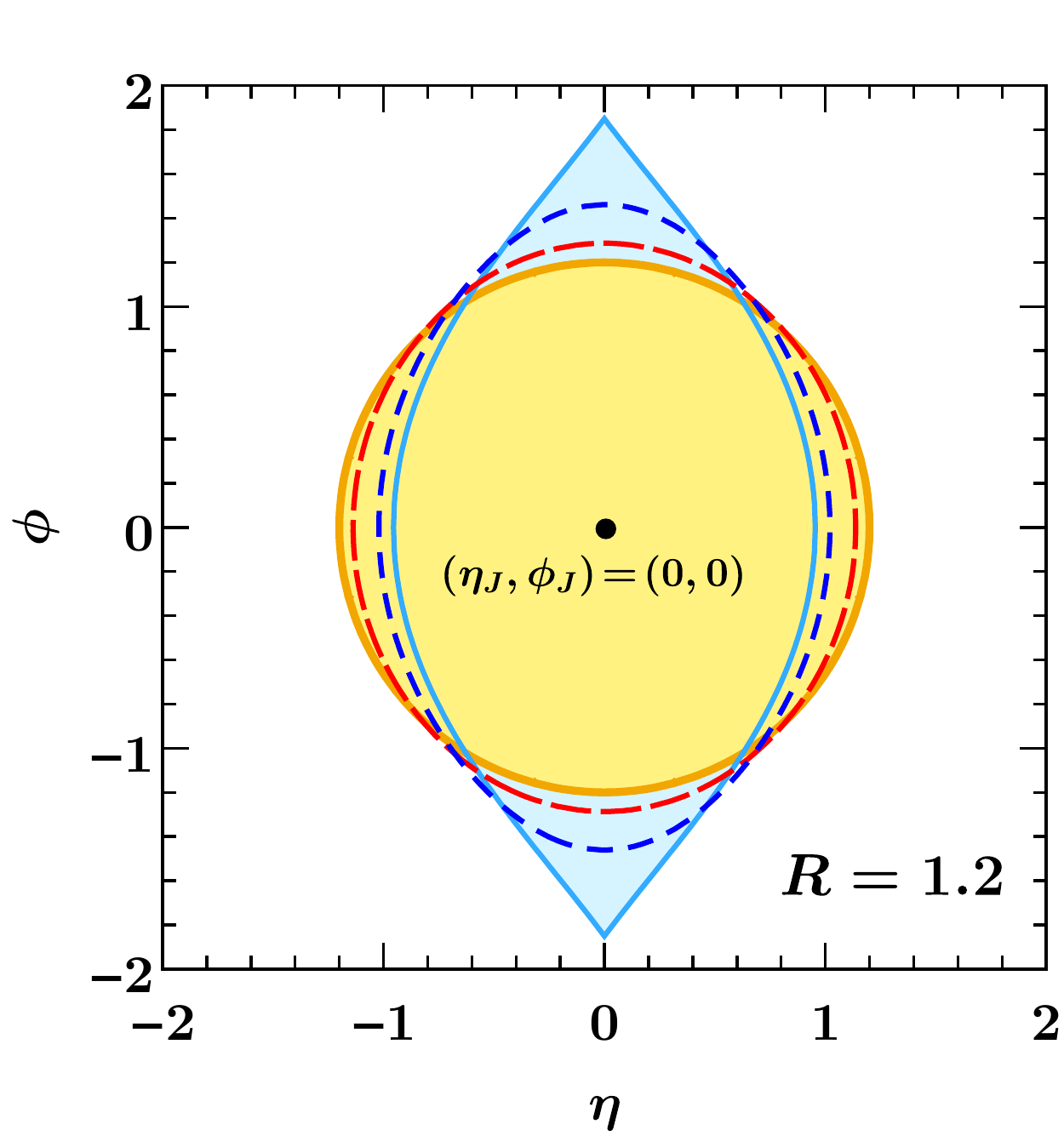}%
\hfill%
\includegraphics[height=5cm]{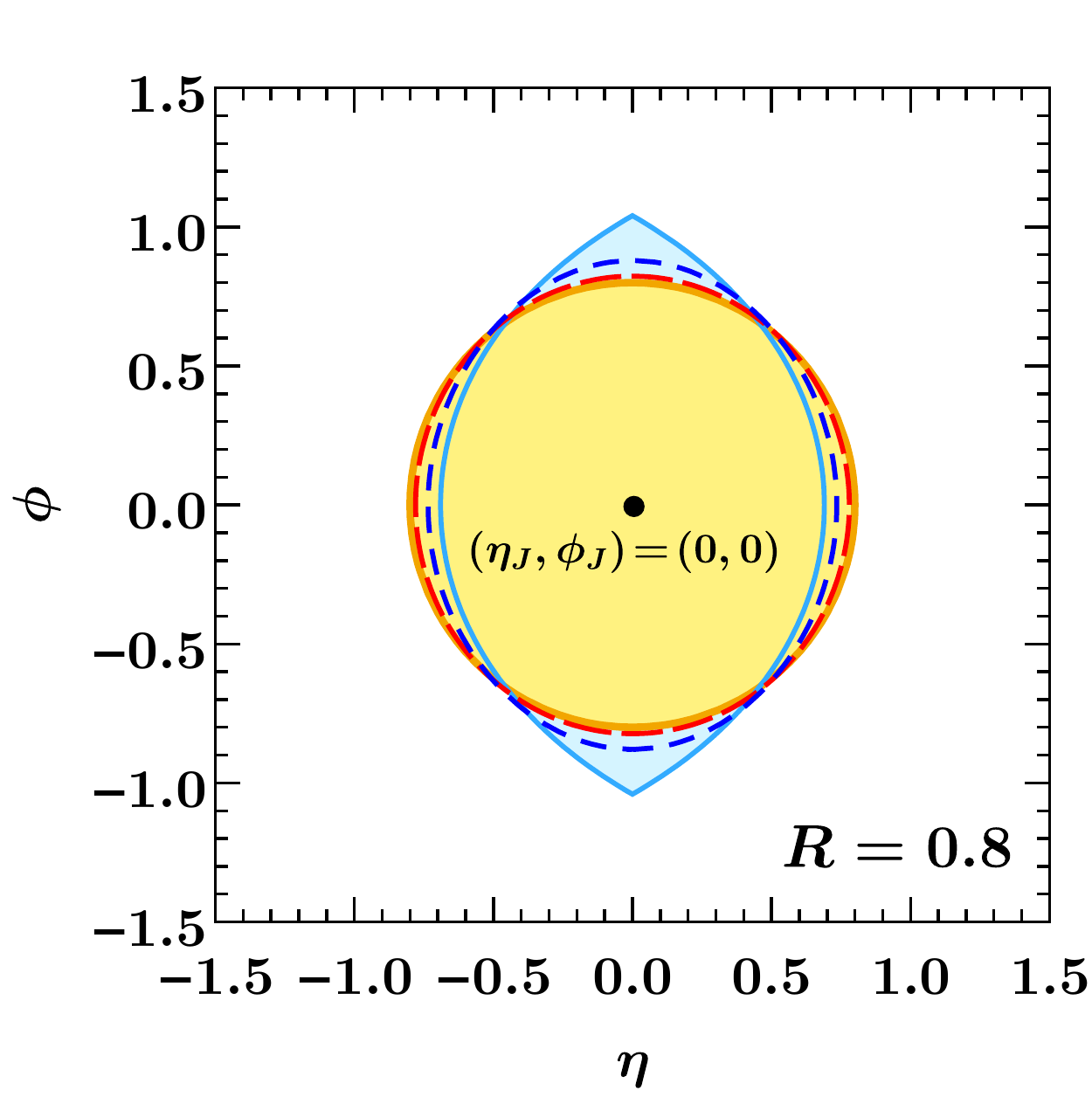}%
\hfill%
\includegraphics[height=5cm]{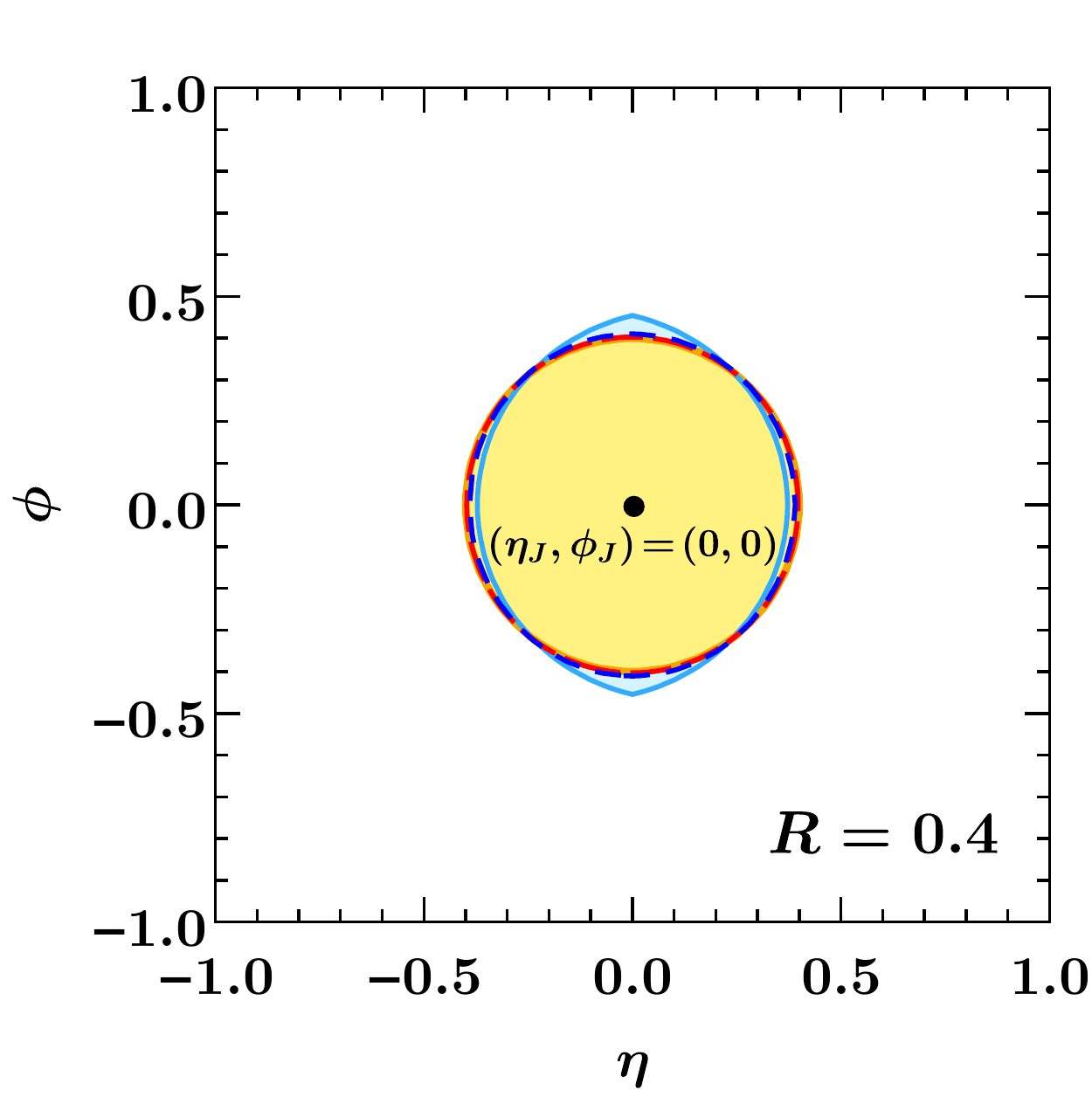}%
\caption{Jet regions (in the limit $\Tau_N \ll p_T^J$) in the $\eta$-$\phi$-plane for different partitionings
for $R=1$ and different $\eta_m=0,1,2$ (top row) and $\eta_m = 0$ and different $R=1.2,0.8,0.4$ (bottom row).
The conical measure, which is equivalent to anti-$k_T$, is shown in yellow, the geometric-$R$ measure in light blue, the modified geometric-$R$ in blue dashed, and the conical geometric measure (XCone default) in red dashed.
\label{fig:jetregions}}
\end{figure}

We consider the following examples of partitionings for comparisons of numerical results:
\begin{itemize}
\item[I:] Conical Measure (equivalent to anti-$k_T$ for isolated jets)~\cite{Thaler:2011gf}:
\begin{align}\label{eq:d_Conical}
d_0(p_i)=1
\,, \qquad
d_{m\ge 1}(p_i)= \frac{R_{im}^2}{R^{2}}
\, .
\end{align}
\item[II:] Geometric-$R$ Measure~\cite{Jouttenus:2013hs}: 
\begin{align}\label{eq:d_GeometricR}
d_0(p_i)= e^{-|\eta_i|}
\,, \qquad
d_{m\ge 1}(p_i)=\frac{n_m \cdot p_i}{\rho_\tau(R,\eta_{m}) \,p_{T i}}= \frac{1}{\rho_\tau(R,\eta_{m})}\,\frac{\mathcal{R}^2_{im}}{2\cosh \eta_{m}}
\, .
\end{align}
\item[III:] Modified Geometric-$R$ Measure~\cite{Stewart:2015waa}: 
\begin{align}\label{eq:d_ModGeometricR}
d_0 (p_i)=  \frac{1}{2 \cosh\eta_i}
\,, \qquad
d_{m\ge 1}(p_i)=\frac{n_m \cdot p_i}{\rho_C(R,\eta_{m})\,p_{T i}}= \frac{1}{\rho_C(R,\eta_{m})}\,\frac{\mathcal{R}^2_{im}}{2\cosh \eta_{m}}
\, .
\end{align}
\item[IV:] Conical Geometric Measure (XCone default)~\cite{Stewart:2015waa}:
\begin{align}\label{eq:d_ConicalGeometric}
d_0 (p_i)=1
\,,\qquad
d_{m\ge 1} (p_i)= \frac{2 \cosh \eta_m  (n_m \cdot p_i)}{R^2 \,p_{T i}}=\frac{\mathcal{R}_{im}^2 }{R^{2}}
\,,
\end{align}
\end{itemize}
where $\rho_\tau$ and $\rho_C$ are discussed below, and the distances in azimuthal angle and rapidity are given by
%%%
\begin{align}\label{eq:DeltaRdef}
\Ra_{im} & \equiv \sqrt{(\eta_i -\eta_{m})^2 +(\phi_i-\phi_{m})^2}\,, \nn \\
 \Rb_{im} & \equiv \sqrt{2 \cosh(\eta_i -\eta_{m}) - 2 \cos(\phi_i-\phi_{m})} \, .
\end{align}
%%%
Since these measures only depend on $\eta_i$ and $\phi_i$, we can obtain explicit jet regions in the $\eta$-$\phi$ plane.
The jet regions for an isolated jet with $R=1$ at different jet rapidities and
different $R$ at central rapidity are shown in \fig{jetregions}.
For small $R$ all distance metrics approach a conical partitioning,
which means in particular that the deviations from this shape are suppressed by powers of $R$. 

For isolated jets the conical distance measure includes all soft radiation within a distance $R$ in $\eta$-$\phi$ coordinates from the jet axis into the jet. Thus, in this case the soft partitioning is equivalent to the one obtained in the anti-$k_T$ algorithm~\cite{Cacciari:2008gp}, which first clusters collinear energetic radiation before clustering soft emissions into the jets (allowing thus for soft-collinear factorization~\cite{Walsh:2011fz}). As explained above, the algorithm for the jet-axes finding is irrelevant for the description of the soft dynamics and the soft function depends only on the soft partitioning with respect to fixed collinear axes. Thus, the soft function for anti-$k_T$ jets and $N$-jettiness jets with the conical measure are identical for isolated jets.

\begin{figure}[t!]
\centering
\hfill
\includegraphics[height=6cm]{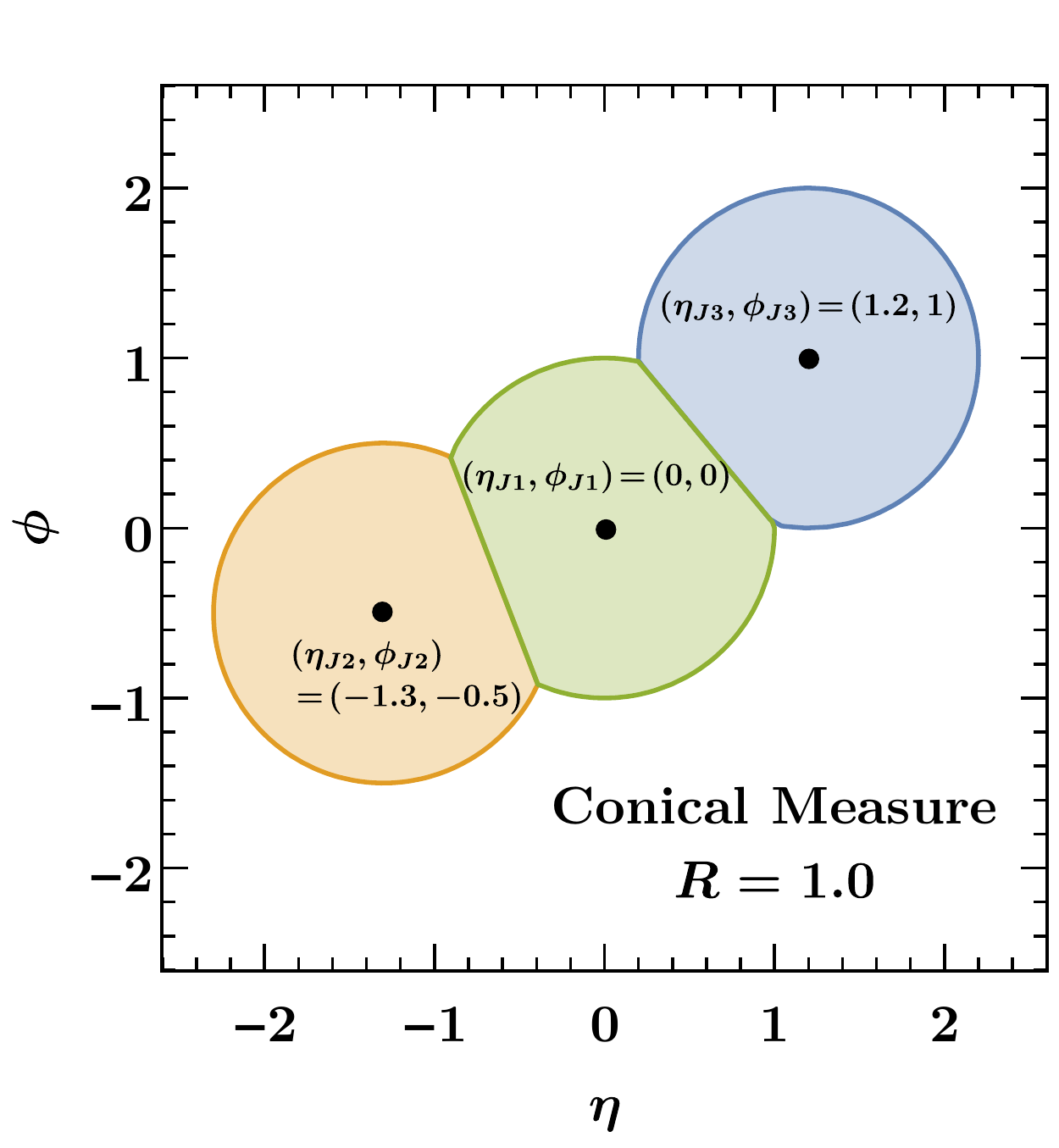}%
\hfill%
\includegraphics[height=6cm]{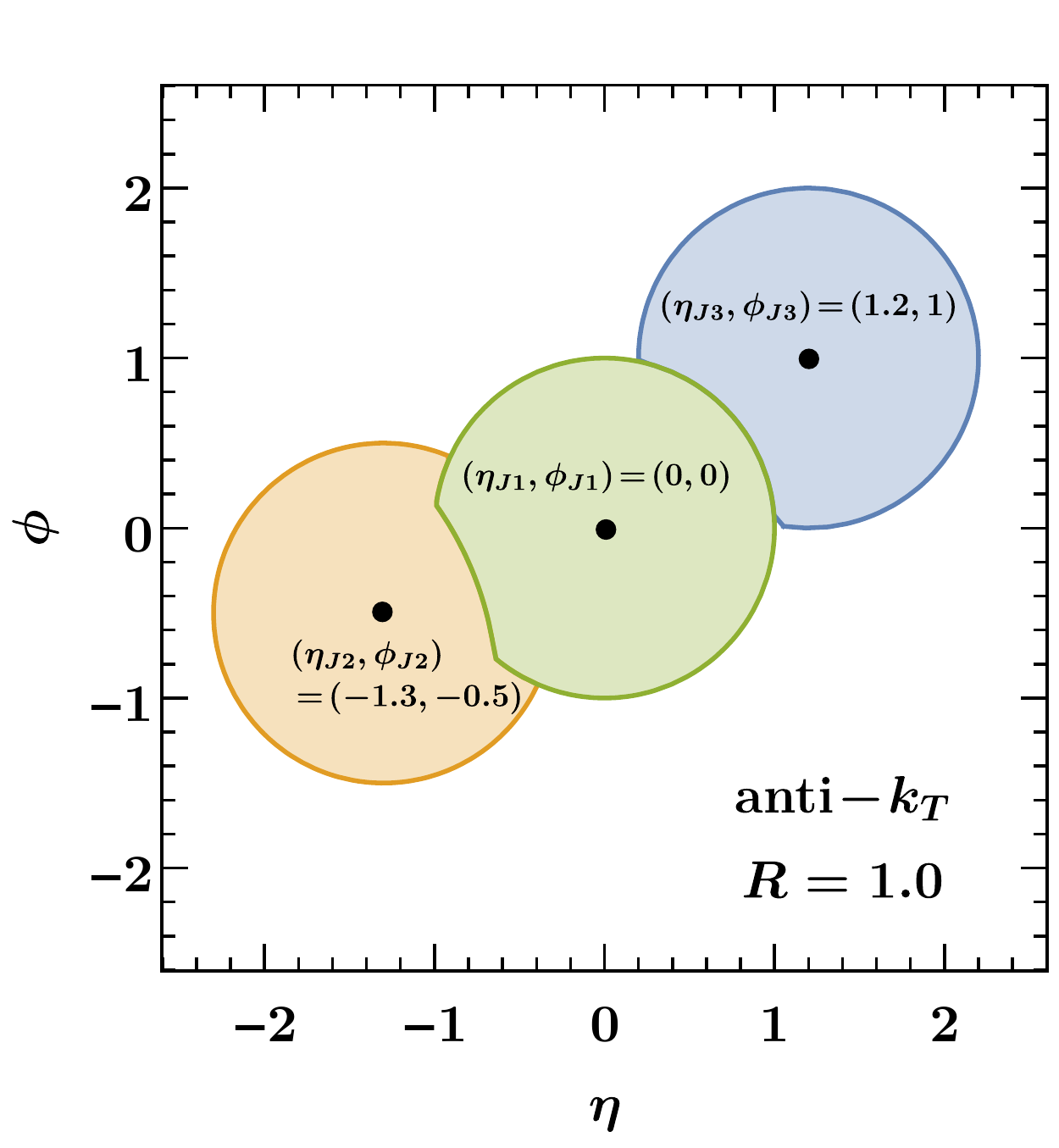}%
\hspace*{\fill}%
\caption{Partitioning (in the limit $\Tau_N \ll p_T^J$) for three overlapping jets with $p^J_{T,1}=2p^J_{T,2} = 4 p^J_{T,3}$ and $R=1$ with distance $>R$ between their axes. The $N$-jettiness partitioning with the conical distance measure is shown on the left and the anti-$k_T$ partitioning on the right.}
\label{fig:ThreeJetAlgorithms}
\end{figure}

For overlapping jets, the anti-$k_T$ and $N$-jettiness partitionings differ.
The distance metrics in the anti-$k_T$ algorithm between soft and the clustered collinear radiation depend also on the transverse momenta of the jets, which starts to matter in the singular region $\Tau_N \ll p_T^J$ once two jets start to overlap, i.e.~for $R_{lm} < 2R$. In this case, anti-$k_T$ assigns soft radiation in the overlap region to the more energetic jet, while the $N$-jettiness partitioning remains purely geometric. This is illustrated in \fig{ThreeJetAlgorithms}, for three jets with different transverse momenta that share common jet boundaries. When the distance between two clusters of energetic collinear radiation drops below $R$, anti-$k_T$ clustering will merge these into a single jet, while the $N$-jettiness partitioning still gives two closeby jets, thus exhibiting a very different behavior.

\begin{figure}
\centering
\hfill%
\includegraphics[height=5.5cm]{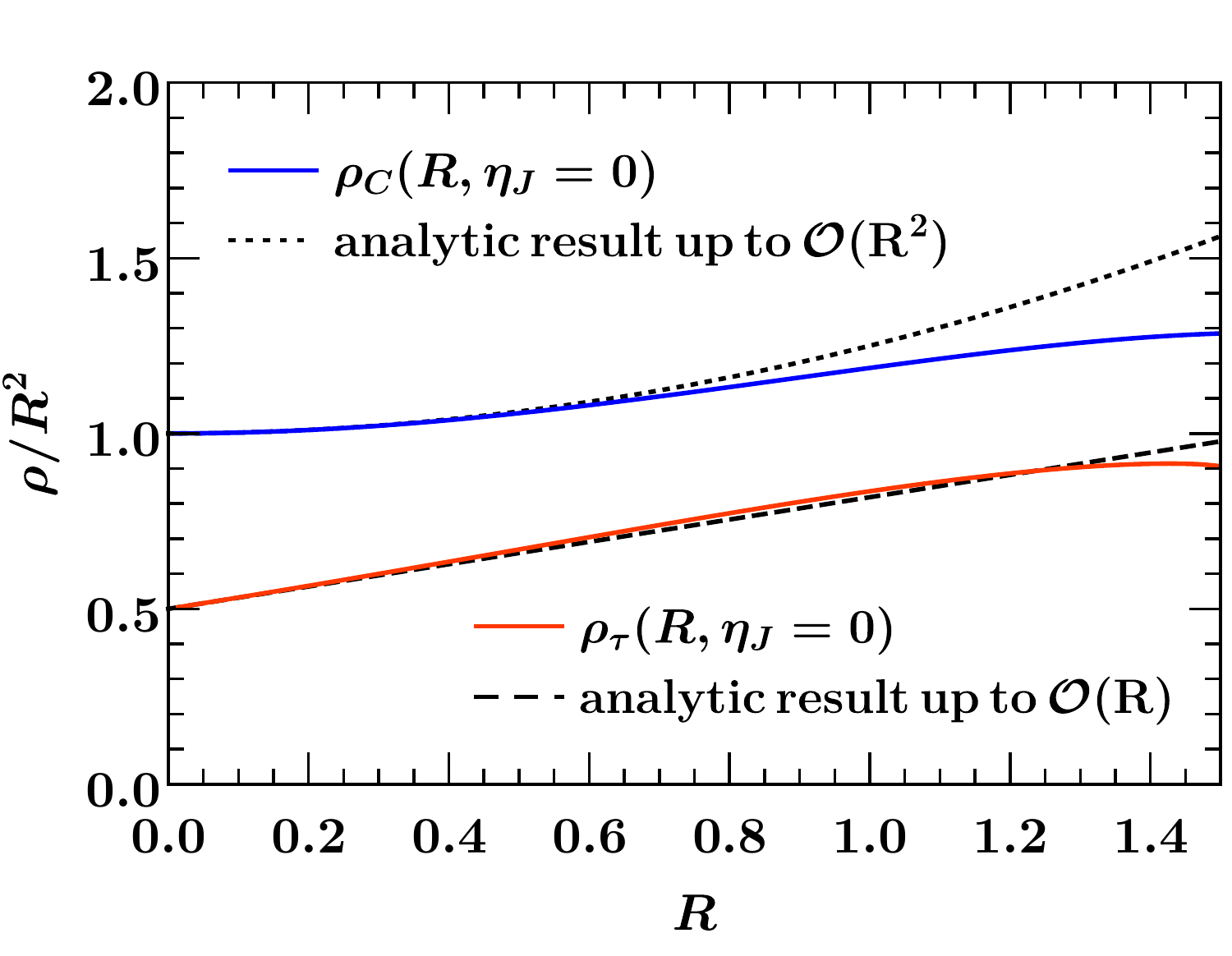}%
\hfill%
\hspace{0.5cm}
\includegraphics[height=5.5cm]{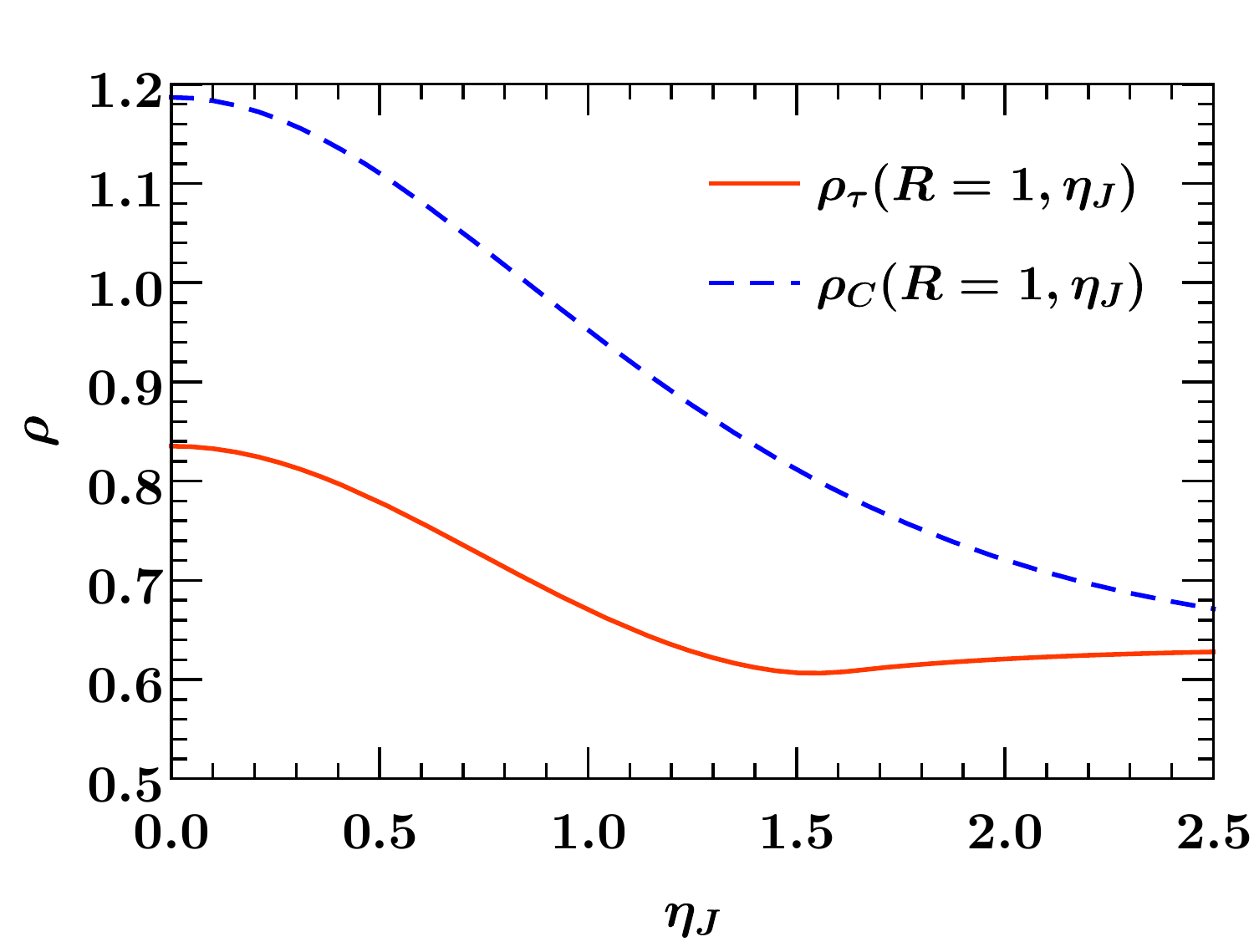}%
\hspace*{\fill}
\caption{Behavior of $\rho_\tau(R,\eta_J)$ and $\rho_C(R,\eta_J)$ for the geometric-$R$ and modified geometric-$R$ measures as functions of $R$ at $\eta_J=0$ (left panel) and of $\eta_J$ for $R=1$ (right panel).
\label{fig:rho}}
\end{figure}

The (modified) geometric-$R$ measures in \eqs{d_GeometricR}{d_ModGeometricR} have the feature that $p_{Ti} d_m(p_i) \sim n_m\cdot p_i$ is linear in the particle momenta $p_i$, as for the pure geometric measure in \eq{Nj} from which they are derived. The geometric-$R$ measure was first used in ref.~\cite{Jouttenus:2013hs} to study the jet mass for $pp \to H +1$ jet, taking advantage of the fact that the soft function for this type of measure was computed in ref.~\cite{Jouttenus:2011wh}. The parameters $\rho_\tau(R,\eta_{m})$ and $\rho_C(R,\eta_{m})$ are determined by requiring the area in the $\eta$-$\phi$-plane for an isolated jet with rapidity $\eta_m$ to be $\pi R^2$, i.e.\ by solving 
%%%
\begin{equation} \label{eq:rhoRdef}
\int_{-\pi}^\pi \! \df\phi \int_{-\infty}^\infty\!\! \df \eta\,
\theta\bigl[d_0(\eta) - d_m(\rho,\eta_m,\eta,\phi)\bigr] =  \pi R^2
\, .
\end{equation}
The solution for $\rho$ in terms of $\eta_m$ and $R$ can be computed analytically in an expansion for small $R$, which gives 
%%%
\begin{align}
\rho_\tau(R,\eta_m) & = R^2 \frac{1\!+\!\tanh|\eta_m|}{2} \biggl\{1+ \frac{2 R}{\pi}\, \theta(R-|\eta_m|) \biggl[\sqrt{1\!-\!\frac{\eta_m^2}{R^2}} -\frac{|\eta_m|}{R} \arccos\frac{\eta_m}{R} \biggr] \!+ \mathcal{O}(R^2)\biggr\}
, \nn \\
\rho_C(R,\eta_m) & = R^2 \biggl\{1+\frac{R^2}{4} \bigl(1-3\tanh^2\! \eta_m\bigr) + \mathcal{O}(R^4)\biggr\}
\,.
\end{align}
%%%
Note that the kink at $\eta_m=0$ leads to $\mathcal{O}(R)$ corrections for $\rho_\tau$ for $|\eta_m|<R$. The full $R$ dependence is obtained numerically. In \fig{rho}, we show $\rho_\tau$ and $\rho_C$ as functions of $R$
for $\eta_m=0$ and as functions of $\eta_m$ for $R=1$.

Compared to the conical measure the shapes of the jet regions are more irregular for the geometric-$R$ measures, as seen in \fig{jetregions}. In particular the beam thrust measure in \eq{d_GeometricR} has a cusp at $\eta =0$ due to the absolute value in the beam distance measure, which is not present for the smooth beam C-parameter measure in \eq{d_ModGeometricR}. Furthermore, we also see a distortion from the circular shape for large jet rapidities towards an elongated shape, which is common to both measures since their beam distance measures become identical in the forward region.

Finally, the conical geometric measure was introduced in ref.~\cite{Stewart:2015waa} and corresponds to the XCone default measure. It is designed to combine the linear dependence of $p_{Ti} d_{m\geq 0}(p_i)$ on the particle momenta of the geometric measures with a nearly conical shape, as can be seen in \fig{jetregions}. One can show that deviations from the circular shape are only of $\mathcal{O}(R^4)$ and still independent of the jet rapidity, since the distance measures in \eq{d_ConicalGeometric} only depend on the differences with respect to the jet coordinates. The jet area is $\pi R^2$ up to very small corrections of $\mathcal{O}(R^6)$, which reach only $\approx 1\%$ even for large $R=1.2$.

%===============================================================================
\subsection{Factorization for different observable choices}
\label{sec:factorization}
%===============================================================================

In this section we display the form of the factorized cross section for $pp \to L+ N$ jets, where $L$ denotes a recoiling color-singlet state, with generic observables in the limit $\Tau_N \ll p_T^J$. The observables can be categorized according to their parametric behavior close to the jet and beam axes into \SCETa-type and \SCETb-type cases. For notational simplicity we assume that the same observable is measured in each jet region (which asymptotically behaves like \eq{Taum_coll} with $\beta\equiv \beta_{m \geq 1}$).
We will mainly focus on the properties of the relevant soft function, which also
encodes all dependence of the singular cross section on the distance measure used for the partitioning.

The scaling of the modes in the effective theory follows in general from the constraints on radiation imposed by the $N$-jettiness measurements $\Taum{m}$ in \eq{Tauidef} with $m=a,b,1,\dots,N$, the jet boundaries determined by the distance measures in \eq{TauN} and potential hierarchies in the hard kinematics. We work in a parametric regime with $\Taum{m} \ll p_T^J$ and without additional hierarchies in the jet kinematics (which corresponds to a generic \SCET setup), i.e.\ assuming hard jets with $p_{T,m}^J \sim Q$, large jet radii $R\sim 1$, well-separated collinear directions $n_l \cdot n_m \sim 1$, and nonhierarchical measurements in the different regions $\Taum{l} \sim \Taum{m}$.
The parametric scaling of the collinear and soft modes is then given by
\begin{alignat}{3}
n_{a,b} \text{-collinear:}
&&
p_{n_{a,b}}^\mu &\sim p_T^J\,(\lambda^{\frac{4}{\gamma}},1,\lambda^{\frac{2}{\gamma}})_{n_{a,b}}
\,, \nn \\
n_{m \geq 1}\text{-collinear:}
&&
\qquad p_{n_{m}}^\mu &\sim p_T^J\,(\lambda^{\frac{4}{\beta}},1,\lambda^{\frac{2}{\beta}})_{n_m}
\,,\nn \\
\text{soft:} && p_{s}^\mu &\sim p_T^J\,(\lambda^2,\lambda^2,\lambda^2)
\, ,
\end{alignat}
where we adopt the scaling $\lambda^2\sim \mathcal{T}_{N}/p_T^J$, and give momenta in terms of lightcone coordinates $p^\mu =(n \cdot p, \bar{n} \cdot p, p_\perp)_n$ with respect to the lightcone direction $n = (1, \hat n)$ and $\bar{n} = (1, -\hat n)$. The properties of the factorization formulas depend on the values of $\beta$ and $\gamma$ and the resulting invariant mass hierarchies between the soft and collinear modes. If $\beta,\gamma \neq 1$ the associated collinear fluctuations live at a different invariant mass scale than the soft modes, leading to a \SCETa-type description. Otherwise at least one collinear mode is separated from the soft modes only in rapidity, giving rise to a \SCETb-type theory involving rapidity divergences for the individual bare quantities and a dependence on an associated rapidity RG scale $\nu$ in the renormalized quantities~\cite{Chiu:2011qc,Chiu:2012ir}. Being fully differential in the hard kinematic phase space $\Phi_N$ and all $N$-jettiness observables $\Taum{m}$, the factorization formulae for the four cases with $\beta, \gamma=1$ and $\beta, \gamma\neq 1$ read:\footnote{We do not include effects from Glauber gluon exchange here. For active-parton scattering their perturbative contributions start at ${\cal O}(\alpha_s^4)$~\cite{Gaunt:2014ska,Zeng:2015iba} and can be calculated and included using the Glauber operator framework of ref.~\cite{Rothstein:2016bsq}. For proton initial states the factorization formulae also do not account for spectator forward scattering effects, since the Glauber Lagrangian of ref.~\cite{Rothstein:2016bsq} has been neglected.}

A) $\gamma \neq 1$, $\beta \neq 1$ (\SCETa beams and \SCETa jets): ($n \in  a,b,1, \dots N$)
\begin{align}  \label{eq:fact1j1b}
\frac{\df\sigma_\kappa(\Phi_N)}
{\df\Taum{a} \cdots \df\Taum{N}}
& = \int \Bigl(\prod_n\df k_n\Bigr) \, {\rm tr} \Bigl[
\widehat H_N^{\kappa}(\Phi_N,\mu) \, \widehat S_N^{\kappa} \Bigl( \bigl\{ \Taum{m}-c_m k_m\bigr\},\{n_m\},\{d_m\},\mu\Bigr) \Bigr]
\nn \\ & \quad \times
\omega_a^{\gamma-1} B_{a}\bigl(\omega_a^{\gamma-1} k_a , x_a, \mu\bigr)  \,\omega_b^{\gamma-1} B_{b}\bigl(\omega_b^{\gamma-1} k_b, x_b, \mu\bigr)
\prod_{j=1}^N \omega_j^{\beta-1} J_j (\omega_j^{\beta-1} k_j,\mu)\, .
\end{align}

B) $\gamma = 1$, $\beta\neq1$ (\SCETb beams and \SCETa jets):
\begin{align}  \label{eq:fact1j2b}
\frac{\df\sigma_\kappa(\Phi_N)}
{\df\Taum{a} \cdots \df\Taum{N}}
& =\int \Bigl(\prod_n\df k_n\Bigr) \, {\rm tr} \Bigl[
\widehat H_N^{\kappa}(\Phi_N,\mu) \, \widehat S_N^{\kappa} \Bigl( \bigl\{ \Taum{m}-c_m k_m\bigr\},\{n_m\},\{d_m\},\mu,\frac{\nu}{\mu}\Bigr) \Bigr] \nn \\
&\quad\times B_{a}\Bigl(k_a , x_a, \mu,\frac{\nu}{\omega_a}\Bigr)  \, B_{b}\Bigl(k_b , x_b, \mu,\frac{\nu}{\omega_b}\Bigr)
\prod_{j=1}^N \omega_j^{\beta-1} J_{j} (\omega_j^{\beta-1} k_j,\mu)\, .
\end{align}

C) $\gamma\neq  1$, $\beta = 1$ (\SCETa beams and \SCETb jets):
\begin{align}  \label{eq:fact2j1b}
\frac{\df\sigma_\kappa(\Phi_N)}
{\df\Taum{a} \cdots \df\Taum{N}}
& = \int \Bigl(\prod_n\df k_n\Bigr)  \, {\rm tr} \Bigl[
\widehat H_N^{\kappa}(\Phi_N,\mu) \, \widehat S_N^{\kappa} \Bigl( \bigl\{ \Taum{m}-c_m k_m\bigr\},\{n_m\},\{d_m\},\mu,\frac{\nu}{\mu}\Bigr) \Bigr] \nn \\
& \quad \times
\omega_a^{\gamma-1} B_{a}\bigl(\omega_a^{\gamma-1} k_a , x_a, \mu\bigr)  \,\omega_b^{\gamma-1} B_{b}\bigl(\omega_b^{\gamma-1} k_b, x_b, \mu\bigr)
\prod_{j=1}^N  J_{j} \Bigl(k_j,\mu,\frac{\nu}{\omega_j}\Bigr) \, .
\end{align}

D) $\gamma = 1$, $\beta = 1$ (\SCETb beams and \SCETb jets):
\begin{align}  \label{eq:fact2j2b}
\frac{\df\sigma_\kappa(\Phi_N)}
{\df\Taum{a} \cdots \df\Taum{N}}
& = \int \Bigl(\prod_n\df k_n\Bigr)  \, {\rm tr} \Bigl[
\widehat H_N^{\kappa}(\Phi_N,\mu) \, \widehat S_N^{\kappa} \Bigl( \bigl\{ \Taum{m}-c_m k_m\bigr\},\{n_m\},\{d_m\},\mu,\frac{\nu}{\mu}\Bigr) \Bigr] \nn \\
&\quad \times B_{a}\Bigl(k_a , x_a, \mu,\frac{\nu}{\omega_a}\Bigr)  \, B_{b}\Bigl(k_b , x_b, \mu,\frac{\nu}{\omega_b}\Bigr)
\prod_{j=1}^N  J_{j} \Bigl(k_j,\mu,\frac{\nu}{\omega_j}\Bigr)\, .
\end{align}
In eqs.~(\ref{eq:fact1j1b})--(\ref{eq:fact2j2b}) the hard function $\widehat{H}^\kappa_N$ encodes the hard interaction process for the partonic channel
\begin{equation}
\kappa_a(q_a)\kappa_b(q_b) \to \kappa_1(q_1)\kappa_2(q_2) \dotsb \kappa_N(q_N) + L(q_L)
\,,\qquad
\kappa =\{\kappa_a,\kappa_b;\kappa_1,\dots, \kappa_N\}
\end{equation}
in terms of the massless (label) momenta $q_m^\mu =\omega_m n_m^\mu/2$, which satisfy partonic (label) momentum conservation
\begin{equation}
q_a^\mu + q_b^\mu = q_1^\mu + \dotsb + q_N^\mu + q_L^\mu
\,,\end{equation}
where $q_L^\mu$ is the total momentum of the recoiling color-singlet final state.
The $x_{a,b}$ and label momenta for the initial states are defined via
\begin{equation}
q_{a,b}^\mu =  \omega_{a,b} \frac{n_{a,b}^\mu}{2} \equiv x_{a,b} E_{\rm cm} \frac{n_{a,b}^\mu}{2}
\,.\end{equation}
The jet functions $J_{m\geq 1}$ and beam functions $B_a$, $B_b$ describe the final-state and initial-state collinear dynamics, respectively, and $\widehat{S}^\kappa_N$ denotes the soft function. $\widehat{H}^\kappa_N$ and $\widehat{S}^\kappa_N$ are matrices in color space. The $c_m$ are the normalization factors of the observable as defined in \eq{Taum_coll}. Due to the requirement $\Tau^{(m)} \ll p_T^J$ the collinear modes do not resolve the jet boundaries, such that the jet functions are of the inclusive type and have been computed at one-loop in ref.~\cite{Larkoski:2014uqa} for arbitrary values $\beta>0$.\footnote{For $\beta=2$ they have been computed before in refs.~\cite{Bauer:2003pi,Fleming:2003gt,Becher:2009th}.} Note that in the jet functions, for cases C and D ($\beta=1$), a rapidity regularization in close correspondence to refs.~\cite{Chiu:2011qc,Chiu:2012ir} leads to an additional dependence on the scale ratio $\nu/\omega_m$.

The factorization for the pure \SCETa case, for $\beta = \gamma=2$, is well studied in the literature~\cite{Stewart:2010tn, Jouttenus:2011wh} and has been applied to phenomenological predictions for single-jet production~\cite{Jouttenus:2013hs}. Also, both cases A and B have been studied in ref.~\cite{Stewart:2015waa}  (with the focus on $\beta=2$). In this work, we present for the first time cases C and D, and we will focus on those in the following discussion. These represent a generalization of the previous cases, and assume that the jet and beam
axes are insensitive to effects due to mutual recoil or to recoil from soft emissions. 

The recoil of the jet axis due to collinear radiation can be relevant for $\beta>1$ (see e.g.~ref.~\cite{Kang:2013nha}), but as discussed in ref.~\cite{Stewart:2015waa}, is avoided by properly aligning the jet axes. For $\beta\leq 1$, the jet axis can in addition recoil against soft radiation, leading to nontrivial perpendicular momentum convolutions between the jet, beam, and soft functions for recoil-sensitive axes (see e.g.\ refs.~\cite{Dokshitzer:1998kz, Larkoski:2014uqa}). Recoil-free jet axes avoiding this issue can be defined, e.g., through a global minimization of $N$-jettiness,
\begin{align}\label{eq:TauN}
\Tau_N & =  \min_{n_1,\dots,n_N} \sum_{i} \sum_{m=a,b,1, \dots, N} \Taum{m}(p_i) \nn \\
&= \min_{n_1,\dots,n_N} \sum_{i} \sum_{m=a,b,1, \dots, N} f_m(\eta_i,\phi_i) \,p_{T i}  \, \prod_{l \neq m} \theta(d_l(p_i)-d_m(p_i))
\,.
\end{align}
%%%
Other sets of axes deviating by only a sufficiently small amount, i.e.~by an angle $\ll \lambda^{2/\beta}$,
yield the same result up to power corrections.

The measurement in the beam region requires a separate discussion, as the beam axes are fixed by the collider setup. 
However, one can still avoid transverse momentum convolutions by making a less granular measurement of the jet energies or transverse momenta, with a procedure analogous to the one discussed in ref.~\cite{Stewart:2015waa}. Momentum conservation in the direction transverse to the beam implies
\begin{align}
\label{eq:pTdelta}
k_T^\mu\equiv p_{T,a}^\mu+p_{T,b}^\mu=q_{T,L}^\mu+\sum_{m= 1}^{N} p_{T,m}^\mu,
\end{align}
where $p_{T,m}$ is the transverse component of the $m$-th jet momentum, so that measurements of the jet transverse momenta (or of the $p_T$ of a recoiling leptonic state) within a bin size $\Delta p_T^J \gg p_T^J \lambda^{2/\gamma}$ for $\gamma >1$ and $\Delta p_T^J \gg p_T^J \lambda^{2}$ for $\gamma\le 1$ allow one to integrate over the unresolved transverse momenta and eliminate residual transverse momentum convolutions.
This leads to the appearance of the common beam functions which are known at one-loop for $\gamma=1$ and $\gamma=2$~\cite{Stewart:2010qs,Berger:2010xi,Becher:2010tm,Becher:2012qa}.

The soft function, which we are primarily interested in here, depends on the measurements $\Taum{m}$ in the different regions, the angles between any collinear directions $n_l\cdot n_m$, and the distance measures $d_m$ involving the jet radius. If either a jet or beam measurement is \SCETb type, it also involves a dependence on the rapidity renormalization scale $\nu$ besides the invariant mass scale $\mu$. The (bare) soft matrix element is defined as
%%%
\begin{align}
&\widehat{S}^\kappa_N \bigl(\{k_m\}, \{n_l\},\{d_m\} \bigr)
= \MAe{0}{\widehat{Y}_\kappa^{\dagger}(\{n_l\}) \prod_m \delta(k_m - \hat{\Tau}^{(m)}) \, \widehat{Y}_\kappa(\{n_l\})}{0}
\,.\end{align}
Here $\hat{\Tau}^{(m)}$ denotes the operator that measures $\Taum{m}$ on all particles in region $m$, i.e.
\begin{align}
\hat{\Tau}^{(m)} \vert X_s\rangle =\sum_{i \in X_s} \Taum{m}(p_i)\prod_{l \neq m}\theta\bigl[d_l(p_i) -d_m(p_i)\bigr] \vert X_s\rangle \, .
\end{align}
The color matrix $\widehat{Y}_\kappa(\{n_l\})$ is a product of $N+2$ soft Wilson lines pointing in the collinear directions $n_a,n_b,n_1,\dots,n_N$. For a given partonic channel, each of these is given in the color representation of the associated external parton with the appropriate path-ordering prescription. In the following, we use a normalization such that the tree level result for $\widehat{S}^\kappa_N$ is diagonal in color space, $\widehat{S}^{\kappa(0)}_N = {\bf 1}_N\prod_m \delta(k_m)$.

The full one-loop soft function for processes with at least one final state jet is so far only known for specific cases. In ref.~\cite{Jouttenus:2011wh} it has been computed for the thrust-like $N$-jettiness with $\beta=\gamma=2$ using them simultaneously for the measurement and partitioning as in \eq{Nj}. In ref.~\cite{Kasemets:2015uus} the one-loop soft function for angularities with $\beta>1$ in $e^+ e^-$ collisions has been calculated also for a common measurement and partitioning. In the following we will extend these calculations to arbitrary angularity measurements (including jet mass) and jet vetoes (including a standard transverse momentum veto) at $pp$-colliders with the separate partitionings as described in \sec{genTau} (including the anti-$k_T$ case).
At one loop, our results with a global measurement in the beam region are identical to those for the corresponding jet-based vetoes.

%%%%%%%%%%%%%%%%%%%%%%%%%%%%%%%%%%%%%%%%%%%%%%%%%%%%%%%%%%%%%%%%%%%%%%%%%%%%%%%%
\section{General hemisphere decomposition at one loop}
\label{sec:GenHemiDecomp}
%%%%%%%%%%%%%%%%%%%%%%%%%%%%%%%%%%%%%%%%%%%%%%%%%%%%%%%%%%%%%%%%%%%%%%%%%%%%%%%%

%%%

\begin{figure}[t!]
\subfigure[]{\includegraphics[scale=0.5]{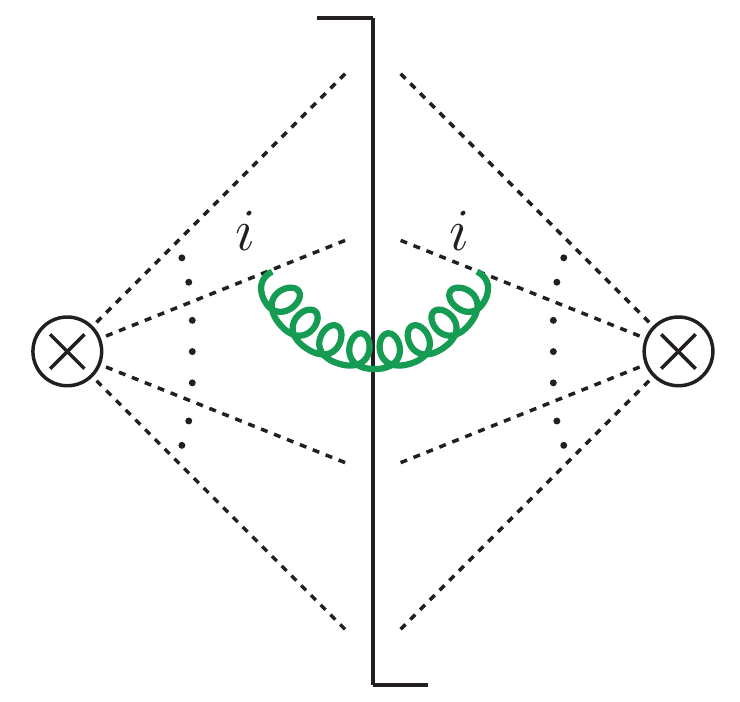}\label{fig:Nsoft-diagram-base-ii}}%
\hfill%
\subfigure[]{\includegraphics[scale=0.5]{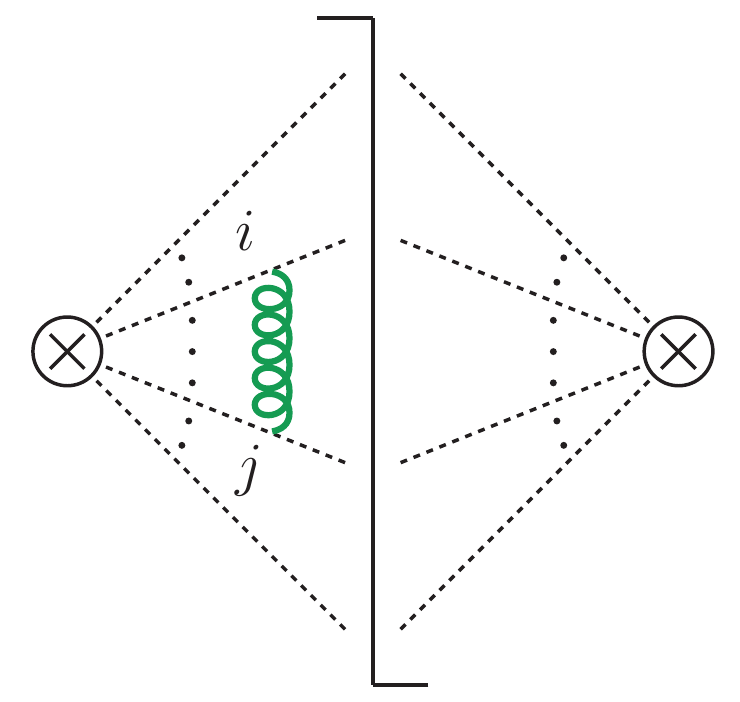}\label{fig:Nsoft-diagram-base-samesideij}}%
\hfill %
\subfigure[]{\includegraphics[scale=0.5]{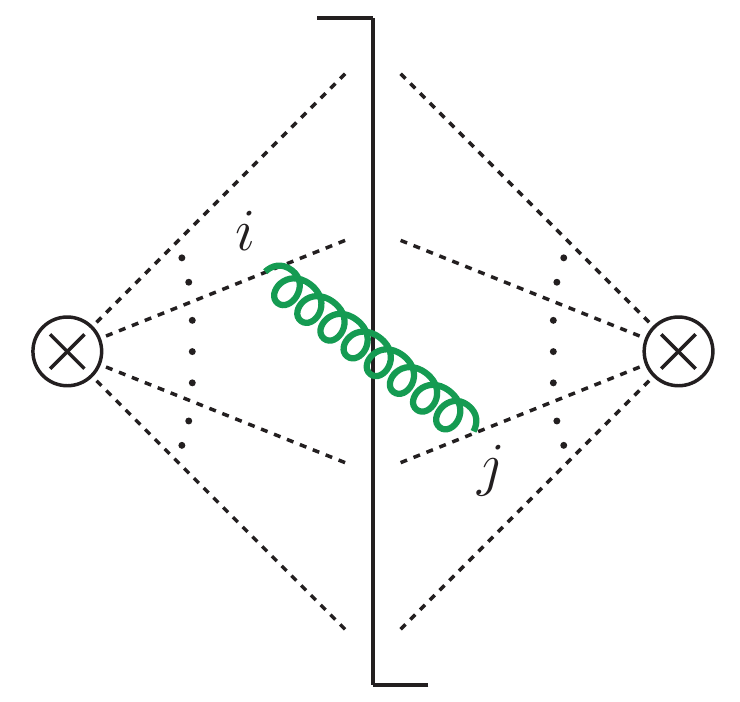}\label{fig:Nsoft-diagram-base-ij}}%
\hfill
\subfigure[]{\includegraphics[scale=0.5]{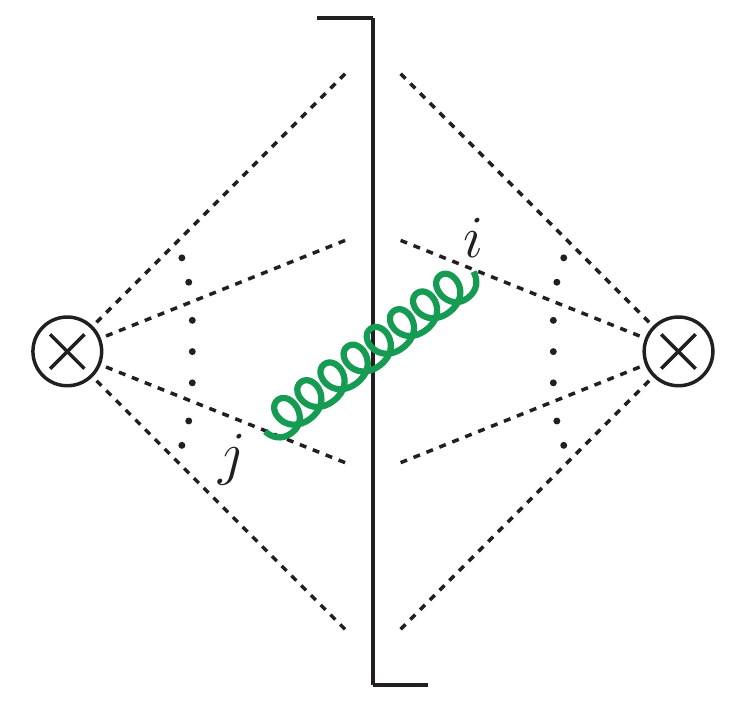}\label{fig:Nsoft-diagram-base-ji}}%
\vspace{-0.5ex}
\caption{One loop contributions to the soft function with multiple collinear legs. The vertical line denotes the final-state cut. Diagrams (a) and (b) vanish in Feynman gauge and dimensional regularization, while (c) and (d) lead to \eq{Sijbare_generalN}.}
\label{fig:NjetSoneloop}
\end{figure}
%%%

The Feynman diagrams for the computation of the one-loop soft function are displayed in \fig{NjetSoneloop}. The virtual diagrams vanish in pure dimensional regularization and the real radiation contribution associated with only one collinear direction vanish in Feynman gauge due to $n_i^2=0$. Thus the one-loop expression is given as a sum over real radiation contributions from different color dipoles each associated with two external hard partons,
%%%
\begin{align}\label{eq:SN_color}
\widehat{S}^{\mathrm{bare}\one}_N(\{k_m\}, \{n_m\},\{d_m\} ) = \sum_{i<j} \bT_i\cdot \bT_j \, S_{ij}(\{k_m\}, \{d_m\} )
\end{align}
with $i,j =a,b,1,\dots, N$ and 
%%%
\begin{align}\label{eq:Sijbare_generalN}
S_{ij} (\{k_m\}, \{d_m\}) &=-2g^2 \Bigl(\frac{e^{\gamma_E}\mu^2}{4\pi}\Bigr)^\eps
\int\! \frac{\df^d p}{(2\pi)^d}\,\biggl(\frac{\nu}{2p_0}\biggr)^\eta \,\frac{n_i\cdot n_j}{(n_i\cdot p)(n_j\cdot p)} \nn \\
& \quad \times 2\pi \delta(p^2)\,\theta(p^0)\, F(\{k_m\}, \{d_m\},p)
\,.
\end{align}
%%%
We have included a factor to account for the regularization of possible rapidity divergences. Since $(\nu/(2p_0))^\eta \to (\nu/\bar{n}_i \cdot p)^\eta$ for $p^\mu \to (\bar{n}_i \cdot p) n^\mu_i/2$, the common expressions for the rapidity regularized jet and beam functions can be used. By contrast, naively applying the Wilson line regulator in refs.~\cite{Chiu:2011qc,Chiu:2012ir} for every single collinear direction would give the factor
%%%
\begin{align}
\Bigl(\frac{\nu}{|\bar{n}_i \cdot p - n_i \cdot p|}\Bigr)^{\frac{\eta}{2}} \times\Bigl(\frac{\nu}{|\bar{n}_j \cdot p - n_j \cdot p|}\Bigr)^{\frac{\eta}{2}}  \quad\stackrel{p^\mu \to (\bar{n}_i \cdot p)\frac{ n^\mu_i}{2}}{\longrightarrow}\quad  \Bigl(\frac{\nu}{\bar{n}_i \cdot p} \Bigr)^\eta \, \frac{1}{|\hat{n}_i \cdot \hat{n}_j|^{\eta/2}} \, .
\end{align}
%%%
The additional factor $ |\hat{n}_i \cdot \hat{n}_j|^{-\eta/2}$  leads to different finite $\mathcal{O}(\eta^0)$ terms, which would lead to a hard function that differs from the standard $\MS$ result, and hence we chose not to use this regulator here. While refs.~\cite{Chiu:2011qc,Chiu:2012ir} chose the spatial $p_3$-component for the regularization, in particular to preserve analyticity properties for virtual corrections, we choose here to only introduce a 
regulator for real radiation corrections, for which the energy component is suitable.\footnote{Rapidity regulators that only act on the real radiation contributions have been used earlier in the literature~\cite{Becher:2011dz} (the regulator we use for our multi-jet situation differs from theirs). An alternative would be a rapidity regulator for the dipole that preserves analyticity and hence can be used for both real and virtual corrections in $S_{ij}$, of the form
\begin{align}
 \Bigl(\frac{\nu\, n_i\cdot n_j}{2|n_i \cdot p -n_j \cdot p|}\Bigr)^{\eta} \,.
\end{align}
This regulator does not have an obvious interpretation as coming from the soft Wilson lines.
}
This is related to a moment of the exponential rapidity regulator used in ref.~\cite{Li:2016axz}.

The function $F$ incorporates the phase-space constraints on the single soft real emission. In terms of the $N$-jettiness measurements $\Taum{m}(p)$ with given distance measures $d_m(p)$ for $m=a,b,1, \dots N$ it reads
\begin{align} 
F(\{k_m\}, \{d_m\},p)  & =  \sum_{m} \delta(k_m - \Taum{m}(p)) \prod_{l \ne m} \delta(k_l) \, \theta(d_l(p) - d_m(p)) \, .
\end{align}
%%%

To compute the integral in \eq{Sijbare_generalN} for arbitrary (one-dimensional) measurements and a general phase-space partitioning we generalize the hemisphere decomposition employed in ref.~\cite{Jouttenus:2011wh}. Our method is based on the fact that the full (IR, UV, rapidity) divergent structure of the soft function contribution $S_{ij}$ is reproduced using arbitrary (IR safe) measurements $\tilde\Tau^{(i)}$, $\tilde\Tau^{(j)}$ that asymptotically satisfy \eq{Taum_coll}, and using arbitrary distance measures $\{\tilde{d}_k\}$, with the only requirement that emissions in the vicinity of the axes $n_i$ and $n_j$ have to be assigned to regions $i$ and $j$, respectively. Having found a combination of measures that allows for an analytic calculation one can then compute the mismatch to the correct measurement and phase-space partitioning in terms of finite (numerical) integrals.

\begin{figure}
\centering
\includegraphics[width=.8\textwidth]{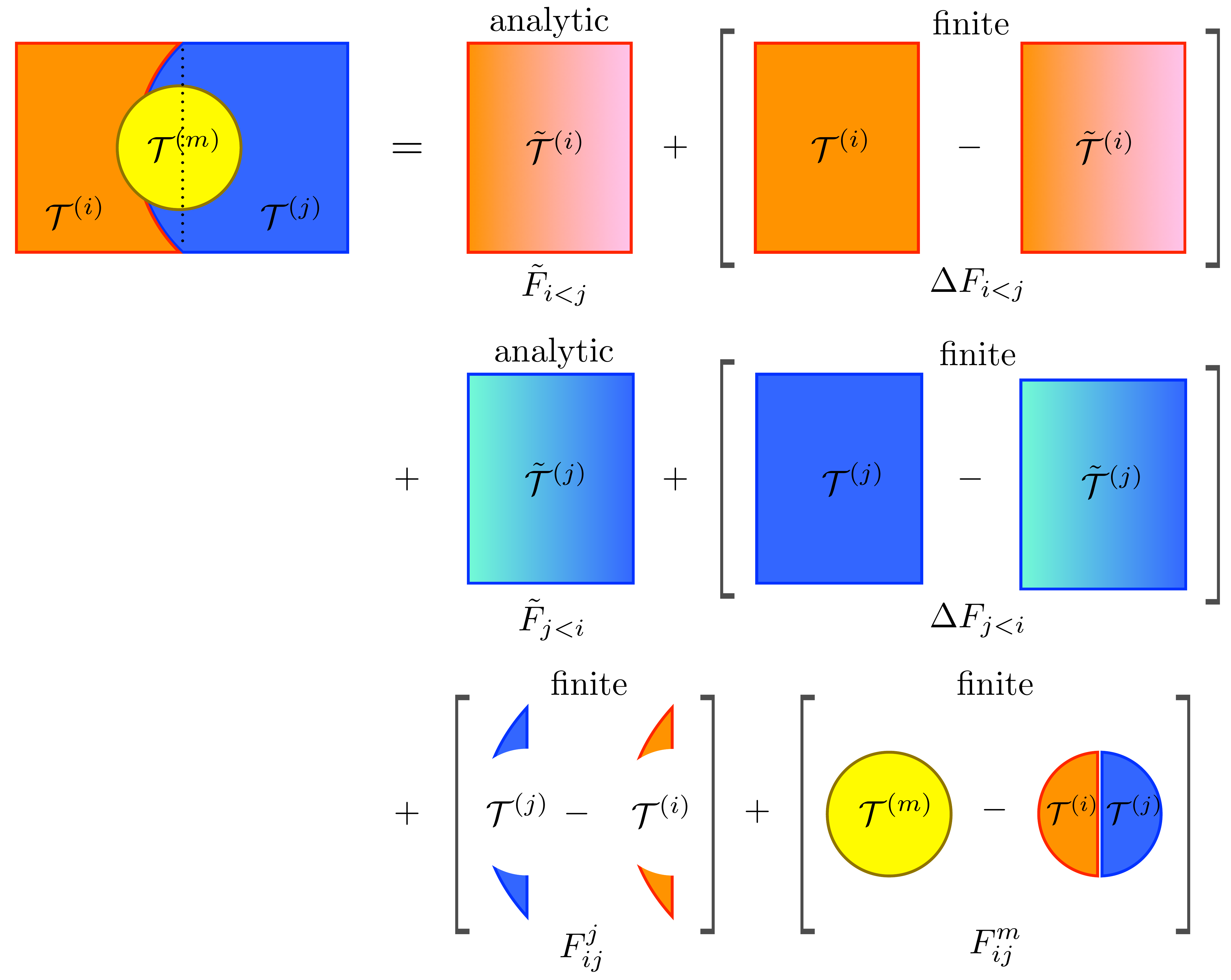}
\caption{Illustration of the hemisphere decomposition of the measurement function in \eq{FNdecomp} into analytic contributions containing all divergent corrections. The remaining finite corrections accounting for the mismatch in measurement or partitioning can be computed by numerical integrations. The color of the filling indicates which variable is measured.
For simplicity we illustrate a case where the correction $F_{ij}^i$ vanishes.
\label{fig:GeneralHemiDecompRect}}
\end{figure}

The most straightforward choice to enable an analytic calculation with the same singular structure as the full result is to employ directly angularities as measurements in the regions $i$, $j$ which are defined by thrust hemispheres, i.e.~to use
\begin{align}\label{eq:Tau_hemi}
\tilde\Tau^{(i)}(p) =  c_i \, (n_i \cdot p)^{\frac{\beta_i}{2}} (\bar{n}_i \cdot p)^{1-\frac{\beta_i}{2}} \, ,\quad \tilde\Tau^{(j)}(p) =c_j (n_j \cdot p)^{\frac{\beta_j}{2}} (\bar{n}_j \cdot p)^{1-\frac{\beta_j}{2}} \, 
\end{align}
with the distance measures 
\begin{align}\label{eq:d_hemi}
\tilde{d}_i(p) = \frac{n_i \cdot p}{\rho_i} \, , \qquad \tilde{d}_j(p)= \frac{n_j \cdot p}{\rho_{j}} \, , \qquad  \tilde{d}_{k \neq i,j}(p) = \infty \, .
\end{align}
We have included factors $\rho_{i},\rho_{j}$ to allow for the possibility of nonequal hemisphere regions $i$ and $j$, which we will exploit in \sec{OneJetCase} to analytically calculate the result in the small-$R$ limit. Taking into account the difference to the actual jet boundaries and measurement, we decompose the measurement function $F$ for the dipole correction $S_{ij}$ as
%%%
\begin{align} \label{eq:FNdecomp}
F(\{k_l\}, \{d_l\},p)  & = \tilde{F}_{i<j}(\{k_l\}, p) + \Delta F_{i<j}(\{k_l\},p)  + \tilde{F}_{j<i}(\{k_l\}, p) + \Delta F_{j<i}(\{k_l\},  p) \nn \\
& \quad + \sum_{m=a,b,1,\dots,N} F_{ij}^m(\{k_l\}, \{d_l\},p)\, ,
\end{align}
with all indices distinguishing separate beam regions $a,b$ and
\begin{align}\label{eq:deltaF}
\tilde{F}_{i<j}(\{k_l\}, p) & =  \delta\bigl(k_i- \tilde{\Tau}^{(i)}(p)\bigr) \,\theta \Bigl(\frac{n_j \cdot p}{\rho_{j}} - \frac{n_i \cdot p}{\rho_i}\Bigr) \,\prod_{l \ne i} \delta(k_l) \, ,
 \nn \\
\Delta F_{i<j}(\{k_l\}, p)& = \Bigl[ \delta\bigl(k_i - \Tau^{(i)}(p)\bigr) - \delta\bigl(k_i-  \tilde{\Tau}^{(i)}(p)\bigr)\Bigr]  \,\theta\Bigl(\frac{n_j \cdot p}{\rho_{j}} - \frac{n_i \cdot p}{\rho_i}\Bigr)\, \prod_{l \ne i} \delta(k_l) \, ,
\nn \\
F_{ij}^i(\{k_l\}, \{d_n\},p) & = \Bigl[ \delta\bigl(k_i - \Tau^{(i)}(p)\bigr)\, \delta(k_j) - \delta\bigl(k_j -  \Tau^{(j)}(p)\bigr)\,\delta(k_i)\Bigr] \nn \\
& \quad \times \theta\Bigl( \frac{n_i \cdot p}{\rho_i}-\frac{n_j \cdot p}{\rho_{j}}\Bigr) \,\theta\bigl(d_j(p)-d_i(p)\bigr) \prod_{l \neq i,j}\theta\bigl(d_l(p)-d_i(p)\bigr) \,\delta(k_l) \, , \nn \\
F_{ij}^{m \neq i,j}(\{k_l\}, \{d_n\},p) & = \Bigl[ \delta\bigl(k_m - \Tau^{(m)}(p)\bigr)\,\delta(k_i) - \delta\bigl(k_i -  \Tau^{(i)}(p)\bigr)\,\delta(k_m)\Bigr] \nn \\ 
& \quad \times \theta\Bigl(\frac{n_j \cdot p}{\rho_{j}} - \frac{n_i \cdot p}{\rho_i}\Bigr)\,\theta\bigl(d_i(p)-d_m(p)\bigr) \prod_{l \neq i}\theta\bigl(d_l(p)-d_m(p)\bigr) \, \delta(k_l)  \nn \\
& \quad +(i \leftrightarrow j) \, .
\end{align}
%%%
The terms $\tilde F_{j<i}$, $\Delta F_{j<i}$, and $F^{j}_{ij}$ in \eq{FNdecomp} are defined in analogy by replacing $i \leftrightarrow j$ in these expressions for $\tilde F_{i<j}$, $\Delta F_{i<j}$ and $F^{i}_{ij}$.
A specific example for this hemisphere decomposition is illustrated in \fig{GeneralHemiDecompRect}. 

The $\tilde F_{i<j}$ denote the measurement of $\tilde\Tau^{(i)}$ in the hemisphere $i$, which can be computed analytically and encodes all divergences. The measurement contribution $\Delta F_{i<j}$ is present if $\Taum{i}$ is not identical to the angularity $\tilde\Tau^{(i)}$. It corrects for this mismatch within the hemisphere boundaries and therefore does not depend on the final partitioning. Since $\Taum{i}$ and $\tilde\Tau^{(i)}$ yield the same collinear and rapidity divergences and also the soft divergences cancel in the difference of the two IR-safe observables this is a finite correction. The remaining pieces $F^k_{ij}$ correct the measurement with the hemisphere boundaries to the actual partitioning given in terms of the distance measures $\{d_h\}$. Here the superscript $m$ indicates that the measurement of $\Taum{m}$ instead of $\Taum{i}$ or $\Taum{j}$ needs to be performed in the associated phase space region where $d_m$ is minimal. For $m= i$ and $m=j$ this corresponds to the boundary mismatch corrections between the regions $i$ and $j$. The only singularities in the phase space mismatch regions are soft IR divergences which cancel between two IR safe measurements, such that the corresponding correction to the soft function is also finite and can be calculated numerically in terms of finite (observable and partitioning dependent) integrals.

%%%
%%%

We decompose the contribution of the $ij$ dipole to the soft function in direct correspondence with \eq{FNdecomp}
%%%
\begin{align}
S_{ij} (\{k_l\}, \{n_k\},\{d_m\})  & =  \tilde{S}_{i<j}(\{k_l\},\hat{s}_{ij})
+ \Delta S_{i<j}(\{k_l\},\hat{s}_{ij})  + \tilde{S}_{j<i}(\{k_l\},\hat{s}_{ij})+ \Delta S_{j<i}(\{k_l\},\hat{s}_{ij})
 \nn \\ & \quad
+  \sum_{m=a,b,1,\dots, N} S_{ij}^m(\{k_l\}, \{d_n\},\hat{s}_{ij}) 
\,,\end{align}
%%%
where the terms on the right-hand side distinguish between two beam regions with separate measurements.  

The expressions for the individual terms follow by replacing the measurement $F(\{k_l\}, \{d_n\}, p)$ in \eq{SN_color} by the corresponding term in \eq{FNdecomp}. The hemisphere corrections to the soft function $\tilde{S}_{i<j}$ and $\tilde{S}_{j<i}$ have been calculated analytically for $\beta_i=2$ in \cite{Jouttenus:2011wh}. For $\beta_i\neq 1$ the result has been given in ref.~\cite{Kasemets:2015uus} in terms of a finite numerical integral. The latter can be evaluated analytically and vanishes for $\rho_i=\rho_j$. This yields the bare result
\begin{align}\label{eq:S_hemi}
\tilde{S}_{i<j}^{\beta_i \neq 1}(\{k_l\},\hat{s}_{ij}) &= \frac{\alpha_s}{4\pi} \frac{1}{\beta_i-1}  \, \prod_{l \neq i} \delta(k_l) \,\biggl\{\frac{8}{\mu \,\xi_{i<j}}\,  \mathcal{L}_1\biggl(\frac{k_i}{\mu
\,\xi_{i<j}}\biggr) - \frac{4}{\eps} \frac{1}{\mu\,\xi_{i<j}} \,\mathcal{L}_0\biggl(\frac{k_i}{\mu\,\xi_{i<j}}\biggr) \\
& \quad+\delta(k_i) \biggl[\frac{2}{\eps^2}-\frac{\pi^2}{6}-(\beta_i-2)(\beta_i-1)\, \theta\Bigl(\frac{\rho_i}{\rho_j}\hat{s}_{\bar{i} j} -1\Bigr) \ln^2 \Bigl(\frac{\rho_i}{\rho_j}\hat{s}_{\bar{i} j}\Bigr)\biggr] +\mathcal{O}(\epsilon)\biggr\} \, , \nn
\end{align}
with the rescaling factor $\xi_{i<j}$ given in terms of the angular term $\hat{s}_{ij}$, with
\begin{align}
\xi_{i<j} \equiv c_i \Bigl(\frac{\rho_i}{\rho_j} \hat{s}_{ij}\Bigr)^{\frac{\beta_i-1}{2}} \, ,\quad \hat{s}_{ij} \equiv \frac{n_i \cdot n_j}{2} = \frac{1-\cos\theta_{ij}}{2} \, , \quad\hat{s}_{\bar{i}j} \equiv \frac{\bar{n}_i\cdot n_j}{2} = \frac{1+\cos\theta_{ij}}{2} \, .
\end{align}
The plus distributions $\cL_n$ are defined as
\begin{align}
 \cL_n(y) \equiv \bigg[\frac{\theta(y) \ln^n y}{y}\bigg]_+
\,.\end{align}
For $\beta_i =1$ the computation is carried out in \app{anal_soft_pieces} which gives the result
\begin{align}\label{eq:S_hemi_b1}
\tilde{S}_{i<j}^{\beta_i = 1}(\{k_l\},\hat{s}_{ij}) &= \frac{\alpha_s}{4\pi}  \, \prod_{l\neq i} \delta(k_l)\,\biggl\{\frac{8}{\mu \,c_i}\,  \mathcal{L}_1\biggl(\frac{k_i}{\mu\,c_i}\biggr) - \frac{8}{\mu\,c_i}   \,\mathcal{L}_0\biggl(\frac{k_i}{\mu \, c_i}\biggr) \biggl[\frac{1}{\eta}+\ln \biggl(\frac{\nu}{\mu} \sqrt{\frac{\rho_i}{\rho_j}\hat{s}_{ij}}\biggl)\biggr]\nn \\
& \quad+\delta(k_i) \biggl[\frac{4}{\eta\,\eps}-\frac{2}{\eps^2}+\frac{4}{\eps} \,\ln\biggl( \frac{\nu}{\mu}\sqrt{\frac{\rho_i}{\rho_j}\hat{s}_{ij}}\biggr)+\frac{\pi^2}{6}+ \theta\Bigl(\frac{\rho_i}{\rho_j}\hat{s}_{\bar{i} j} -1\Bigr) \ln^2 \Bigl(\frac{\rho_i}{\rho_j}\hat{s}_{\bar{i} j}\Bigr)\biggr]\nn \\
& \quad +\mathcal{O}(\eta,\epsilon)\biggr\} \, .
\end{align}
The hemisphere results $\tilde{S}_{j<i}$ are given by simply replacing $i \leftrightarrow j$
in \eqs{S_hemi}{S_hemi_b1}.

We will now explicitly display the corrections to the hemisphere results in \eqs{S_hemi}{S_hemi_b1} in terms of finite integrals that can be computed numerically. Depending on the specific partitioning and $N$-jettiness measurement, different integration variables can be appropriate, e.g.~the rapidity $\eta$ and azimuthal angle $\phi$ in the lab frame (i.e.~coordinates with respect to the beam axis) or the relative rapidity $\eta'$ and azimuthal angle $\phi'$ in a boosted frame where the collinear directions $n_i$ and $n_j$ are back-to-back. The former is usually more convenient for the conical (anti-$k_T$) distance measure in \eq{d_Conical} since the integration boundaries are just circles in the $\eta$-$\phi$ plane, while the geometric measures in eqs.~(\ref{eq:d_GeometricR})--(\ref{eq:d_ConicalGeometric}) involve naturally the momentum projections $n_i \cdot p$, $n_j \cdot p$ for which the variables $\eta'$, $\phi'$ are usually more practical (see  refs.~\cite{Jouttenus:2011wh,Kasemets:2015uus}). For definiteness we use here beam coordinates, since our general $N$-jettiness measurements for $pp \to N$ jets in \eq{Tauidef} and also the distance measures in eqs.~(\ref{eq:d_Conical})--(\ref{eq:d_ConicalGeometric}) are displayed in terms of those, and since our main focus will be the anti-$k_T$ case. First we write the momentum projections in eqs.~(\ref{eq:Sijbare_generalN}),~(\ref{eq:Tau_hemi}) and (\ref{eq:d_hemi}) as
 \begin{align}
n_k \cdot p = p_T \,g_k(\eta,\phi) \, , \qquad \bar{n}_k \cdot p = p_T \,g_{\bar{k}}(\eta,\phi)
 \end{align}
 with 
 \begin{align}\label{eq:g_m}
  g_a(\eta,\phi) & \equiv g_0(\eta,\phi) = e^{-\eta} \, , \nn \\
  g_b(\eta,\phi) & \equiv g_{\bar{0}}(\eta,\phi) = e^{\eta} \, , \nn \\
  g_{m>0} (\eta,\phi) & =\frac{\cosh (\eta -\eta_{m}) -\cos (\phi -\phi_{m})}{\cosh \eta_{m}} \, , \nn \\
  g_{\bar{m}>0} (\eta,\phi) & = \frac{\cosh (\eta +\eta_{m}) +\cos (\phi -\phi_{m})}{\cosh \eta_{m}} \, .
 \end{align}
Keeping only the $\epsilon$-dependence in the phase space integration of \eq{Sijbare_generalN} which is required to regulate the soft singularities, we can write the correction terms as  
 \begin{align}
 \Delta S_{i<j} (\{k_l\},\hat{s}_{ij}) & =- \frac{\alpha_s}{\pi^2}\, \mu^{2\eps} \int_{0}^\infty \frac{\df p_T}{p_T^{1+2\eps}}  \int_{-\pi}^\pi \df \phi  \int_{-\infty}^\infty \df \eta \, \frac{\hat{s}_{ij}}{g_i(\eta,\phi) \, g_j(\eta,\phi)} \, \Delta F_{i<j}(\{k_l\},p)  + \ord{\eps} 
 \,,\end{align}
 and similarly for $S^{m}_{ij}$. We can then use that
\begin{align} \label{eq:abpTintegral}
&\mu^{2 \epsilon} \int_{0}^\infty \frac{\df p_T}{p_T^{1+2 \epsilon}} \bigl[ \delta(k_i) \, \delta\bigl(k_m - p_T f_m(\eta,\phi)\bigr) - \delta\bigl(k_i - p_T f_i(\eta,\phi)\bigr) \, \delta(k_m)\bigr]
\nn \\
&\qquad = \delta(k_i) \,\frac{1}{\mu} \cL_0 \Bigl( \frac{k_m}{\mu} \Bigr) - \frac{1}{\mu} \,\cL_0 \Bigl( \frac{k_i}{\mu} \Bigr) \,\delta(k_m) - \ln \biggl( \frac{f_m(\eta,\phi)}{f_i(\eta,\phi)}\biggr) \,\delta(k_i)\, \delta(k_m) +\mathcal{O}(\epsilon)\, .
\end{align}
To obtain the correction $\Delta S_{i<j}$ we replace in \eq{abpTintegral} $k_m \to k_i$, $f_m \to \tilde{f}_i = c_i \,g^{\beta_i/2}_i g^{1-\beta_i/2}_{\bar{i}}$ giving
\begin{align}
\Delta S_{i<j}(\{k_l\},\hat{s}_{lm}) & = \frac{\alpha_s}{\pi}  \,I_{1,i<j}(f_i,\hat{s}_{ij}) \,\prod_{l} \delta(k_l)
\end{align}
in terms of the angle dependent integral $I_{1,i<j}$ which depends only on the observable $\Tau^{(i)}$ (via $f_i$) and the angle $\hat{s}_{ij}$,
\begin{align}\label{eq:I1_hemi}
I_{1,i<j} (f_i,\hat{s}_{ij}) &= \frac{\hat{s}_{ij} }{\pi}\int_{-\pi}^\pi \df \phi \int_{-\infty}^\infty \df \eta  \,\ln \biggl( \frac{f_i(\eta,\phi)}{c_i [g_i(\eta,\phi)]^{\beta_i/2}\,[g_{\bar{i}}(\eta,\phi)]^{1-\beta_i/2}}\biggr) \, \frac{1}{g_i(\eta,\phi) \, g_j(\eta,\phi)} \nn \\
& \quad \times  \theta\Bigl(\frac{g_j(\eta,\phi)}{\rho_j}-\frac{g_i(\eta,\phi)}{\rho_i}\Bigr) \, .
\end{align}
Similar expressions appear also in ref.~\cite{Banfi:2014sua} in computations of soft corrections for general event shapes in $e^+ e^-$-collisions. 
Finally, the non-hemisphere correction $ S^m_{ij}$ can be written as (see also refs.~\cite{Jouttenus:2011wh,Kasemets:2015uus})
\begin{align}\label{eq:S_lm^n}
 S^m_{ij}(\{k_l\}, \{d_n\},\hat{s}_{ij}) & = \frac{\alpha_s}{\pi} \biggl\{ \biggl[\delta(k_m) \, \frac{1}{\mu} \cL_0 \Bigl( \frac{k_i}{\mu} \Bigr) - \frac{1}{\mu} \cL_0 \Bigl( \frac{k_m}{\mu} \Bigr) \,\delta(k_i)\biggr] I^m_{0,ij}(\{d_l\},\hat{s}_{ij}) \prod_{l\neq i,m} \delta(k_l) \nn \\
 & \quad + I^m_{1,ij}(\{d_l\},f_i,f_m,\hat{s}_{ij})\,\prod_{l} \delta(k_l)\biggr\} +(i \leftrightarrow j) \, ,
\end{align}
in terms of the integrals $I_{0,ij}^{m}$ (and $I_{0,ji}^{m}$), which depends on the partitioning and the angle $\hat{s}_{ij}$, and the integrals $I_{1,ij}^{m}$ (and $I_{1,ji}^{m}$), which in addition depend on the measurements $\Tau^{(i)}$ ($\Tau^{(j)}$) and $\Tau^{(m)}$. These are given by
\begin{align}
I^m_{0,ij}(\{d_l\},\hat{s}_{ij}) & = \frac{ \hat{s}_{ij}}{\pi}\int_{-\pi}^\pi \df \phi \int_{-\infty}^\infty \df \eta \,   \frac{1}{g_i(\eta,\phi) \, g_j(\eta,\phi)} \nn \\
 & \quad \times  \theta\Bigl(\frac{g_j(\eta,\phi)}{\rho_j}-\frac{g_i(\eta,\phi)}{\rho_i}\Bigr) \prod_{l \neq m}\theta\bigl(d_l(\eta,\phi)-d_m(\eta,\phi)\bigr) \, , \\
 I^m_{1,ij}(\{d_l\},f_i,f_m,\hat{s}_{ij}) & = \frac{ \hat{s}_{ij}}{\pi}\int_{-\pi}^\pi \df \phi \int_{-\infty}^\infty \df \eta \,  \ln \biggl( \frac{f_m(\eta,\phi)}{f_i(\eta,\phi)}\biggr) \, \frac{1}{g_i(\eta,\phi) \, g_j(\eta,\phi)} \nn \\
 & \quad \times  \theta\Bigl(\frac{g_j(\eta,\phi)}{\rho_j}-\frac{g_i(\eta,\phi)}{\rho_i}\Bigr) \prod_{l \neq m}\theta\bigl(d_l(\eta,\phi)-d_m(\eta,\phi)\bigr) \, .
\end{align}

The above expressions allow for a determination of the $N$-jet soft function at one-loop for arbitrary measurements and distance measures. In practice, evaluating these integrals can be quite tedious, since the phase-space constraints can lead to slow or unstable numerical evaluations. For the one-jet case and distance measures we consider next we solve for the integration limits allowing for fast and precise numerical integrations.

%%%%%%%%%%%%%%%%%%%%%%%%%%%%%%%%%%%%%%%%%%%%%%%%%%%%%%%%%%%%%%%%%%%%%%%%%%%%%%%%
\section{$L+1$ jet production at hadron colliders}
\label{sec:OneJetCase}
%%%%%%%%%%%%%%%%%%%%%%%%%%%%%%%%%%%%%%%%%%%%%%%%%%%%%%%%%%%%%%%%%%%%%%%%%%%%%%%%

%===============================================================================
\subsection{Setup}
%===============================================================================

As a concrete example for the comparison of numerical results we discuss the case $pp \to L + 1 $ jet. Choosing $\phi_J = 0$ without loss of generality the lightcone direction of the jet is given by
%%%
\begin{align}
n^\mu_J=  (1,\hat n_J) = \Bigl(1,\frac{1}{\cosh \eta_J},0, \tanh \eta_J\Bigr)
\,,\end{align}
In this case we partition the phase space only into a single jet and a beam region and the observable is given by
%%%
\begin{align} \label{eq:Tau_def}
\Tau_1 = \sum_i \left\{
\begin{tabular}{ll}
$\Tau_B(p_i),$ & {\rm for }  $d_B(p_i) < d_J(p_i)$ ,\\
$\Tau_J(p_i),$ &  {\rm for } $d_J(p_i) < d_B(p_i)$  .
\end{tabular}\right.
\end{align}
%%%
For $\Tau_B \equiv \Taum{0}$ and $\Tau_J \equiv \Taum{1}$ we use the parameterizations in \eq{Tauidef} to specify the observable. As jet observables we consider angularities defined by
%%%
\begin{align} \label{eq:fJchoices}
\text{Angularity}\,\,\, \Tau_J^\beta: \quad  f^\beta_{J}(\eta_i,\phi_i) = {\mathcal{R}_{iJ}^\beta}
\end{align}
%%%
where $\mathcal{R}_{iJ}$ denotes the distance of the emission $i$ with respect to the jet axis as defined in \eq{DeltaRdef}. Among these is for $\beta=2$ the observable $\Tau_J^{\beta=2}(p_i)=2\cosh \eta_J  (n_J \cdot p_i)$ corresponding directly to the measurement of the jet mass, $m_J^2 \simeq p_T^J \Tau_J^{\beta=2}$, as exploited in refs.~\cite{Jouttenus:2013hs,Stewart:2014nna,Kolodrubetz:2016dzb}. In contrast to \eq{Tau_hemi}, which is the more common definition in $e^+ e^-$ collisions, we have defined the angularities in a way which is invariant under boosts along the beam direction and corresponds to the measurement for the Conical Geometric case in ref.~\cite{Stewart:2015waa} with the specification $\gamma = 1$ (including the XCone default and the Recoil-Free default). For $\beta=1$ the definition in \eq{fJchoices} also corresponds to the default way to study $N$-subjettiness~\cite{Thaler:2010tr}.

As measurements of the beam region observable (or jet vetoes) we discuss 
%%%
\begin{align} \label{eq:fBchoices}
\text{beam thrust}\,\,\, \Tau_B^\tau \,\, (\gamma=2):  \quad &f_{B}^{\tau}(\eta) = e^{-|\eta|}  \, ,\nn \\
\text{C-parameter}\,\,\, \Tau_B^C\,\, (\gamma=2):  \quad &f_{B}^{C}(\eta)  = \frac{1}{2 \cosh \eta} \, , \nn \\
\text{transverse energy} \,\,\, \Tau_B^{p_T}\,\, (\gamma=1):  \quad &f_{B}^{p_T}(\eta_i) = 1 \, .
\end{align}
%%%
These choices include both \SCETa-type observables (beam thrust and C-parameter) and \SCETb-type observables (transverse energy). Thus, with the various choices for $\Tau_B$ and $\Tau_J$, we cover all possible combinations of observable types for which the factorization was discussed in \sec{factorization}.

%===============================================================================
\subsection{Computation of the soft function}
%===============================================================================

The color space for the soft function $S^\kappa_1$ with three external collinear directions is one-dimensional and we write the one-loop expression in analogy to  \eq{SN_color} as
\begin{align}
S^{\kappa \one}_1(\{k_j\},\{d_j\},\eta_J) & = \bT_a\cdot \bT_b \, S_{ab}(\{k_j\},\{d_j\},\eta_J)+ \bT_a\cdot \bT_J \, S_{aJ}(\{k_j\},\{d_j\},\eta_J) \nn \\
& \quad +\bT_b\cdot \bT_J \, S_{bJ}(\{k_j\},\{d_j\},\eta_J) \, ,
\end{align}
 where $S_{bJ}$ can be inferred from $S_{aJ}$ due to symmetry,
 \begin{align}
 S_{bJ}(\{k_j\},\{d_j\},\eta_J)  = S_{aJ}(\{k_j\},\{d_j\},-\eta_J) \, .
 \end{align}
 For a pure gluonic channel $\kappa = \{g,g;g\}$ the color factors are
 \begin{align}
 \bT_a\cdot \bT_b = \bT_a\cdot \bT_J =\bT_b\cdot \bT_J = -\frac{C_A}{2} \, ,
 \end{align}
 while for the channel $\kappa = \{g,q;q\}$ (and in analogy for its permutations)
  \begin{align}
  \bT_a\cdot \bT_b = \bT_a\cdot \bT_J = -\frac{C_A}{2} \, , \quad  \bT_b\cdot \bT_J = \frac{C_A}{2}-C_F \, ,
  \end{align}
 The expressions for the Feynman diagrams of the corrections $S_{ab}$ and $S_{aJ}$ are given by \eq{Sijbare_generalN} with $N=1$.
 
\begin{figure}
\centering
\includegraphics[width=\textwidth]{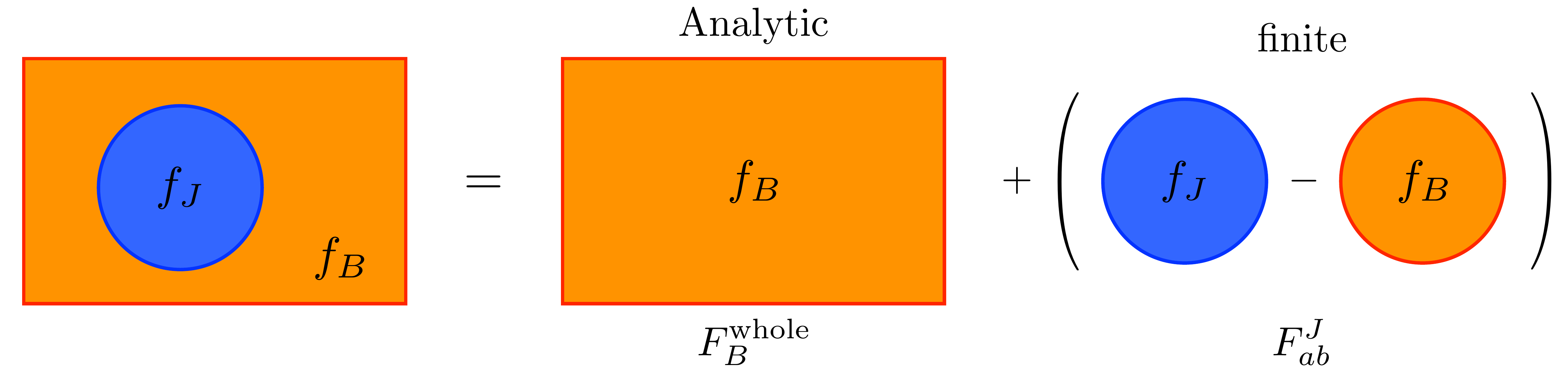}
\caption{The hemisphere decomposition adapted to the case of a beam-beam dipole ($i=a$, $j=b$). The circle indicates the jet region defined by $d_B(p)>d_J(p)$.\label{fig:Small_R_Decomp_Beam}}
\end{figure}

Following the hemisphere decomposition in \sec{GenHemiDecomp}, for the beam-beam dipole correction $S_{ab}$ the full hemisphere corrections, i.e.~without considering the jet region, can be computed analytically for the measurements in \eq{fBchoices}. Thus the contributions $\tilde{F}_{a<b}$, $\tilde{F}_{b<a}$, $\Delta F_{a<b}$ and $\Delta F_{b<a}$ in \eq{FNdecomp} can be represented by a single function $F_B^{\text{whole}}$ encoding the full measurement of the beam region observable $\Tau_B$ in the whole phase space. We therefore write the measurement function $F$ as\footnote{Compared to \sec{GenHemiDecomp} we perform here the decomposition for a single beam region.}
 %%%
 \begin{align} \label{eq:abFdecomp}
 F(\{k_j\},\{d_j\},\eta_J, p) & = F_B^{\text{whole}}(\{k_j\},p) + F_{ab}^{J}(\{k_j\},\{d_j\},\eta_J, p)
 \nn \, , \\
 F_B^{\text{whole}}(\{k_j\},p) & = \delta\bigl(k_B - p_T f_B(\eta)\bigr)\, \delta(k_J)
 \, ,\nn\\
 F_{ab}^{J}(\{k_j\},\{d_j\},\eta_J,p) & =\Bigl[ \delta(k_B) \,\delta\bigl(k_J - p_T f_J(\eta,\phi)\bigr) - \delta\bigl(k_B - p_T f_B(\eta)\bigr)\, \delta(k_J) \Bigr]  \nn \\
 & \quad \times \theta\bigl(d_B(\eta) - d_J(\eta,\phi)\bigr) 
\,,\end{align}
 %%%
 which is illustrated in \fig{Small_R_Decomp_Beam}.
 The analytic corrections $S_{ab}^{\text{whole}}$ corresponding to $F_B^{\text{whole}}$ can be easily obtained from \eqs{S_hemi}{S_hemi_b1} (and using \eq{I1_hemi} for the C-parameter), see also e.g.~refs.~\cite{Fleming:2007xt,Hoang:2014wka,Chiu:2012ir},
 \begin{align} \label{eq:Sabanal}
 S_{ab}^{{\rm whole}, \tau}(\{k_j\}) & = \frac{\alpha_s}{4\pi} \, \delta(k_J)\,\biggl\{\frac{16}{\mu} \cL_1\biggl(\frac{k_B}{\mu}\biggr) -\frac{8}{\mu\epsilon}\cL_0\biggl(\frac{k_B}{\mu}\biggr)+\biggl[\frac{4}{\eps^2}- \frac{\pi^2}{3}\biggr] \delta(k_B) +\mathcal{O}(\epsilon)\biggr\} \, , 
 \nn \\
 S_{ab}^{{\rm whole}, C}(\{k_j\}) &= \frac{\alpha_s}{4\pi} \, \delta(k_J)\,\biggl\{\frac{16}{\mu} \cL_1\biggl(\frac{k_B}{\mu}\biggr) -\frac{8}{\mu\epsilon}\cL_0\biggl(\frac{k_B}{\mu}\biggr)+\biggl[\frac{4}{\eps^2}- \pi^2\biggr] \delta(k_B) +\mathcal{O}(\epsilon)\biggr\} \, ,
 \nn \\
 S_{ab}^{{\rm whole}, p_T}(\{k_j\})& =\frac{\alpha_s}{4\pi} \, \delta(k_J)\,\biggl\{ \frac{16}{\mu}\,  \mathcal{L}_1\biggl(\frac{k_B}{\mu}\biggr) - \frac{16}{\mu}   \,\mathcal{L}_0\biggl(\frac{k_B}{\mu}\biggr) \biggl[\frac{1}{\eta}+\ln \biggl(\frac{\nu}{\mu} \biggl)\biggr] \nn 
 \\
 & \quad + \delta(k_B) \biggl[\frac{8}{\eta\,\eps}-\frac{4}{\eps^2}+\frac{8}{\eps} \,\ln\biggl( \frac{\nu}{\mu}\biggr)+\frac{\pi^2}{3}\biggr]+\mathcal{O}(\eta,\epsilon)\biggr\} \, .
 \end{align}
 The remaining correction $S_{ab}^{J}$ due to the angularity measurement in the jet region is of $\mathcal{O}(R^2)$, i.e.~the jet area, and is given by
 %%%
 \begin{align} \label{eq:abnum}
 S_{ab}^{J}(\{k_j\},\{d_j\},\eta_J) &= \frac{\alpha_s}{\pi} \bigg\{I^J_{0,ab}(\{d_j\},\eta_J) \biggl[\delta(k_J)\, \frac{1}{\mu} \cL_0 \Bigl( \frac{k_B}{\mu} \Bigr) - \frac{1}{\mu} \cL_0 \Bigl( \frac{k_J}{\mu} \Bigr) \,\delta(k_B) \biggr]\nn \\
 & \quad + I^J_{1,ab}(\{d_j\},\{f_j\},\eta_J) \,\delta(k_B) \,\delta(k_J)  \bigg\}
 \,,
 \nn \\
 I^J_{0,ab}(\{d_j\},\eta_J) & = \frac{1}{\pi} \int^\pi_{-\pi} \df \phi \int_{-\infty}^{\infty} \df \eta \, \theta\bigl(d_B(\eta) - d_J(\eta,\phi)\bigr) \, ,\nn \\
 I^J_{1,ab}(\{d_j\},\{f_j\},\eta_J)  & = \frac{1}{\pi}\int^\pi_{-\pi} \df \phi \int_{-\infty}^{\infty}  \df \eta\,\ln \biggl( \frac{f_J(\eta,\phi)}{f_B(\eta)} \biggr)  \,\theta\bigl(d_B(\eta) - d_J(\eta,\phi)\bigr) \, .
 \end{align}
 %%%
  $I^J_{0,ab}$ corresponds just to the jet area in the $\eta$-$\phi$ plane and is identical to $R^2$ for the conical and the geometric-$R$ measures, while for the conical geometric measure there are deviations of $\mathcal{O}(R^6)$. 

%%%
%%%

In order to compute the integrals for the beam-jet dipoles, one can follow the hemisphere decomposition as presented in \sec{GenHemiDecomp} which yields numerical corrections of $\mathcal{O}(1)$ and logarithmically enhanced terms for small $R$. However, we will present here a more efficient adaption of this decomposition exploiting the fact that for the measurements considered in this section the soft function can be computed analytically in an expansion in terms of the jet radius $R$. As already discussed in ref.~\cite{Kolodrubetz:2016dzb} this provides a fairly good approximation for not too large values of $R$. In the following we will compute numerically only deviations from these results, such that the numerical integrals will scale with powers of $R$ thus avoiding large cancellations for $R \ll 1$.\footnote{We have checked that the numerical results from the two alternative decompositions agree.} 

\begin{figure}
\centering
\includegraphics[width=\textwidth]{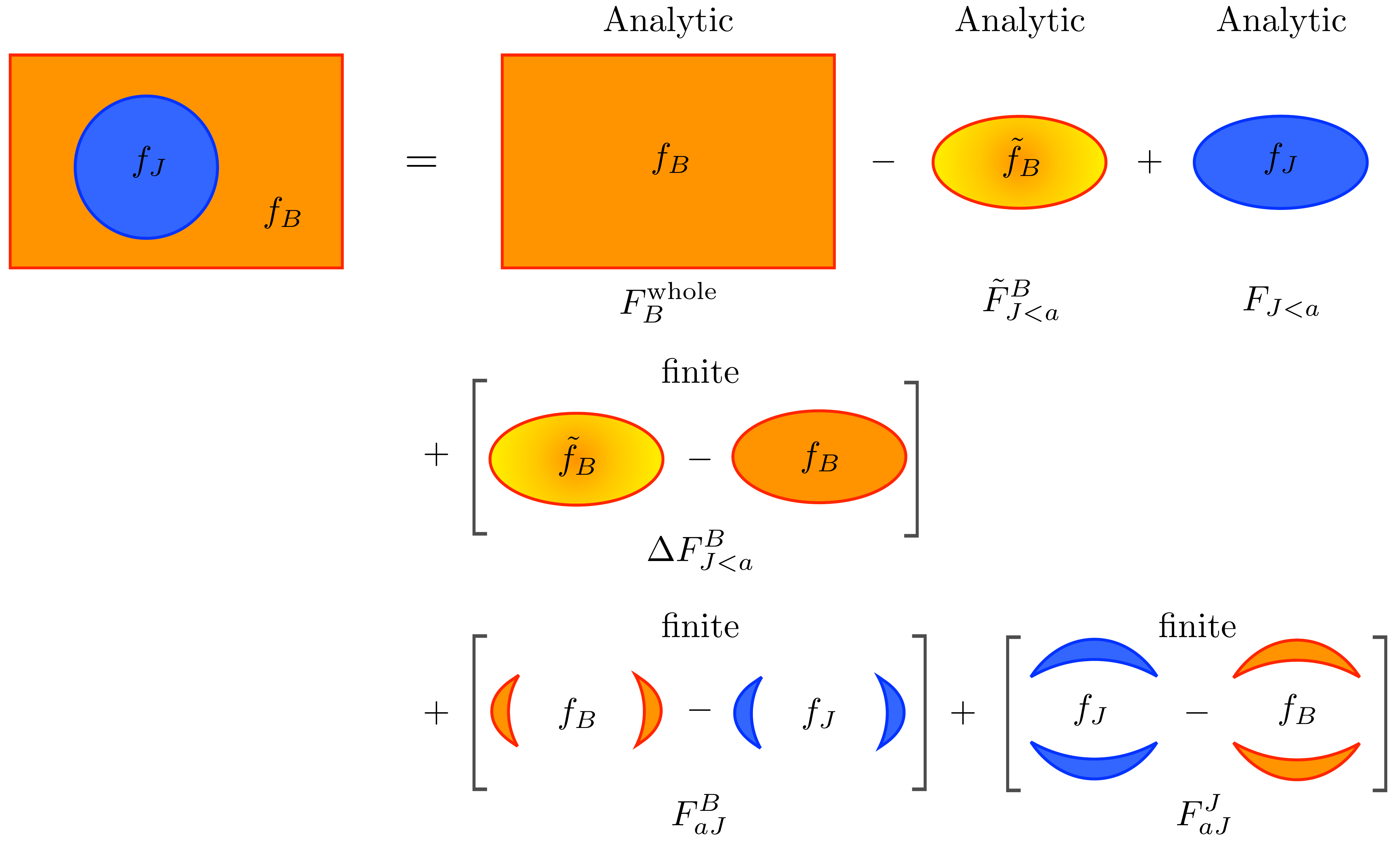}
\caption{The general adapted version of the hemisphere decomposition for the case of a beam-jet dipole ($i=a$, $j=J$). The first line represents terms which can be calculated analytically, while the second and third line contain finite, numerical corrections which vanish in the small $R$ limit. \label{fig:Small_R_Decomp_Rect}}
\end{figure}

First, we can choose in \eq{deltaF} the parameter $\rho_J$ such that for $R \ll 1$ it yields a conical shape for the jet region with an active area $\pi R^2$. In this limit all distance measures considered here lead to the same partitioning as shown in \fig{jetregions} with deviations being suppressed by $R$. Using \eq{g_m} the associated condition for the parameter $\rho_J$ reads for the $aJ$-dipole (with $\rho_a =1$)
\begin{align} \label{eq:rhoJRdef}
\int_\pi^\pi  \df\phi \int_{-\infty}^\infty \df \eta \,\theta\biggl[\rho_J \,e^{-\eta} - \frac{\mathcal{R}_{iJ}^2}{2 \cosh \eta_J} \biggr] \stackrel{\!}{=}  \pi R^2 \, .
\end{align}
Expanding the phase space constraint in the small-$R$ limit gives an analytic relation for $\rho_J$,
\begin{align} \label{eq:leadingrhoRJ}
\rho_J(R) =\rho^R_J\bigl[1+ \mathcal{O}(R)\bigr] \quad {\rm with} \, \, \, \rho^R_J = R^2 \,\frac{1+\tanh{\eta_J}}{2} \, .
\end{align} 
The soft function corrections due to the measurement of angularities in the jet hemisphere can be computed analytically. If the corrections due to the measurement of the beam region observable in the beam hemisphere can also be computed analytically, all remaining numerical corrections will be automatically small for $R \ll 1$. This is the case for the transverse energy veto, where \eq{S_hemi_b1} provides an exact hemisphere result for arbitrary $\rho$. However, for a general veto (including beam thrust and C-parameter) we have not obtained an analytic hemisphere result. To avoid large numeric corrections from the term $\Delta F_{a<J}$ in \eq{deltaF}, we can instead decompose the hemisphere measurement function $F_{a<J}$ into a piece without constraints due to a jet region and its measurement, calculated analytically in ref.~\cite{Kolodrubetz:2016dzb}, and a subtraction term in the jet hemisphere (with the measurement of the beam region observable), which can be computed in a series expansion in $R$. For the correction $S_{aJ}$ we thus write $F$ as 
%%%
\begin{align} \label{eq:aJFdecomp}
F(\{k_l\},\{d_n\},\eta_J, p) & = F_{a<J}(\{k_l\},R,\eta_J,p) + F_{J<a}(\{k_j\},R,\eta_J,p) + \sum_{m=J,B} F^m_{aJ}(\{k_l\}, \{d_n\},\eta_J,p) \nn \\
& = F_B^{\text{whole}}(\{k_l\},p) -\tilde{F}_{J<a}^B(\{k_l\},R,\eta_J,p)+ F_{J<a}(\{k_l\},R,\eta_J,p) 
\nn \\
& \quad + \Delta F_{J<a}^{B}(\{k_l\},R,\eta_J,p) + \sum_{m=J,B} F^m_{aJ}(\{k_l\}, \{d_n\},\eta_J,p) \, ,
\end{align}
where 
\begin{align}\label{eq:F_onejet}
F_B^{\text{whole}}(\{k_l\},\eta_J,p) & =  \delta\bigl(k_B - p_T f_B(\eta)\bigr) \,\delta(k_J) \, , \nn 
\\
\tilde{F}_{J<a}^{B}(\{k_l\},R,\eta_J,p) & = \delta\bigl(k_B - p_T \tilde{f}_B(\eta-\eta_J)\bigr) \,  \delta(k_J) \, \theta\Bigl(n_a \cdot p -  \frac{n_J \cdot p}{\rho_J^R }\Bigr) \, ,
\nn \\
F_{J<a}(\{k_l\},R,\eta_J,p) & =\delta(k_B) \,\delta\bigl(k_J - p_T f_J(\eta,\phi)\bigr) \, \theta\Bigl(n_a \cdot p -  \frac{n_J \cdot p}{\rho_J^R}\Bigr) \, ,
\nn \\
\Delta F_{J<a}^{B}(\{k_l\},R,\eta_J,p) & =  \Bigl[ \delta\bigl(k_B - p_T \tilde{f}_B(\eta-\eta_J)\bigr) - \delta\bigl(k_B - p_T f_B(\eta)\bigr) \Bigr]  \, \delta(k_J)  \nn \\
& \quad \times \theta\Bigl(n_a \cdot p -  \frac{n_J \cdot p}{\rho_J^R }\Bigr) \, ,
\nn \\
F^B_{aJ}(\{k_l\}, \{d_n\},\eta_J,p) & = \Bigl[\delta\bigl(k_B - p_T f_B(\eta)\bigr)  \,\delta(k_J)- \delta(k_B) \, \delta\bigl(k_J - p_T f_J(\eta,\phi)\bigr) \Bigr]
\nn \\
& \quad \times  \theta\bigl(d_J(\eta,\phi)-d_B(\eta)\bigr) \, \theta\Bigl(n_a \cdot p-\frac{n_J \cdot p}{\rho_J^R }\Bigr) \, , \nn \\
F^J_{aJ}(\{k_l\}, \{d_n\},\eta_J,p) & = \Bigl[ \delta(k_B) \, \delta\bigl(k_J - p_T f_J(\eta,\phi)\bigr) - \delta\bigl(k_B - p_T f_B(\eta)\bigr)  \,\delta(k_J) \Bigr]
\nn \\
& \quad \times  \theta\bigl(d_B(\eta) - d_J(\eta,\phi)\bigr) \, \theta\Bigl(\frac{n_J \cdot p}{\rho_J^R }-n_a \cdot p\Bigr) \,.
\end{align}
%%%
Here the expanded measurement of the beam region observable in the jet region is denoted by $\tilde{\Tau}_B=p_T \tilde{f}_B(\eta-\eta_J)$ with
\begin{align}\label{eq:tilde_fB}
\tilde{f}_B(\eta-\eta_J)  \equiv f_B(\eta_J)\, e^{\eta_J-\eta} =\frac{n_a \cdot p}{p_T} \, f_B(\eta_J) \, e^{\eta_J} \, .
\end{align}
The corresponding decomposition of the soft function is given by
\begin{align}\label{eq:SaJ_decomp}
S_{aJ}(\{k_l\},\{d_n\},\eta_J, p) 
& = S_{aJ}^{\text{whole}}(\{k_l\},\eta_J,p) -\tilde{S}_{J<a}^{B}(\{k_l\},\eta_J)+ S_{J<a}(\{k_l\},R,\eta_J) 
\nn \\
& \quad + \Delta S_{J<a}^{B}(\{k_l\},R,\eta_J) + \sum_{m=J,B} 
 S_{aJ}^m(\{k_l\}, \{d_n\}, \eta_J) \,,
\end{align}
where each individual term is given by replacing the measurement $F(\{k_l\}, \{d_n\}, p)$ in \eq{SN_color} by the corresponding term in \eq{aJFdecomp}. This decomposition is illustrated in \fig{Small_R_Decomp_Rect}. We now discuss the different pieces in turn, giving the associated results.

%%%

The term $F_B^{\text{whole}}$ corresponds to the measurement of the beam observable within the complete phase space without constraints due to the jet region. In the context of $pp \to L +1 $ jet this correction was calculated in \cite{Kolodrubetz:2016dzb} for the measurements in \eq{fBchoices} and denoted by $S_B$ therein.\footnote{For an energy veto at $e^+e^-$ collisions the associated ``inclusive" correction to the one-loop soft function has been first computed in~\cite{Ellis:2010rwa}. For $pp \to$ dijets also the correction from the jet-jet dipole can be calculated for a $p_T$-veto~\cite{Hornig:2016ahz}.} The bare corrections are given by
\begin{align} \label{eq:S_BC}
S^{{\rm whole},\tau}_{aJ} (\{k_l\},\eta_J) &= \frac{\alpha_s} {4 \pi}\, \delta(k_J)   \biggl\{ 16 \eta_J \,\theta(-\eta_J) 
\frac{1}{\mu} \,\cL_0\Bigl(\frac{k_B}{\mu}\Bigr)+ \delta(k_B) \Bigl[-\frac{8 \eta_J}{\eps}  \,\theta(-\eta_J) 
\nn \\
&  \quad -4\,\Li_2\bigl(e^{-2|\eta_J|}\bigr)-8 \eta_J^2 \, \theta(-\eta_J) \Bigr]  +\ord{\eps} \biggr\} \, ,
\nn \\
S^{{\rm whole},C}_{aJ}  (\{k_l\},\eta_J)&= \frac{\alpha_s} {4 \pi}\, \delta(k_J) \biggl\{ 8 \ln\Bigl(\frac{1+\tanh \eta_J}{2}\Bigr) \frac{1}{\mu} \,\cL_0\Bigl(\frac{k_B}{\mu}\Bigr)
+ \delta(k_B)  \biggl[-\frac{4}{\eps} \ln\Bigl(\frac{1+\tanh \eta_J}{2}\Bigr)
\nn \\
&  \quad +4\,\Li_2\Bigl(\frac{1+\tanh \eta_J}{2}\Bigr)+2 \ln^2\Bigl(\frac{1-\tanh \eta_J}{2}\Bigr) -8 \ln^2(2\cosh\eta_J)-\frac{2\pi^2}{3}\biggr] \nn \\
& \quad +\ord{\eps} \biggr\} \, ,
\nn \\
S^{{\rm whole},p_T}_{aJ}  (\{k_l\},\eta_J)&= \frac{\alpha_s} {4 \pi}\, \delta(k_J)\biggl\{
\frac{1}{\mu} \,\cL_0\Bigl(\frac{k_B}{\mu}\Bigr)\biggl[-\frac{8}{\eta} +\frac{4}{\eps}-8\ln \Bigl(\frac{\nu e^{-\eta_J}}{\mu}\Bigr)\biggr] \nn \\
& \quad + \delta(k_B)  \biggl[\frac{4}{\eta \, \eps}-\frac{4}{\eps^2}+\frac{4}{\eps} \ln \Bigl(\frac{\nu e^{-\eta_J}}{\mu}\Bigr)+\frac{\pi^2}{3} \biggr]+\ord{\eta,\eps} \biggr\}
\, .\end{align}

The measurement of the beam region observable leads to a different divergent behavior for radiation collinear to the jet axis than for the jet measurement. This requires the computation of the analytic piece $-\tilde{F}_{J<a}^{B}$ (in the jet hemisphere) to correct for this mismatch. For its calculation we employ a measurement $\tilde{\Tau}_B$ which is linear in the momentum component $n_a \cdot p$ and identical to the beam observable $\Tau_B$ in the vicinity of $n_J$ (i.e.~for $\eta \to \eta_J$), see \eq{tilde_fB}. In dimensional regularization the associated correction gives just the result for the hemisphere contribution in \cite{Jouttenus:2011wh} (with an appropriate rescaling factor),
\begin{align}\label{eq:S_hemiB}
\tilde{S}^{B}_{J<a} (\{k_l\},R,\eta_J) &= \frac{\alpha_s}{4 \pi} \, \delta(k_J)\biggl\{\frac{8  R}{\mu \,f_B(\eta_J)} \,\cL_1 \biggl( \frac{k_B  R}{\mu \, f_B(\eta_J)}\biggr) -\frac{4}{\epsilon} \,\frac{R}{\mu\,f_B(\eta_J)} \,\cL_0 \biggl( \frac{k_B R}{\mu \,f_B(\eta_J)}\biggr) \nn \\
& \quad +\biggl[\frac{2}{\epsilon^2}- \frac{\pi^2}{6}\biggr] \delta(k_B)\biggr\}  \, .
\end{align}

The term $F_{J<a}$ corresponds to the measurement of the jet observable in the rescaled jet hemisphere. The results for the angularities defined in \eq{fJchoices} can be obtained analytically from the hemisphere results in \eqs{S_hemi}{S_hemi_b1} and a finite correction coming from \eq{I1_hemi}. The latter accounts for the difference of the boost invariant jet angularity in \eq{fJchoices} from the generic definition in \eq{Tau_hemi} and is calculated in \app{anal_soft_pieces}. In total we obtain
\begin{align}\label{eq:S_hemi2}  
 S^{\beta \neq 1}_{J<a}(\{k_l\},R,\eta_J) &= \frac{\alpha_s}{4\pi} \, \frac{\delta(k_B)}{\beta\!-\!1}  \biggl\{\frac{8}{\mu \,R^{\beta-1}}\,  \mathcal{L}_1\biggl(\frac{k_J}{\mu 
 \,R^{\beta-1}}\biggr) \!-\! \frac{4}{\eps} \frac{1}{\mu\,R^{\beta-1}} \,\mathcal{L}_0\biggl(\frac{k_J}{\mu\,R^{\beta-1}}\biggr) \\
  & \quad+\delta(k_J) \biggl(\frac{2}{\eps^2}-\frac{\pi^2}{6}-2(\beta-1)(\beta-2)\,\theta(R-1)\ln^2 R\biggr) +\mathcal{O}(\epsilon)\biggr\}  \, , \nn\label{eq:S_hemi2_b1} \\
 S^{\beta = 1}_{J<a}(\{k_l\},R,\eta_J) &= \frac{\alpha_s}{4\pi}  \, \delta(k_B) \biggl\{\frac{8}{\mu}\,  \mathcal{L}_1\biggl(\frac{k_J}{\mu}\biggr) - \frac{8}{\mu}   \,\mathcal{L}_0\biggl(\frac{k_J}{\mu}\biggr) \biggl[\frac{1}{\eta}+\ln \biggl(\frac{\nu R}{2\mu\cosh\eta_J}\biggl)\biggr]  
\\
     &\quad+\delta(k_J) \biggl(\frac{4}{\eta\, \eps}-\frac{2}{\eps^2}+\frac{4}{\epsilon}\ln \biggl(\frac{\nu R}{2\mu\cosh\eta_J}\biggl)+\frac{\pi^2}{6}+2\,\theta(R-1) \ln^2 R\biggr) \nn \\
     & \quad +\mathcal{O}(\eta,\epsilon)\biggr\}  \nn
\, . 
\end{align}

The analytic contributions  in the small R limit are given by
\begin{align}
S_{aJ} (\{k_l\},R,\eta_J)=S^{\rm whole}_{aJ} (\{k_l\},\eta_J)+\tilde{S}^{B}_{J<a} (\{k_l\},R,\eta_J) +S_{J<a} (\{k_l\},R,\eta_J) + {\cal O}(R^{1,2})
\end{align}
where the displayed terms are ${\cal O}(R^0)$ corrections and depend only logarithmically on $R$. They
are independent of the specific partitioning (jet definition), and for $R \ll 1$ yield the full result up to power corrections.  In the context of an effective theory for a small jet radius the soft radiation is factorized into different types of soft modes~\cite{Chien:2015cka,Becher:2015hka,Hornig:2016ahz,Kolodrubetz:2016dzb}. The measurement $F_B^{\text{whole}}$ applies to wide-angle soft radiation, which does not resolve the jet region but depends on the Wilson line of the jet. The corrections $\tilde{S}^{B}_{J<a}$ and $S_{J<a}$ correspond to the results for the matrix elements of ``soft-collinear" and ``collinear-soft" modes, respectively, in the nomenclature of ref.~\cite{Chien:2015cka}. These are boosted and constrained by the jet boundary. In the limit $R \ll 1$  the beam-jet dipoles give the same results, $S_{aJ}=S_{bJ}$, and the Wilson lines from the beams $a$ and $b$ fuse giving a total color factor $\bfT_J \cdot (\bfT_a+\bfT_b) = -\bfT^2_J$~\cite{Stewart:2014nna}.

The measurement corrections $\Delta F_{J<a}^B$, $F^B_{aJ}$ and $F^J_{aJ}$ can be in general not computed analytically, but are again finite corrections that allow for a numerical evaluation. The term $\Delta F_{J<a}^B$ corrects the subtraction in the jet hemisphere from the measurement in the beam region with $\tilde{f}_B$ to the correct observable $f_B$. As in \sec{GenHemiDecomp} we can write this correction in terms of an integral in $\eta$-$\phi$ coordinates,
%%%
\begin{align} \label{eq:aJdiffnum}
\Delta S^B_{J<a} (\{k_i\},R,\eta_J) &= \frac{\alpha_s}{\pi}  \,\Delta I^B_{1,aJ}(f_B,R,\eta_J)   \,\delta(k_B) \, \delta(k_J) 
\,,
\end{align}
%%%
with
%%%
\begin{align} \label{eq:I1diff}
 \Delta I^B_{1,aJ}(f_B,R,\eta_J) & = \frac{1}{2\pi} \int_{-\pi}^\pi \df \phi \int_{-\infty}^{\infty}\df \eta\,\frac{e^{\eta - \eta_J}}{\cosh(\eta-\eta_J) -\cos \phi} \, \ln \biggl( \frac{e^{\eta_J} f_B(\eta_J)}{e^{\eta} f_B(\eta)} \biggr) \,\theta\Bigl(n_a \cdot p - \frac{n_J \cdot p}{\rho_J^R}\Bigr) \nn \\
 & =   \theta(R-1)\biggl[\int^{R-1}_{0}\df x \, h_{1}(f_B,\eta_J,x)  + \int^{R+1}_{R-1}\df x \,h_{2}(f_B,R,\eta_J,x)\biggr] \nn \\
 & \quad + \theta(1-R)\int^{R+1}_{1-R}\df x \,h_{2}(f_B,R,\eta_J,x) \, ,
\end{align}
where we have defined the integration variable $x \equiv e^{\eta-\eta_J}$ and 
\begin{align}
h_1(f_B,\eta_J,x) & = \frac{2x}{\abs{x^2-1}} \, \ln\biggl( \frac{f_B(\eta_J)}{x f_B(\eta_J + \ln x)} \biggr)   \, , \nn \\
h_2(f_B,R,\eta_J,x) & = \biggl[1-\frac{2}{\pi} \arctan\biggl(\frac{|x-1|}{x+1} \sqrt{\frac{(1+x)^2-R^2}{R^2-(x-1)^2}}\biggr) \biggr]\,h_1(f_B,\eta_J,x) \,.
\end{align}
This correction depends also only on the specific shape of the hemisphere for a given value of $R$, but not on the general partitioning. Since the full integrand does not exhibit singular behavior close to the jet axis (i.e. for $\eta \to \eta_J$ and $\phi \to 0$), it scales with the jet area for a smooth measurement in the beam region, i.e.~$\Delta I^B_{1,aJ}$ is $\mathcal{O}(R^2)$.\footnote{We have checked numerically that for the transverse momentum veto with $f_B(\eta)=1$ the integral $\Delta I^B_{1,aJ}$ vanishes for $R\leq 1$ and gives $-4 \ln^2 R$ for $R>1$ as implied by the full analytic hemisphere result in \eq{S_hemi_b1}.}

The terms $F^B_{aJ}$ and $F^J_{aJ}$ correct for the difference between the actual jet definition (through the partitioning) and the employed jet hemisphere with scaling parameter $\rho_J^R$. Their contribution to the soft function directly corresponds to \eq{S_lm^n}. $S^B_{aJ}$ is given by
%%%
\begin{align} \label{eq:aJedgenum}
S^B_{aJ}(\{k_l\},\{d_n\},\eta_J) &= \frac{\alpha_s}{\pi} \bigg\{ I^B_{0,aJ}(\{d_n\},\eta_J) \bigg[\delta(k_B) \, \frac{1}{\mu} \cL_0 \Bigl( \frac{k_J}{\mu} \Bigr) - \frac{1}{\mu} \cL_0 \Bigl( \frac{k_B}{\mu} \Bigr) \,\delta(k_J) \bigg] \nn
\\
& \quad +  I^B_{1,aJ}(\{d_n\},\{f_n\},\eta_J)  \,  \delta(k_B) \,\delta(k_J) \bigg\}
\,,
\end{align}
%%%
where the relevant integrals depend now on the specific distance measures and are given by
%%%
\begin{align} \label{eq:Iedge}
I^B_{0,aJ}(\{d_n\},\eta_J)  &= \frac{1}{2\pi} \int_{-\pi}^\pi \df \phi \int_{-\infty}^{\infty}\df \eta\,\frac{e^{\eta - \eta_J}}{\cosh(\eta-\eta_J) -\cos \phi} \, \\ 
& \quad  \times  \theta \bigl(d_J(\eta,\phi) - d_B(\eta)\bigr)\,\theta\Bigl(R^2 e^{\eta_J-\eta}-2\cosh(\eta-\eta_J) +2\cos\phi\Bigr)
 \, , \nn \\
I^B_{1,aJ}(\{d_n\},\{f_n\},\eta_J) &= \frac{1}{2\pi} \int_{-\pi}^\pi \df \phi \int_{-\infty}^{\infty}\df \eta\,\frac{e^{\eta - \eta_J}}{\cosh(\eta-\eta_J) -\cos \phi} \, \ln \biggl( \frac{f_B(\eta)}{f_J(\eta,\phi)} \biggr)\,     \nn \\
& \quad \times\theta \bigl(d_J(\eta,\phi) - d_B(\eta)\bigr)\,\theta\Bigl(R^2 e^{\eta_J-\eta}-2\cosh(\eta-\eta_J) +2\cos\phi\Bigr) \, . \nn
\end{align}
%%%
In analogy, $S^J_{aJ}$ is given by
%%%
\begin{align} \label{eq:aJedgenum2}
S^J_{aJ}(\{k_l\},\{d_n\},\eta_J) &= \frac{\alpha_s}{\pi} \bigg\{ I^J_{0,aJ}(\{d_n\},\eta_J) \bigg[\delta(k_J) \, \frac{1}{\mu} \cL_0 \Bigl( \frac{k_B}{\mu} \Bigr) - \frac{1}{\mu} \cL_0 \Bigl( \frac{k_J}{\mu} \Bigr) \,\delta(k_B) \bigg] \nn
\\
& \quad +  I^J_{1,aJ}(\{d_n\},\{f_n\},\eta_J)  \,  \delta(k_B) \,\delta(k_J) \bigg\} \, ,
\end{align}
%%%
with 
%%%
\begin{align} \label{eq:Iedge2}
I^J_{0,aJ}(\{d_n\},\eta_J)  &= \frac{1}{2\pi} \int_{-\pi}^\pi \df \phi \int_{-\infty}^{\infty}\df \eta\,\frac{e^{\eta - \eta_J}}{\cosh(\eta-\eta_J) -\cos \phi} \, \\ 
& \quad  \times  \theta \bigl(d_B(\eta,\phi) - d_J(\eta)\bigr)\,\theta\Bigl(2\cosh(\eta-\eta_J) -2\cos\phi-R^2 e^{\eta_J-\eta}\Bigr)
 \, , \nn \\
I^J_{1,aJ}(\{d_n\},\{f_n\},\eta_J) &= \frac{1}{2\pi} \int_{-\pi}^\pi \df \phi \int_{-\infty}^{\infty}\df \eta\,\frac{e^{\eta - \eta_J}}{\cosh(\eta-\eta_J) -\cos \phi} \, \ln \biggl( \frac{f_J(\eta,\phi)}{f_B(\eta,\phi)} \biggr)\,     \nn \\
& \quad \times\theta \bigl(d_B(\eta,\phi) - d_J(\eta)\bigr)\,\theta\Bigl(2\cosh(\eta-\eta_J) -2\cos\phi-R^2 e^{\eta_J-\eta}\Bigr) \, . \nn
\end{align}
These integrals scale individually as $\mathcal{O}(R)$, but yield in total $\mathcal{O}(R^2)$ contributions, as explained in \app{scalingR}.\footnote{This holds only for a smooth measurement in the beam region. For the beam thrust veto and $|\eta_J|<R$ the resulting total correction is of $\mathcal{O}(R)$ due to the kink at $\eta=0$.}
We will discuss in \app{num_evaluation} how the numerical evaluation of these integrals can be carried out efficiently by explicitly determining the integration domains. While a full analytic calculation of these does not seem feasible in general, it is possible to compute them in an expansion for $R \ll R_0$ (where $R_0$ denotes the generic convergence radius where the expansion breaks down). We calculate the terms at $\mathcal{O}(R^2)$ in \app{R_expansion}. Such an expansion has been also applied in \cite{Dasgupta:2012hg,Liu:2014oog} for the inclusive jet mass spectrum where it was found that $\mathcal{O}(R^4)$ corrections have a negligible impact for phenomenologically relevant values of $R$.

%===============================================================================
\subsection{Summary of corrections}
\label{sec:anal_softs}
%===============================================================================

To give a transparent overview of all corrections we display in the following the structure of the full (renormalized) soft functions for all combinations $\beta \neq 1$, $\beta=1$ and $\gamma=1,2$. Since \eqs{S_hemi}{S_hemi_b1} encode the full $\mu$- and $\nu$-dependence of the soft function, one can directly read off the counterterms for the soft function absorbing all $1/\epsilon$- and $1/\eta$-divergences. These result in the well-known one-loop anomalous dimensions for the associated soft function defined by 
\begin{align}
\mu \frac{\df}{\df \mu} S^{\kappa}_{1} (\{k_i\},\{d_i\}, \eta_J,\mu,\nu) &= \int \df k_B' \, \df k_J' \, \gamma^\kappa_{S_1} (\{k_i-k_i'\},\eta_J,\mu,\nu)\, S^{\kappa}_{1} (\{k_i'\},\{d_i\}, \eta_J,\mu,\nu) \, , \nn \\
\nu \frac{\df}{\df \nu} S^{\kappa}_{1} (\{k_i\},\{d_i\}, \eta_J,\mu,\nu) & =  \int \df k_B' \, \df k_J' \, \gamma^\kappa_{S_1,\nu} (\{k_i-k_i'\},\mu)\, S^{\kappa}_{1} (\{k_i'\},\{d_i\}, \eta_J,\mu,\nu) \, .
\end{align}
The $\nu$-anomalous dimension is only present for $\beta=1$ or $\gamma=1$. The explicit one-loop expressions for all cases read
\begin{align}
\gamma^{\kappa(1)}_{S_1,\beta \neq 1, \gamma =2} (\{k_i\},\eta_J,\mu)& =  \frac{\alpha_s(\mu)}{4\pi}\, 2\Gamma_0\biggl\{
 \bfT_J^2 \,\frac{1}{\beta-1}\,\frac{1}{\mu}  \,\cL_0\Bigl(\frac{k_J}{\mu}\Bigr)\, \delta (k_B) \nn \\
 & \quad + (\bfT_a^2+ \bfT_b^2) \,\frac{1}{\mu}  \,\cL_0\Bigl(\frac{k_B}{\mu}\Bigr)\, \delta(k_J)  +  (\bfT_a^2-\bfT_b^2)\,\eta_J\,\delta(k_J)\,\delta(k_B) \biggr\} \, , \nn \\
\gamma^{\kappa(1)}_{S_1,\beta \neq 1, \gamma =1} (\{k_i\},\eta_J,\mu,\nu)& = \frac{\alpha_s(\mu)}{4\pi}\, 2\Gamma_0\biggl\{\bfT_J^2\,\frac{1}{\beta-1}\,\frac{1}{\mu}  \,\cL_0\Bigl(\frac{k_J}{\mu}\Bigr)\delta (k_B) \nn \\
 & \quad +  \Bigl[-(\bfT_a^2+\bfT_b^2)\ln\Bigl(\frac{\nu}{\mu}\Bigr)+  (\bfT_a^2-\bfT_b^2)\,\eta_J\Bigr]\,\delta(k_J)\,\delta(k_B) \biggr\} \, , \nn \\
\gamma^{\kappa(1)}_{S_1,\beta =1, \gamma =2}(\{k_i\},\eta_J,\mu,\nu) & = \frac{\alpha_s(\mu)}{4\pi}\, 2\Gamma_0\biggl\{
 (\bfT_a^2+ \bfT_b^2) \,\frac{1}{\mu}  \,\cL_0\Bigl(\frac{k_B}{\mu}\Bigr)\, \delta(k_J)\nn \\
 & \quad +  \Bigl[-\bfT_J^2\ln\Bigl(\frac{\nu}{2\mu \cosh \eta_J}\Bigr)+  (\bfT_a^2-\bfT_b^2)\,\eta_J\Bigr]\,\delta(k_J)\,\delta(k_B)  \biggr\}\, , \nn \\
\gamma^{\kappa(1)}_{S_1,\beta =1, \gamma =1} (\{k_i\},\eta_J,\mu,\nu) & = \frac{\alpha_s(\mu)}{4\pi}\, 2\Gamma_0\,\delta(k_J)\,\delta(k_B)\biggl\{-(\bfT_a^2+\bfT_b^2+\bfT_J^2)\ln\Bigl(\frac{\nu}{\mu}\Bigr)\nn \\
& \quad +\bfT_J^2\ln(2 \cosh \eta_J)+  (\bfT_a^2-\bfT_b^2)\,\eta_J \biggr\}\, , \end{align}
for the $\mu$-anomalous dimensions with $\Gamma_0 =4 $ being the coefficient of the one-loop cusp anomalous dimension. The $\nu$-anomalous dimensions are given by
\begin{align}
\gamma^{\kappa(1)}_{S_1,\nu,\beta \neq 1, \gamma =1} (\{k_i\},\mu)& = \frac{\alpha_s(\mu)}{4\pi}\, 2\Gamma_0 (\bfT_a^2+\bfT_b^2)\,\frac{1}{\mu}  \,\cL_0\Bigl(\frac{k_B}{\mu}\Bigr)\, \delta(k_J) \, ,  \\
\gamma^{\kappa(1)}_{S_1,\nu,\beta = 1, \gamma =2} (\{k_i\},\mu)& = \frac{\alpha_s(\mu)}{4\pi}\, 2\Gamma_0 \,\bfT_J^2\,\frac{1}{\mu}  \,\cL_0\Bigl(\frac{k_J}{\mu}\Bigr)\, \delta(k_B) \, , \nn \\
\gamma^{\kappa(1)}_{S_1,\nu,\beta = 1, \gamma =1} (\{k_i\},\mu)& = \frac{\alpha_s(\mu)}{4\pi}\, 2\Gamma_0 \,\biggl\{(\bfT_a^2+\bfT_b^2)\,\frac{1}{\mu}  \,\cL_0\Bigl(\frac{k_B}{\mu}\Bigr)\, \delta(k_J)+\bfT_J^2\,\frac{1}{\mu}  \,\cL_0\Bigl(\frac{k_J}{\mu}\Bigr)\, \delta(k_B) \biggr\}\, . \nn
\end{align}

For $\beta \neq 1$ and $\gamma =2$, i.e.~\SCETa jet and beams, the renormalized result for the one-loop soft function reads 
 \begin{align} \label{eq:S_b1j1}
 & S^{\kappa(1)}_{1,\beta \neq 1,\gamma=2} (\{k_i\},\{d_i\}, \eta_J,\mu) = \frac{\alpha_s(\mu)}{4\pi}\biggl\{
 \bfT_a \cdot \bfT_b  \biggl[\frac{16}{\mu} \,\cL_1\Bigl(\frac{k_B}{\mu}\Bigr) \,\de(k_J) 
 \nn \\ & \quad
 + s_{ab,B}(\{d_i\},\eta_J) \Bigl(\frac{1}{\mu}\cL_0\Bigl(\frac{k_B}{\mu}\Bigr) \delta(k_J) - \frac{1}{\mu}\cL_0\Bigl(\frac{k_J}{\mu}\Bigr)\delta(k_B)  \Bigr) +  s_{ab,\delta}(\{d_i\},\{f_i\},\eta_J)\,\delta(k_B) \,\de(k_J)\biggr] 
 \nn \\ & \quad
 + \bfT_a \cdot \bfT_J
 \biggl[
    \frac{1}{\beta-1}\,\frac{8}{\mu}\,\cL_1\Bigl(\frac{k_J}{\mu}\Bigr) \,\delta(k_B)+ \frac{8}{\mu}\,\cL_1\Bigl(\frac{k_B}{\mu}\Bigr) \,\delta(k_J) 
    \nn \\ & \quad
     + s_{aJ,B}(\{d_i\},\eta_J) \,\frac{1}{\mu}\, \cL_0\Bigl(\frac{k_B}{\mu}\Bigr) \,\delta(k_J)    
    + s_{aJ,J}(\{d_i\},\eta_J)\, \frac{1}{\mu}\, \cL_0\Bigl(\frac{k_J}{\mu}\Bigr) \,\delta(k_B)
    \nn \\ & \quad     
    + s_{aJ,\delta}(\{d_i\},\{f_i\},\eta_J) \, \delta(k_J) \, \delta(k_B)  \biggr]
    + \bfT_b \cdot \bfT_J\biggl[\eta_J\leftrightarrow -\eta_J\biggr] \biggr\} \, ,
 \end{align}
 For $\beta \neq 1$ and $\gamma =1$, i.e.~a \SCETa jet and \SCETb beams, the result reads
  \begin{align} \label{eq:S_b2j1}
 &  S^{\kappa(1)}_{1,\beta \neq 1,\gamma=1} (\{k_i\},\{d_i\}, \eta_J,\mu,\nu) = \frac{\alpha_s(\mu)}{4\pi}\biggl\{
  \bfT_a \cdot \bfT_b  \biggl[\frac{16}{\mu} \,\cL_1\Bigl(\frac{k_B}{\mu}\Bigr) \,\delta(k_J)-\frac{16}{\mu} \,\cL_0\Bigl(\frac{k_B}{\mu}\Bigr) \ln \Bigl(\frac{\nu}{\mu}\Bigr) \,\de(k_J) 
  \nn \\ & \quad
  + s_{ab,B}(\{d_i\},\eta_J) \Bigl(\frac{1}{\mu}\cL_0\Bigl(\frac{k_B}{\mu}\Bigr) \delta(k_J) - \frac{1}{\mu}\cL_0\Bigl(\frac{k_J}{\mu}\Bigr)\delta(k_B)  \Bigr) +  s_{ab,\delta}(\{d_i\},\{f_i\},\eta_J)\,\delta(k_B) \,\de(k_J)\biggr] 
  \nn \\ & \quad
  + \bfT_a \cdot \bfT_J
  \biggl[\frac{1}{\beta-1}\, \frac{8}{\mu}\,\cL_1\Bigl(\frac{k_J}{\mu}\Bigr) \,\delta(k_B) + \frac{8}{\mu}\,\cL_1\Bigl(\frac{k_B}{\mu}\Bigr) \,\delta(k_J)  -\frac{8}{\mu}\,\mathcal{L}_0\Bigl(\frac{k_B}{\mu}\Bigr) \ln \Bigl(\frac{\nu}{\mu}\Bigr) \,\delta(k_J)
     \nn \\ & \quad
      + s_{aJ,B}(\{d_i\},\eta_J)\,\frac{1}{\mu}\, \cL_0\Bigl(\frac{k_B}{\mu}\Bigr) \,\delta(k_J)    
     + s_{aJ,J}(\{d_i\},\eta_J)\, \frac{1}{\mu}\, \cL_0\Bigl(\frac{k_J}{\mu}\Bigr) \,\delta(k_B)
     \nn \\ & \quad     
     + s_{aJ,\delta}(\{d_i\},\{f_i\},\eta_J) \, \delta(k_J) \, \delta(k_B)  \biggr]
     + \bfT_b \cdot \bfT_J\biggl[\eta_J\leftrightarrow -\eta_J\biggr] \biggr\} \, ,
  \end{align}
  For $\beta = 1$ and $\gamma =2$, i.e.~a \SCETb jet and \SCETa beams, the result reads
   \begin{align} \label{eq:S_b1j2}
  &  S^{\kappa(1)}_{1,\beta =1,\gamma=2} (\{k_i\},\{d_i\}, \eta_J,\mu,\nu) = \frac{\alpha_s(\mu)}{4\pi}\biggl\{
   \bfT_a \cdot \bfT_b  \biggl[\frac{16}{\mu} \,\cL_1\Bigl(\frac{k_B}{\mu}\Bigr) \,\de(k_J) 
   \nn \\ & \quad
   + s_{ab,B}(\{d_i\},\eta_J) \Bigl(\frac{1}{\mu}\cL_0\Bigl(\frac{k_B}{\mu}\Bigr) \delta(k_J) - \frac{1}{\mu}\cL_0\Bigl(\frac{k_J}{\mu}\Bigr)\delta(k_B)  \Bigr) +  s_{ab,\delta}(\{d_i\},\{f_i\},\eta_J)\,\delta(k_B) \,\de(k_J)\biggr] 
   \nn \\ & \quad
   + \bfT_a \cdot \bfT_J
   \biggl[\frac{8}{\mu}\,  \mathcal{L}_1\biggl(\frac{k_J}{\mu}\biggr) \,\delta(k_B) - \frac{8}{\mu}   \,\mathcal{L}_0\biggl(\frac{k_J}{\mu}\biggr) \ln \biggl(\frac{\nu}{2\mu\cosh\eta_J}\biggl) \,\delta(k_B) +
      \frac{8}{\mu}\,\cL_1\Bigl(\frac{k_B}{\mu}\Bigr) \,\delta(k_J) 
      \nn \\ & \quad
       +s_{aJ,B}(\{d_i\},\eta_J)\,\frac{1}{\mu}\, \cL_0\Bigl(\frac{k_B}{\mu}\Bigr) \,\delta(k_J)    
      +  s_{aJ,J}(\{d_i\},\eta_J)\, \frac{1}{\mu}\, \cL_0\Bigl(\frac{k_J}{\mu}\Bigr) \,\delta(k_B)
      \nn \\ & \quad     
      + s_{aJ,\delta}(\{d_i\},\{f_i\},\eta_J) \, \delta(k_J) \, \delta(k_B)  \biggr]
      + \bfT_b \cdot \bfT_J\biggl[\eta_J\leftrightarrow -\eta_J\biggr] \biggr\} \, ,
   \end{align}
     For $\beta = 1$ and $\gamma =1$, i.e.~\SCETb jet and beams, the result reads
    \begin{align} \label{eq:S_b2j2}
  &  S^{\kappa(1)}_{1,\beta = 1,\gamma=1} (\{k_i\},\{d_i\}, \eta_J,\mu,\nu) = \frac{\alpha_s(\mu)}{4\pi}\biggl\{
   \bfT_a \cdot \bfT_b  \biggl[\frac{16}{\mu} \,\cL_1\Bigl(\frac{k_B}{\mu}\Bigr) \,\delta(k_J)-\frac{16}{\mu} \,\cL_0\Bigl(\frac{k_B}{\mu}\Bigr) \ln \Bigl(\frac{\nu}{\mu}\Bigr) \,\de(k_J) 
   \nn \\ & \quad
   + s_{ab,B}(\{d_i\},\eta_J) \Bigl(\frac{1}{\mu}\cL_0\Bigl(\frac{k_B}{\mu}\Bigr) \delta(k_J) - \frac{1}{\mu}\cL_0\Bigl(\frac{k_J}{\mu}\Bigr)\delta(k_B)  \Bigr) +  s_{ab,\delta}(\{d_i\},\{f_i\},\eta_J)\,\delta(k_B) \,\de(k_J)\biggr] 
   \nn \\ & \quad
   + \bfT_a \cdot \bfT_J
   \biggl[\frac{8}{\mu}\,  \mathcal{L}_1\biggl(\frac{k_J}{\mu}\biggr) \,\delta(k_B) - \frac{8}{\mu}   \,\mathcal{L}_0\biggl(\frac{k_J}{\mu}\biggr) \ln \biggl(\frac{\nu}{2\mu\cosh\eta_J}\biggl) \,\delta(k_B)  + \frac{8}{\mu}\,\cL_1\Bigl(\frac{k_B}{\mu}\Bigr) \,\delta(k_J)\nn\\
  &\quad  -\frac{8}{\mu}\,\mathcal{L}_0\Bigl(\frac{k_B}{\mu}\Bigr) \ln \Bigl(\frac{\nu}{\mu}\Bigr)\,\delta(k_J)
       + s_{aJ,B}(\{d_i\},\eta_J)\,\frac{1}{\mu}\, \cL_0\Bigl(\frac{k_B}{\mu}\Bigr) \,\delta(k_J)    
      + s_{aJ,J}(\{d_i\},\eta_J)\, \frac{1}{\mu}\, \cL_0\Bigl(\frac{k_J}{\mu}\Bigr) \,\delta(k_B)
      \nn \\ & \quad     
      + s_{aJ,\delta}(\{d_i\},\{f_i\},\eta_J) \, \delta(k_J) \, \delta(k_B)  \biggr]
      + \bfT_b \cdot \bfT_J\biggl[\eta_J\leftrightarrow -\eta_J\biggr] \biggr\} \, ,
    \end{align}
Using the analytic results in eqs.~(\ref{eq:Sabanal}),~(\ref{eq:S_BC}),~(\ref{eq:S_hemiB})~and (\ref{eq:S_hemi2}) the coefficients of the distributions are given by 
\begin{align}\label{eq:soft_coeffs}
s_{ab,B} (\{d_i\},\eta_J)  & = 4I^J_{0,ab} (\{d_i\},\eta_J) \simeq 4R^2\, , \nn \\
s_{ab,\delta} (\{d_i\},f^\tau_B,f_J^\beta,\eta_J)  & =  -\frac{\pi^2}{3} + 4 I^J_{1,ab} (\{d_i\},f^\tau_B,f_J^\beta,\eta_J)  \, , \nn \\
s_{ab,\delta}(\{d_i\},f^C_B,f_J^\beta,\eta_J)  & = -\pi^2 +  4 I^J_{1,ab} (\{d_i\},f^C_B,f_J^\beta,\eta_J)  \, , \nn \\
s_{ab,\delta} (\{d_i\},f^{p_T}_B,f_J^\beta,\eta_J) & = \frac{\pi^2}{3} +  4 I^J_{1,ab} (\{d_i\},f_B^{p_T},f_J^\beta,\eta_J)\, , \nn \\
s_{aJ,B}(\{d_i\},\eta_J)  & = 8 (\eta_J+ \ln R) - 4 I^B_{0,aJ}(\{d_i\},\eta_J) +4 I^J_{0,aJ}(\{d_i\},\eta_J) \, , \nn \\
s_{aJ,J}(\{d_i\},\eta_J)  & = -8\ln R+4 I^B_{0,aJ}(\{d_i\},\eta_J) - 4I^J_{0,aJ}(\{d_i\},\eta_J)  \, , \nn \\
s_{aJ,\delta} (\{d_i\},f^{\tau}_B,f_J^\beta,\eta_J) & = -4\,\Li_2\bigl(e^{-2|\eta_J|}\bigr)+4 \eta_J^2 [\theta(\eta_J) - \theta(-\eta_J)] 
   \nn \\ & \quad 
  +2  \ln^2 R\bigl[2\beta-(\beta-2)\theta(R-1)\bigr]  + 8 |\eta_J| \ln R- \frac{\pi^2}{6}\,\frac{\beta}{\beta-1}  \,\delta_{\beta \neq 1} \nn \\
  & \quad  +4\Delta I_{1,aJ}^{B}(f^{\tau}_B,R,\eta_J)+4 \sum_{m=B,J}  I^m_{1,aJ}(\{d_i\},f^{\tau}_B,f_J^\beta,\eta_J)\, , \nn \\
s_{aJ,\delta} (\{d_i\},f^C_B,f_J^\beta,\eta_J) & = 4 \,\Li_2\Bigl(\frac{1\!+\!\tanh \eta_J}{2}\Bigr) - 2 \ln^2\Bigl(\frac{1\!+\!\tanh \eta_J}{2}\Bigr) +4\eta_J^2 + 8 \ln R \, \ln (2\cosh \eta_J) \nn  \\
& \quad +2  \ln^2 R\bigl[2\beta-(\beta-2)\theta(R-1)\bigr] - \frac{\pi^2}{6}\Bigl[4+\frac{\beta}{\beta-1} \,\delta_{\beta \neq 1}\Bigr]   \nn \\
& \quad +4\Delta I_{1,aJ}^{B}(f^{C}_B,R,\eta_J)+4\sum_{m=B,J} I^m_{1,aJ}(\{d_i\},f^{C}_B,f_J^\beta,\eta_J)\, , \nn \\
s_{aJ,\delta} (\{d_i\},f^{p_T}_B,f_J^\beta,\eta_J) & =2  \ln^2 R\bigl[2\beta-(\beta-2)\theta(R-1)\bigr] +\frac{\pi^2}{6}\Bigl[2-\frac{\beta}{\beta-1} \,\delta_{\beta \neq 1}\Bigr] \nn \\
& \quad +4\sum_{m=B,J} I^m_{1,aJ}(\{d_i\},f^{p_T}_B,f_J^\beta,\eta_J)\, ,
\end{align}
where $\delta_{\beta\neq1}=1$ for $\beta \neq 1$ and zero otherwise. The numerical integrals $I^J_{0,ab}$ and $I^J_{1,ab}$ are defined in \eq{abnum}, $I_{0,aJ}^B$ and $I_{1,aJ}^B$ are defined in \eq{Iedge}, $I_{0,aJ}^J$ and $I_{1,aJ}^J$ are defined in \eq{Iedge2} and $\Delta I_{1,aJ}^{B}(f_B,R,\eta_J)$ is given in \eq{I1diff}.

As one can see from \eq{soft_coeffs} the soft function contains Sudakov double logarithms $\ln R$ and $\ln e^{\eta_J}$ which deteriorate the perturbative expansion of the soft function for a small jet radius and forward jets and may require an all-order resummation. This can be achieved by additional factorization of the soft function in the framework of \SCETp theories as discussed e.g.~in refs.~\cite{Bauer:2011uc,Larkoski:2015kga,Pietrulewicz:2016nwo,Chien:2015cka,Becher:2015hka,Kolodrubetz:2016dzb}.

%===============================================================================
\subsection{Full numerical results}
\label{sec:num_softs}
%===============================================================================

\begin{figure}
\centering
\includegraphics[height=3.8cm]{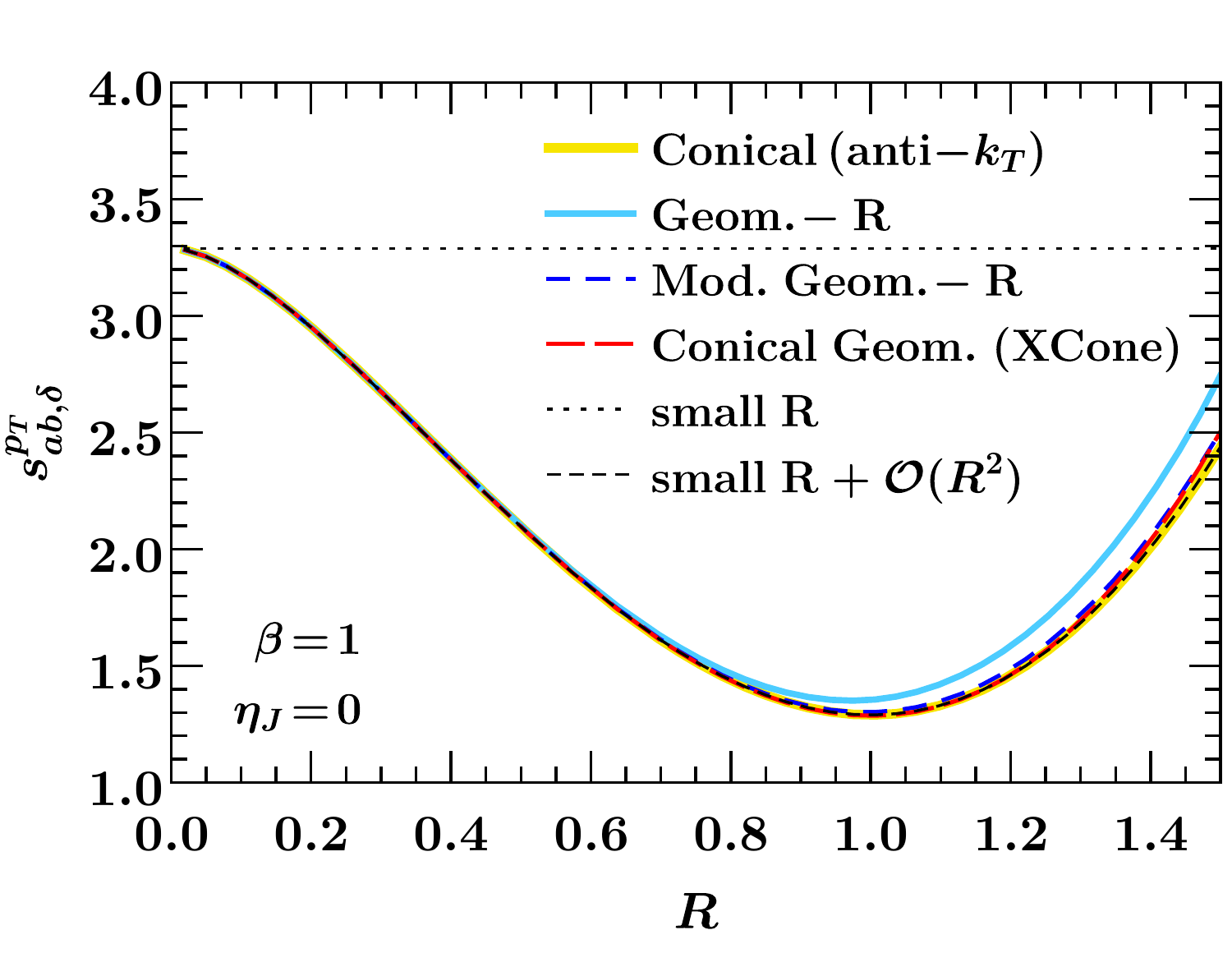} \hspace{0.2cm}
\includegraphics[height=3.8cm]{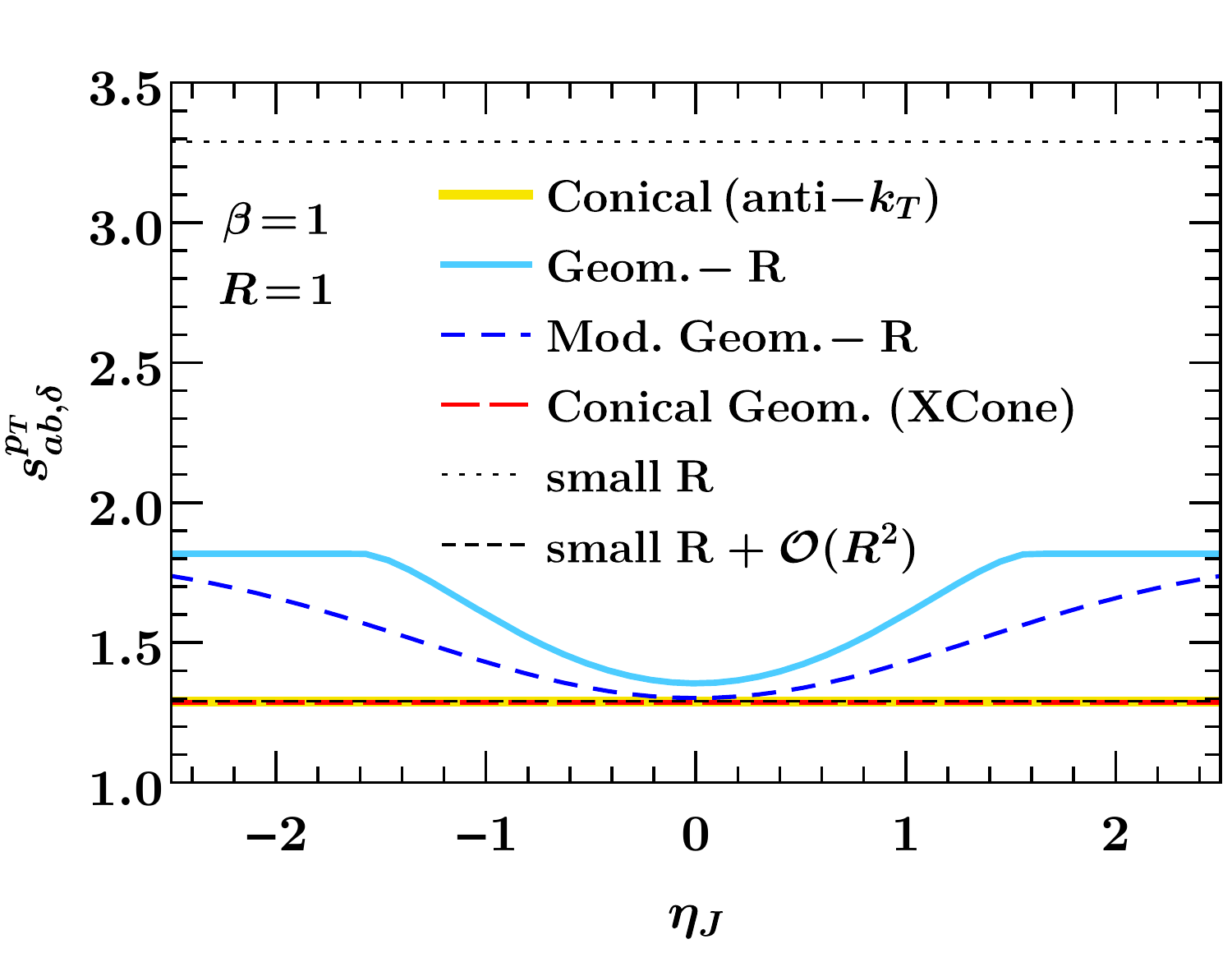} \\
\includegraphics[height=3.8cm]{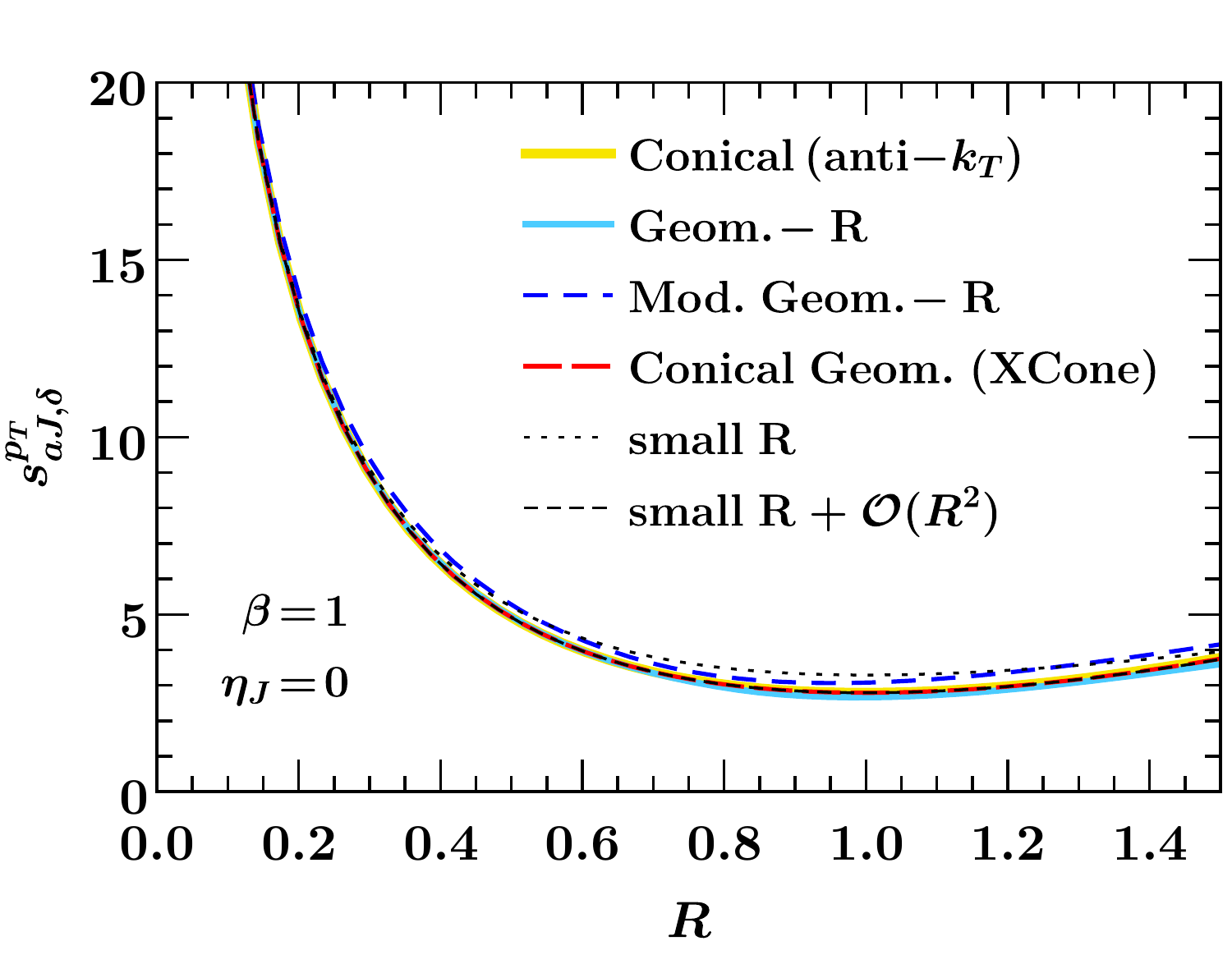} \hspace{0.2cm}
\includegraphics[height=3.8cm]{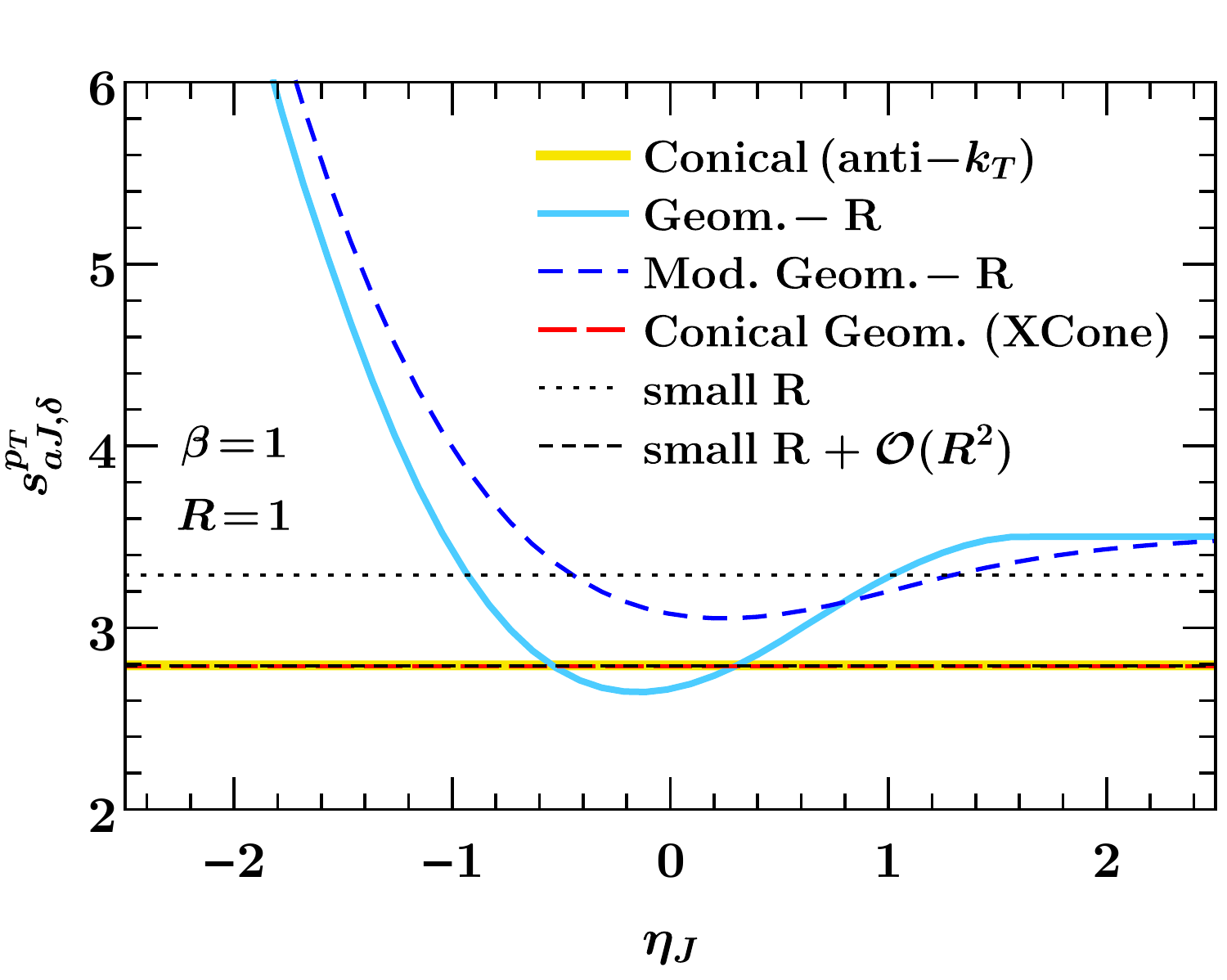}
\caption{The coefficients $s_{ab,\delta}$ (top row) and $s_{aJ,\delta}$ (bottom row) for a jet angularity with $\beta=1$, for the various distance measures and with the small $R$ results for a $p_T$ jet veto in the beam region, for $\eta_J=0$ in terms of $R$ (left column) and for $R=1$ in terms of $\eta_J$ (right column). \label{fig:S_beta1}} 
\end{figure}

We now compare the contributions to the soft function, shown through plots of the various coefficients $s_{ab}$, $s_{aJ}$ of the distributions defined in \eq{soft_coeffs}. Our main focus is on the jet mass measurement ($\beta=2$) but we also show a few results for a jet angularity measurement with $\beta=1$ in \fig{S_beta1}. We consider the various partitionings described in \sec{jetregions} and beam region observables in \eq{fBchoices}.

\begin{figure}
\centering
\includegraphics[height=3.8cm]{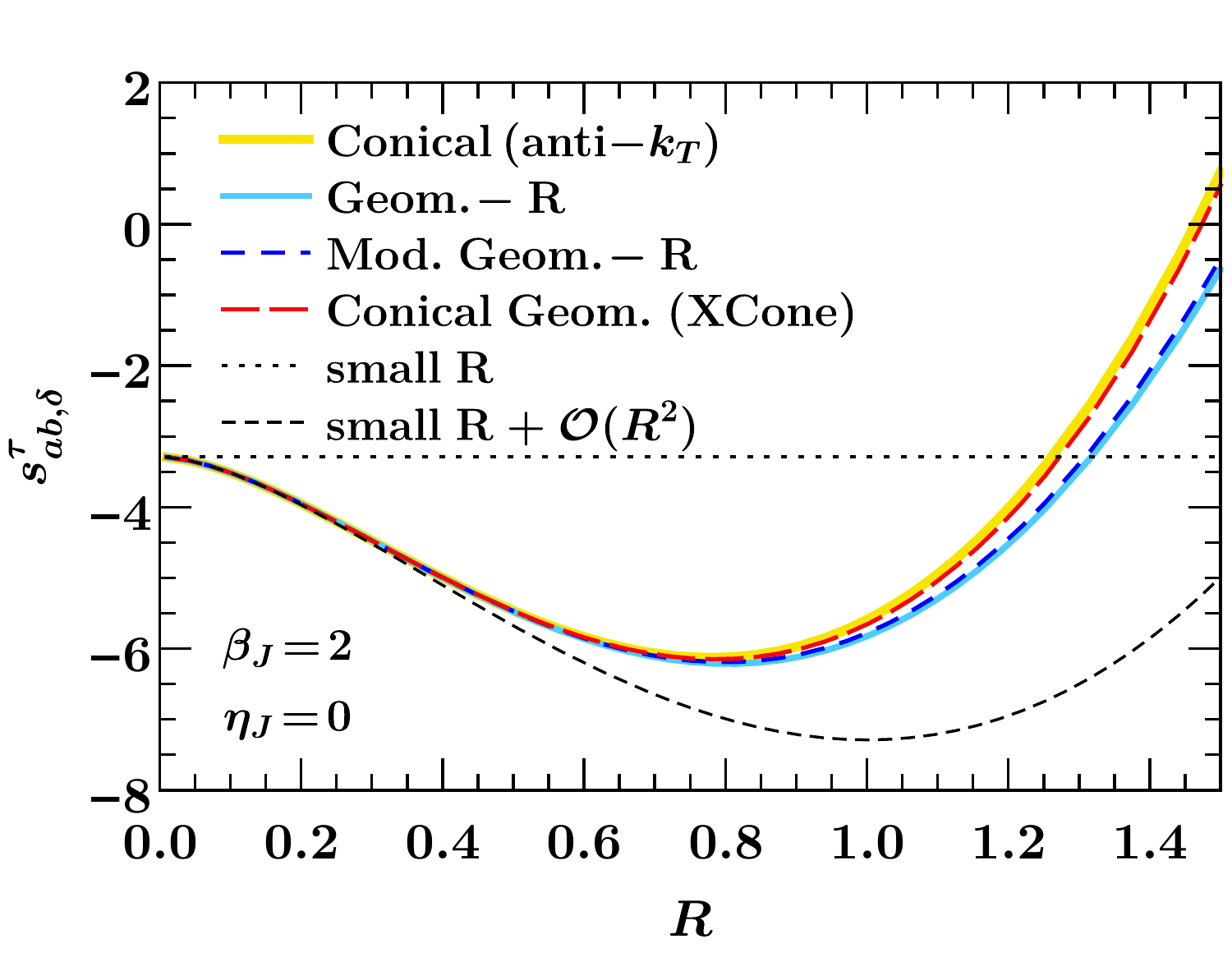}%
\hfill%
\includegraphics[height=3.8cm]{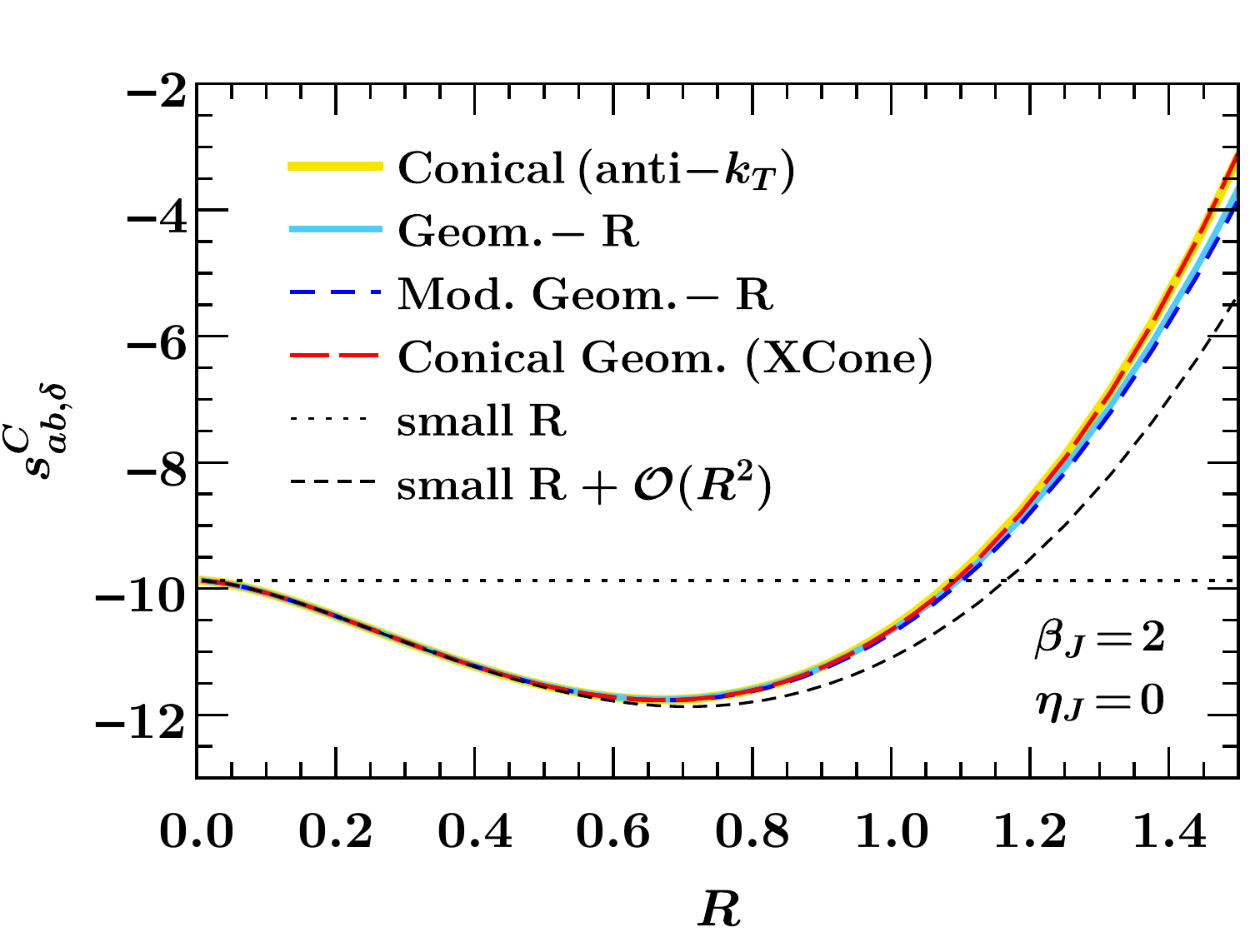}%
\hfill%
\includegraphics[height=3.8cm]{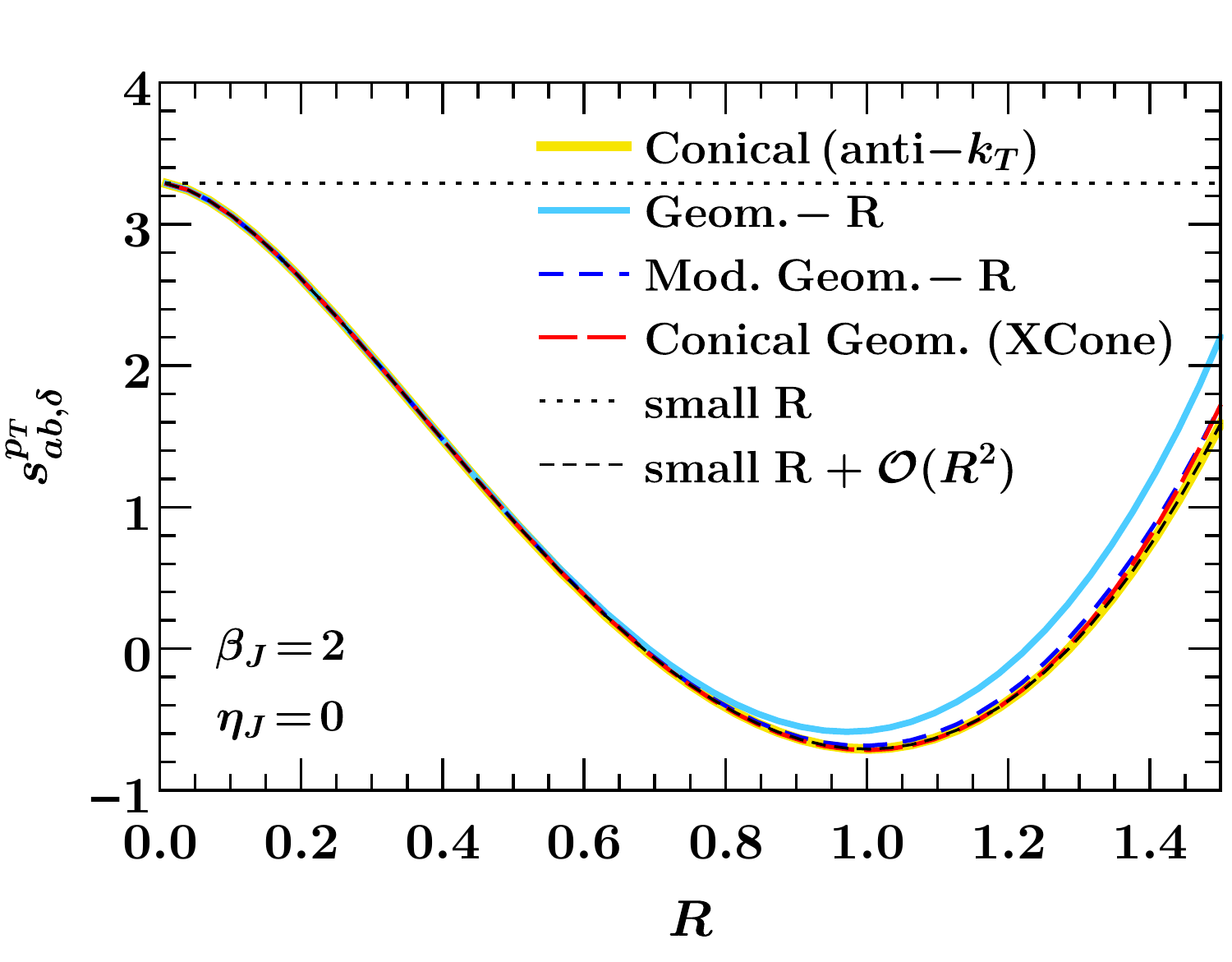}%
\\
\centering
\includegraphics[height=3.8cm]{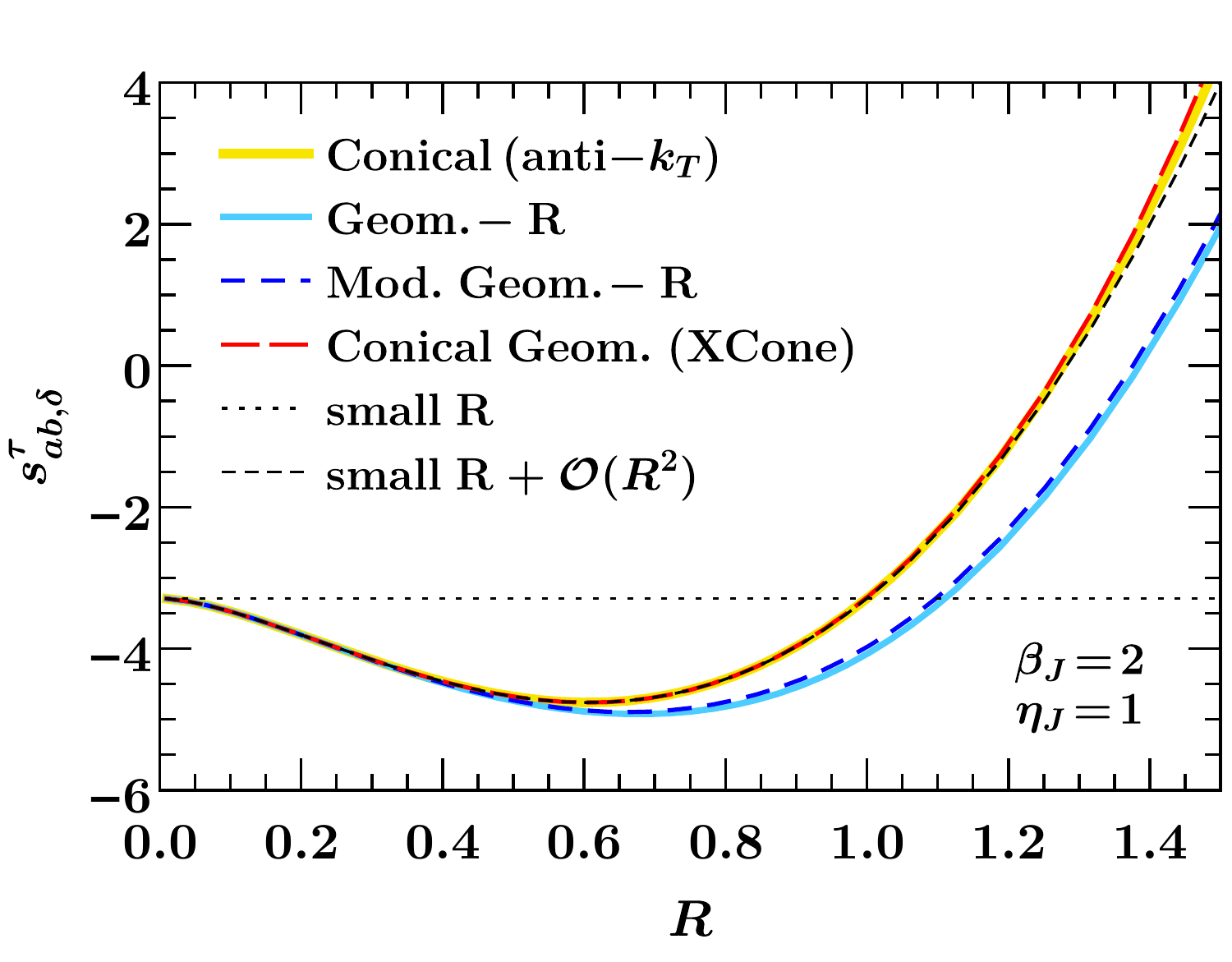}%
\hfill%
\includegraphics[height=3.8cm]{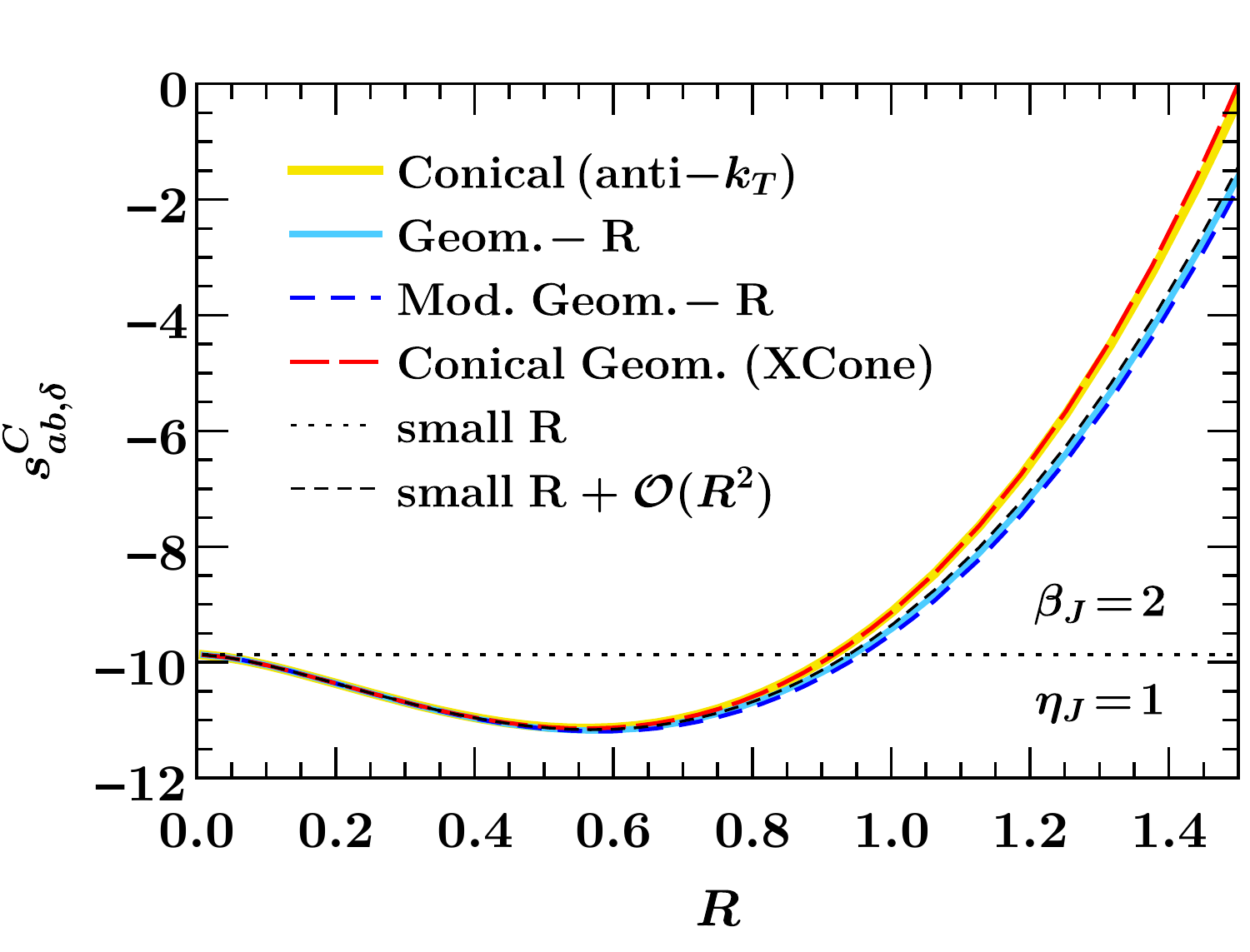}%
\hfill%
\includegraphics[height=3.8cm]{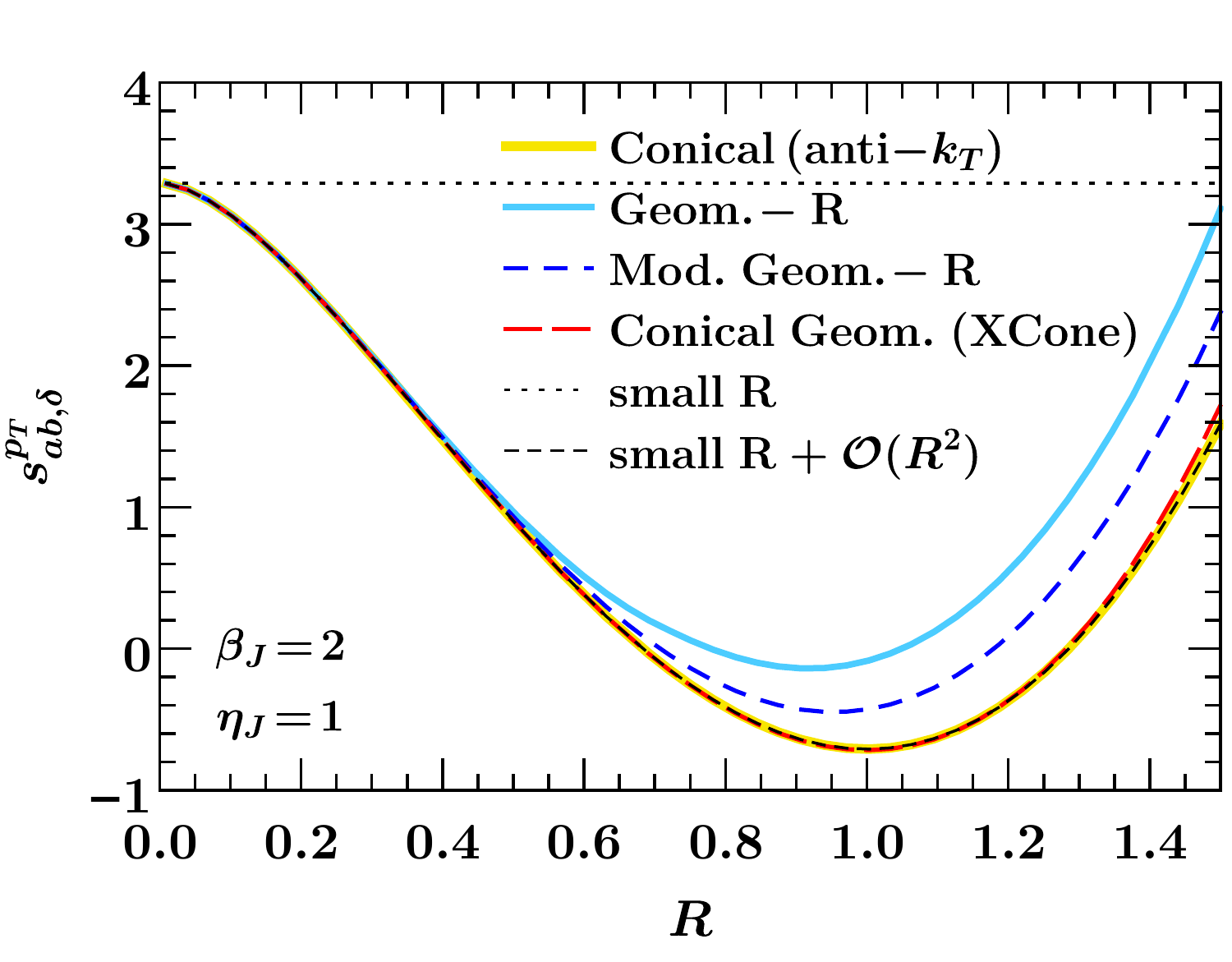}%
\caption{The coefficient $s_{ab,\delta}$ for the various distance measures and with the small $R$ results for beam thrust (left column), C-parameter (middle column) and $p_T$ (right column) for a jet mass measurement ($\beta=2$) for $\eta_J=0$ (top row) and $|\eta_J| =1$ (bottom row) as function of $R$. For the $p_T$ measurement including the analytic corrections at $\mathcal{O}(R^2)$ yield already the exact result for anti-$k_T$.\label{fig:Sab12}} 
\end{figure}

\begin{figure}
\centering
\includegraphics[height=3.8cm]{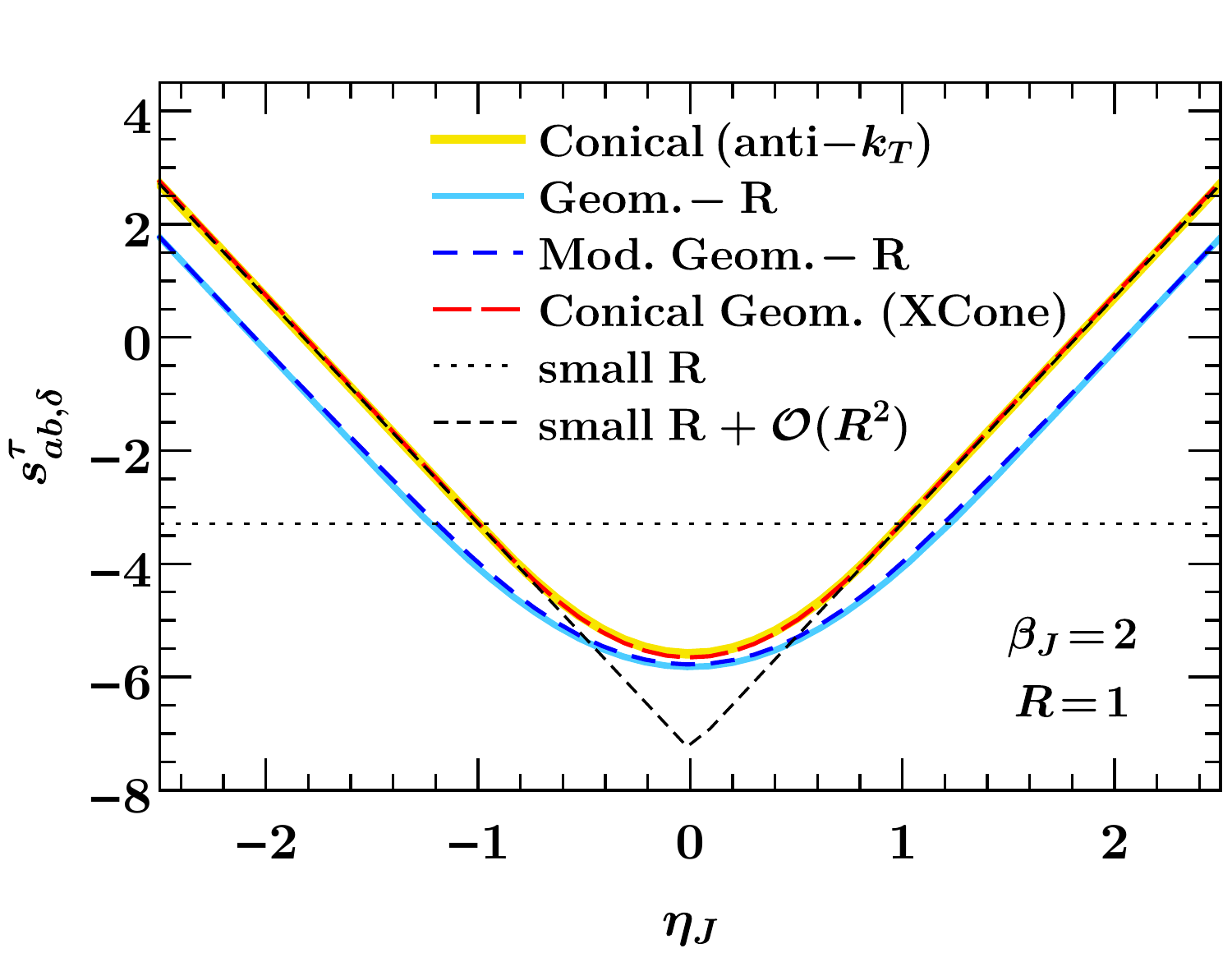}%
\hfill%
\includegraphics[height=3.8cm]{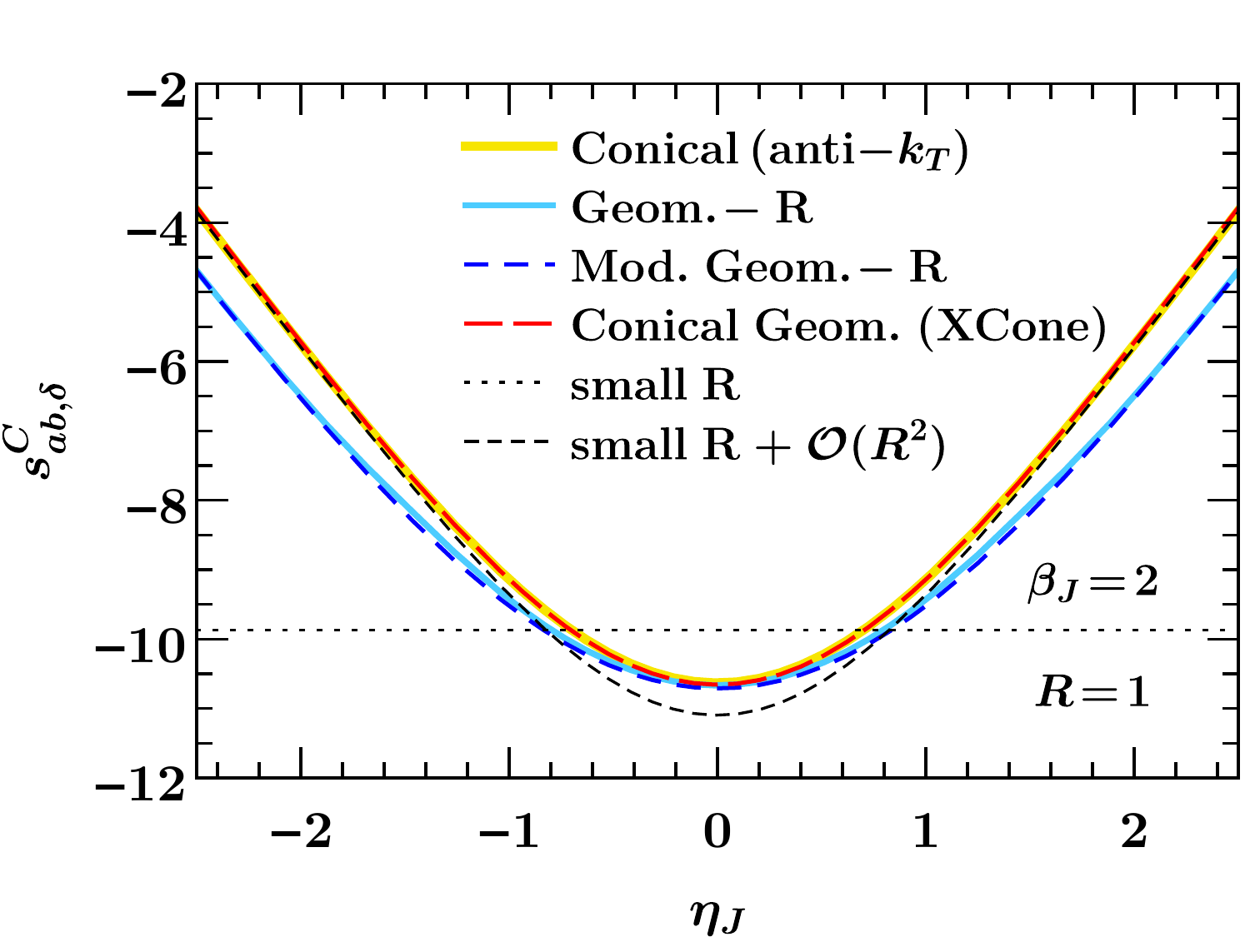}%
\hfill%
\includegraphics[height=3.8cm]{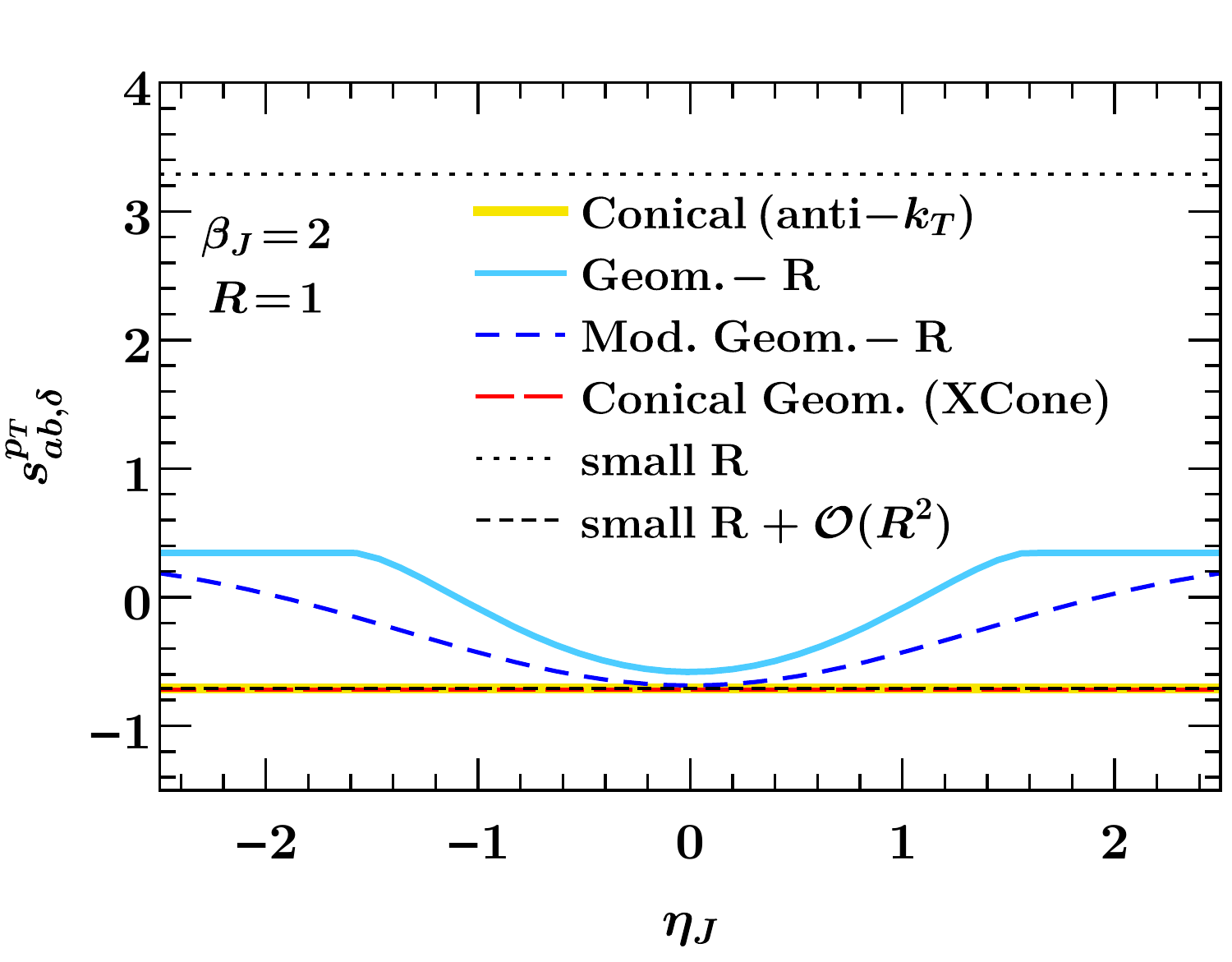}%
\caption{Same as \fig{Sab12}, but for $R=1$ as function of $\eta_J$.
\label{fig:Sab1_eta}} 
\end{figure}

\begin{figure}
\centering
\includegraphics[height=3.8cm]{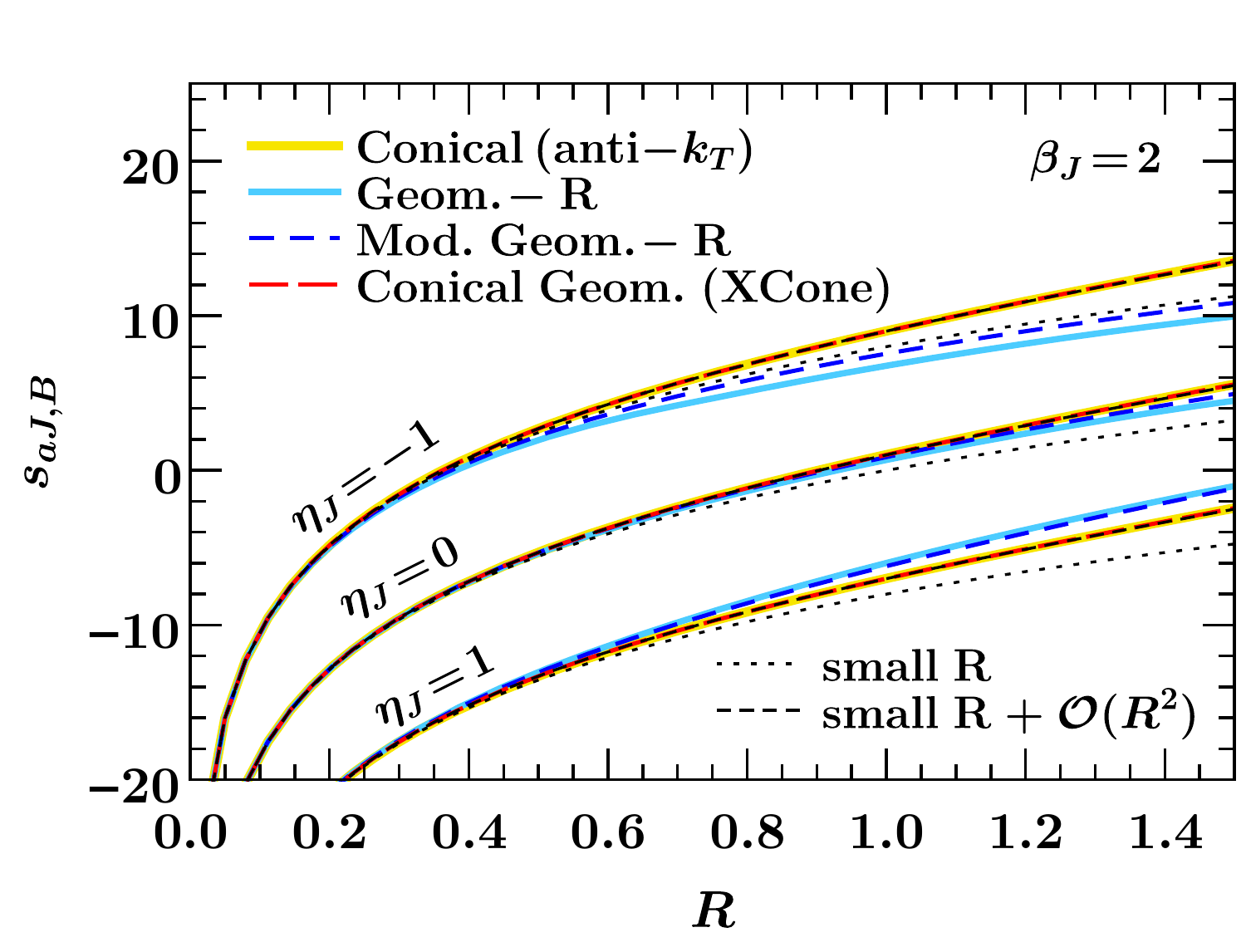}%
\hfill%
\includegraphics[height=3.8cm]{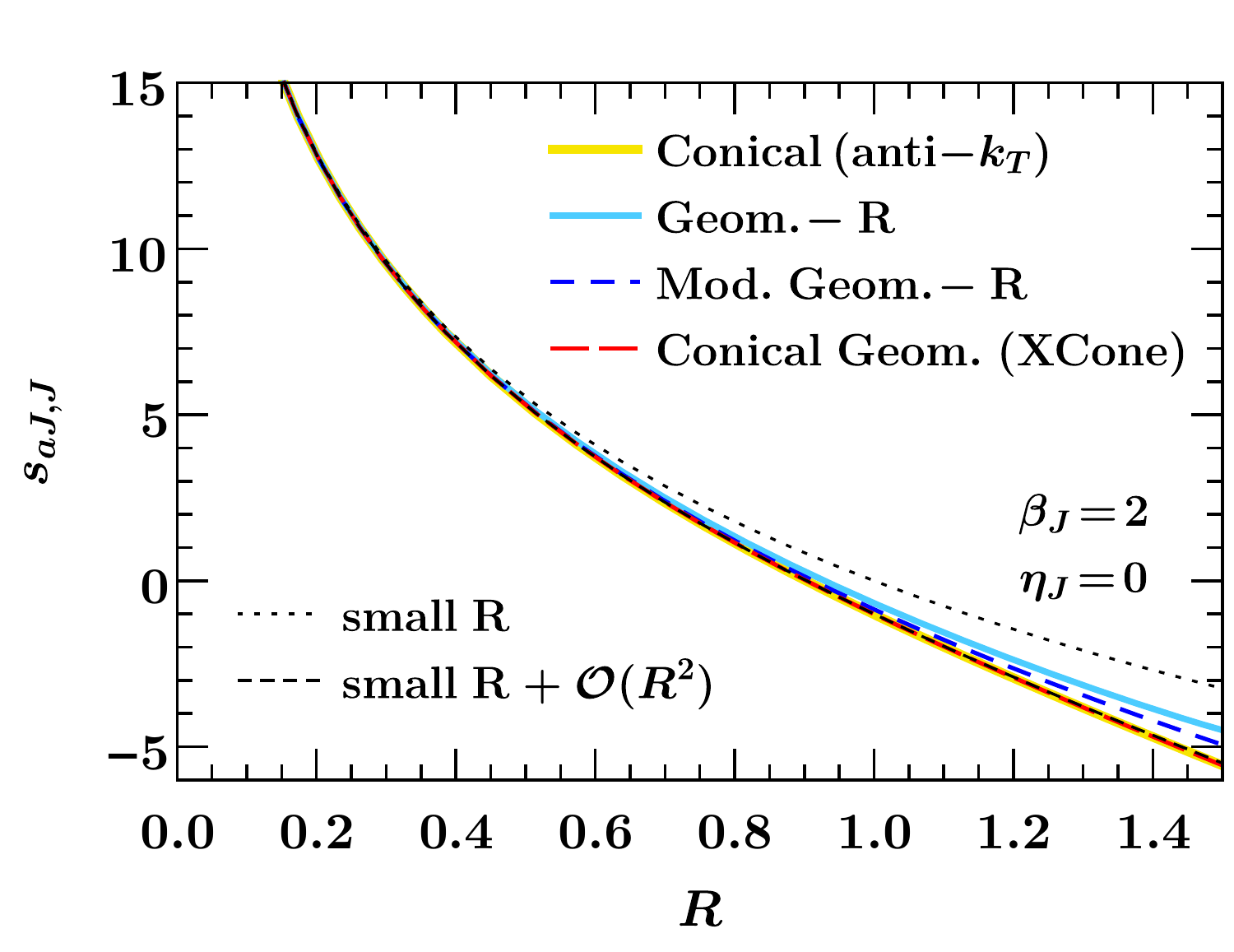}%
\hfill%
\includegraphics[height=3.8cm]{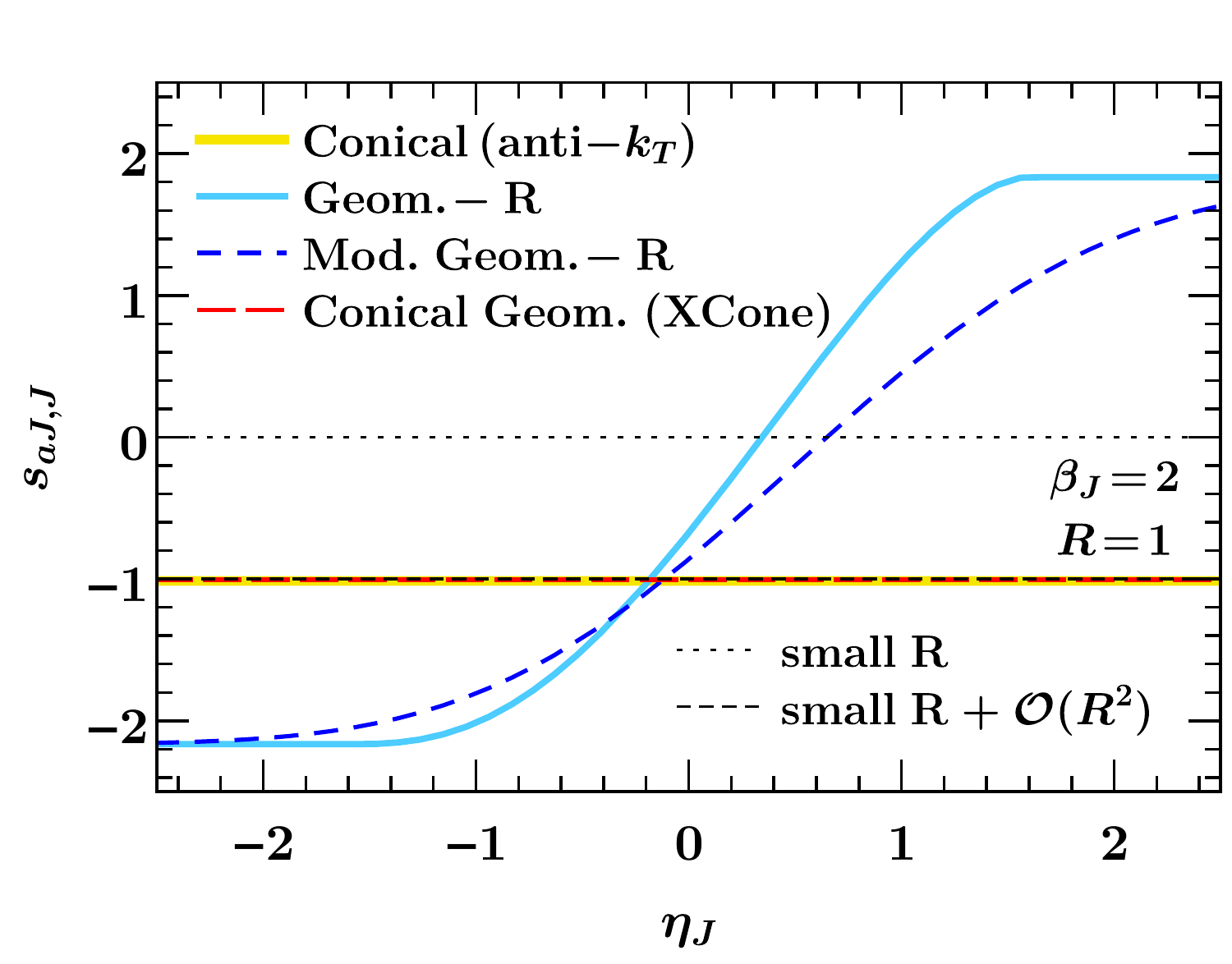}%
\caption{The coefficients $s_{aJ,B}$ and $s_{aJ,J}$ for the various distance measures and with the small $R$ results. These are independent of the specific measurements in the beam and jet regions. Shown are $s_{aJ,B}$ for $\eta_J =-1,0,1$ in terms of $R$ (left), $s_{aJ,J}$ for $\eta_J=0$ as function of $R$ (middle) and for $R=1$ as function of $\eta_J$ (right). \label{fig:SaJ0}} 
\end{figure}

\begin{figure}
\centering
\includegraphics[height=3.8cm]{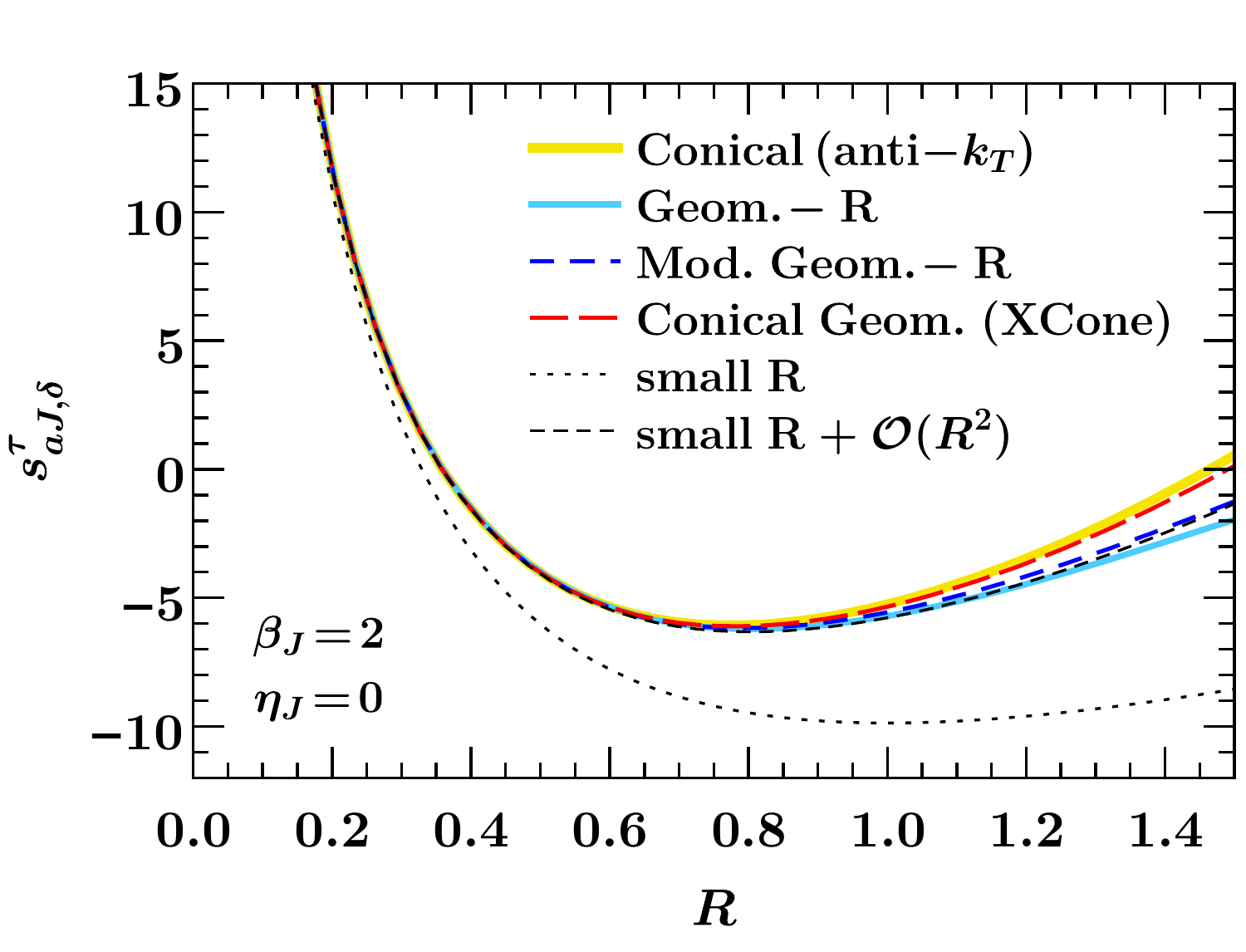}%
\hfill%
\includegraphics[height=3.8cm]{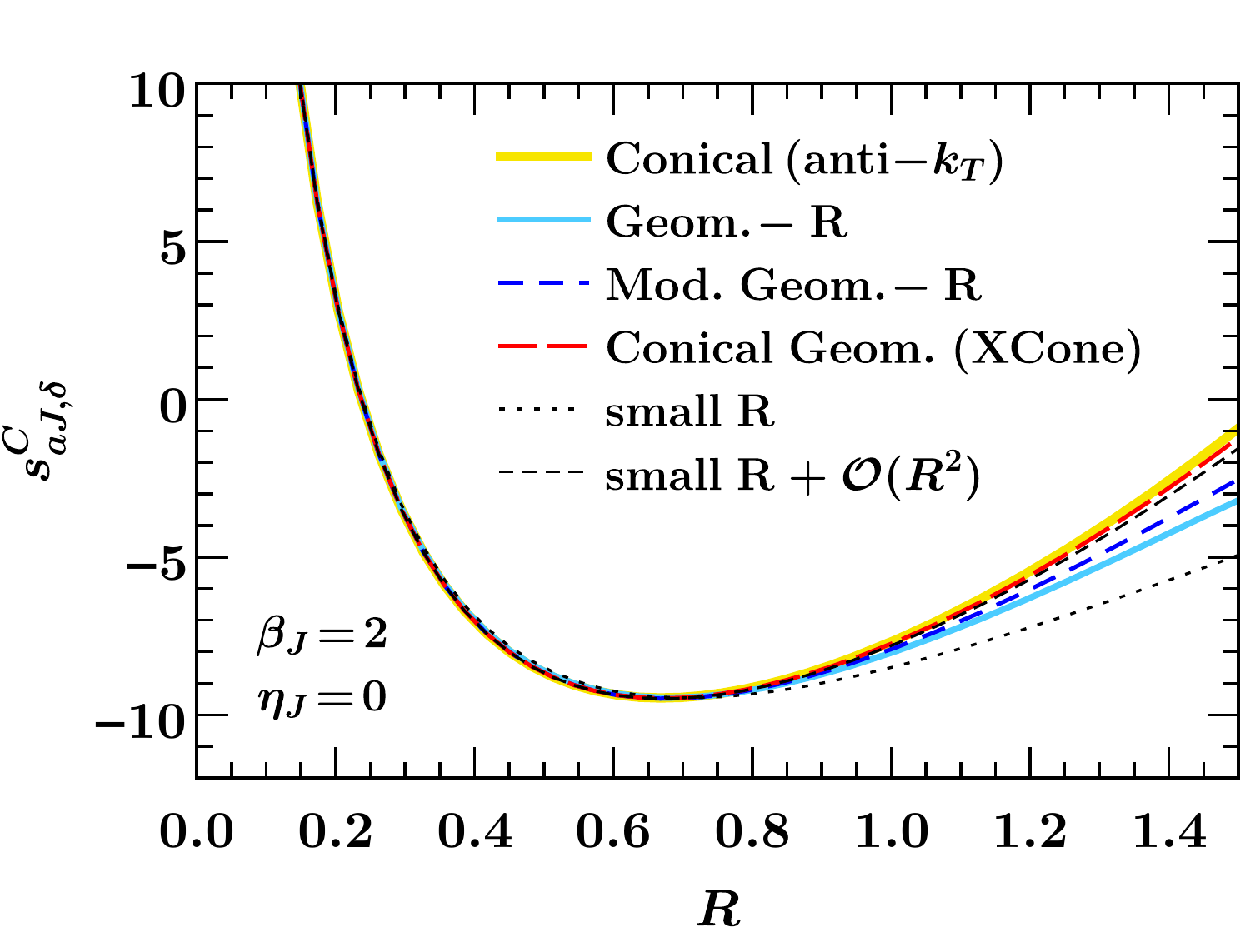}%
\hfill%
\includegraphics[height=3.8cm]{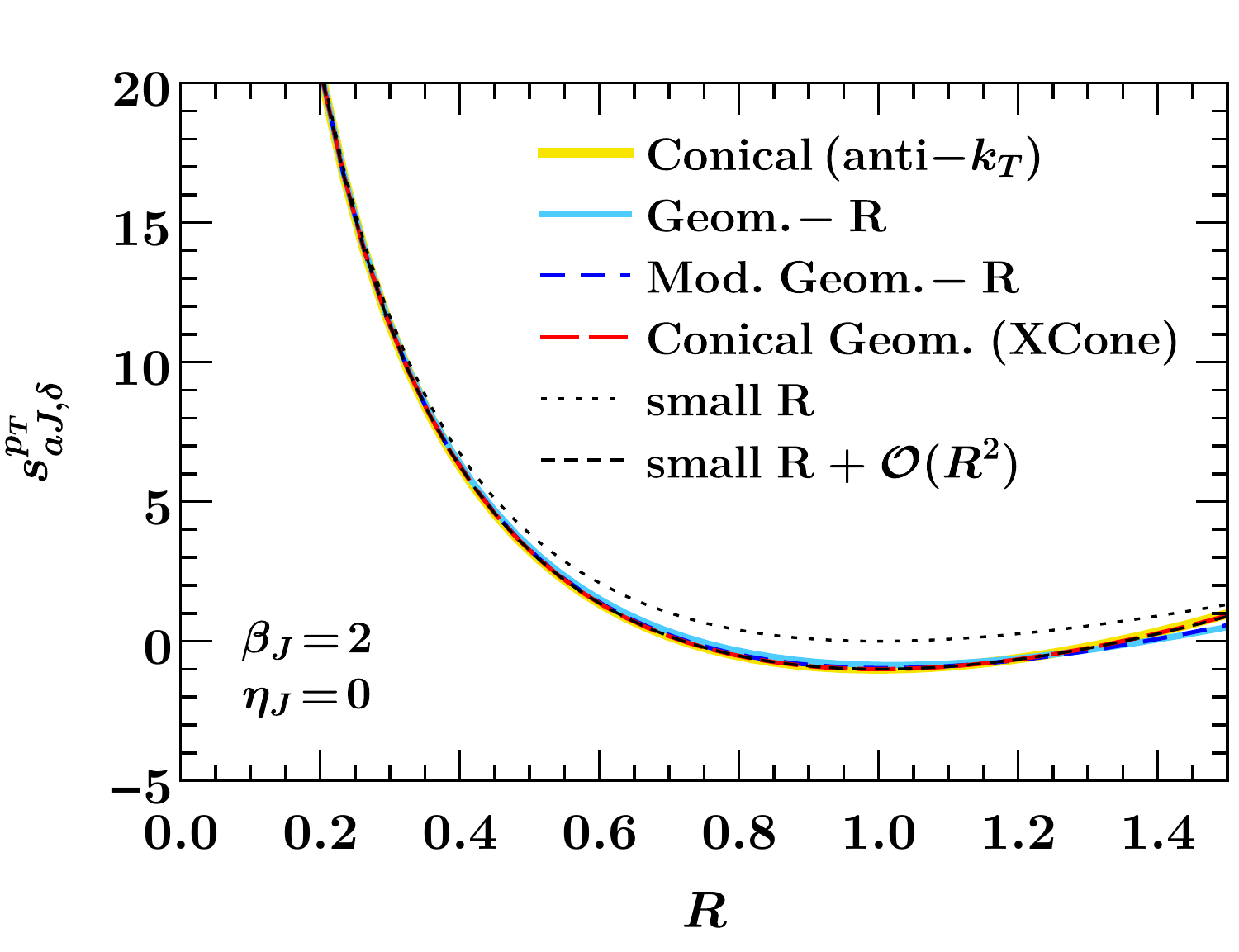}%
\\
\includegraphics[height=3.8cm]{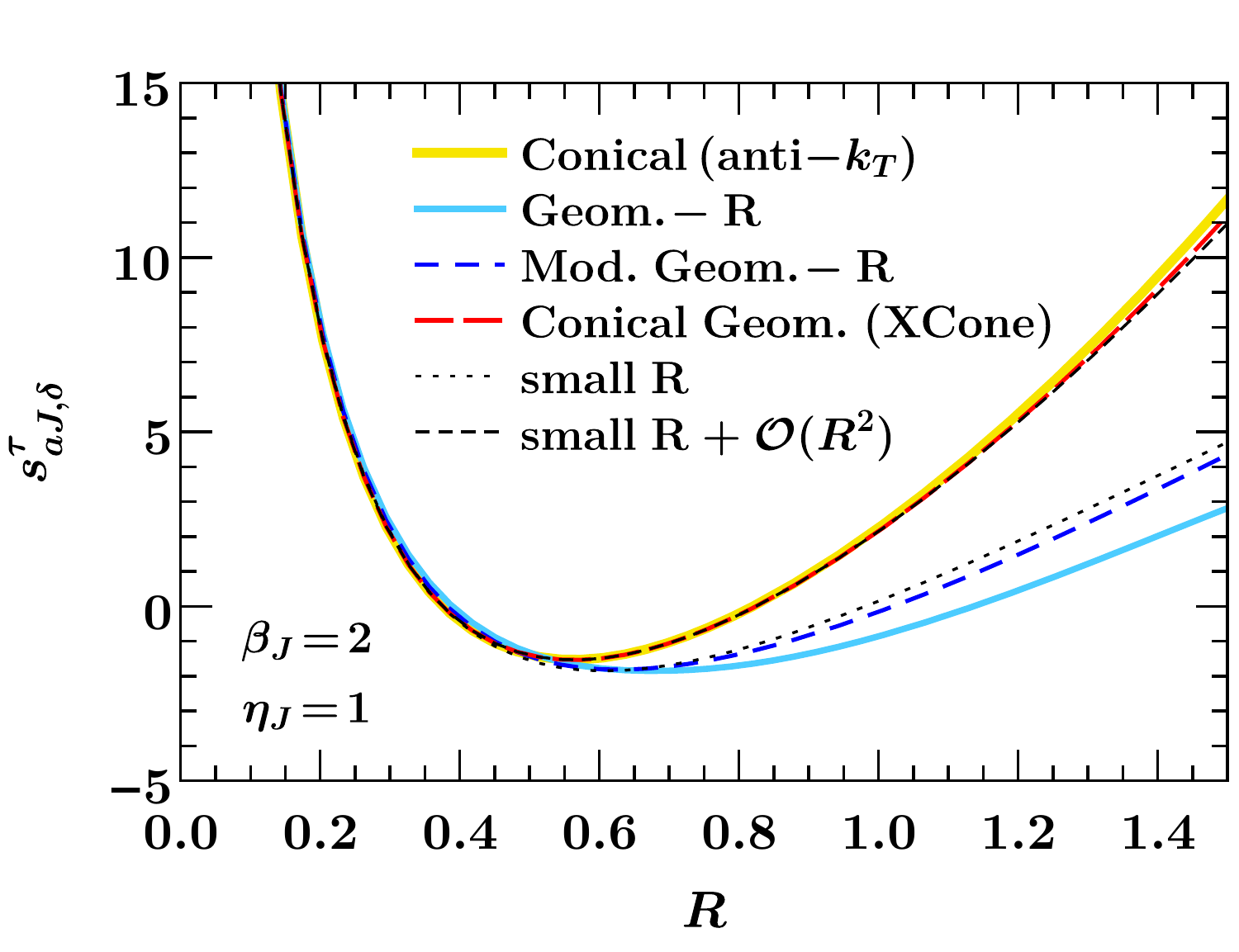}%
\hfill%
\includegraphics[height=3.8cm]{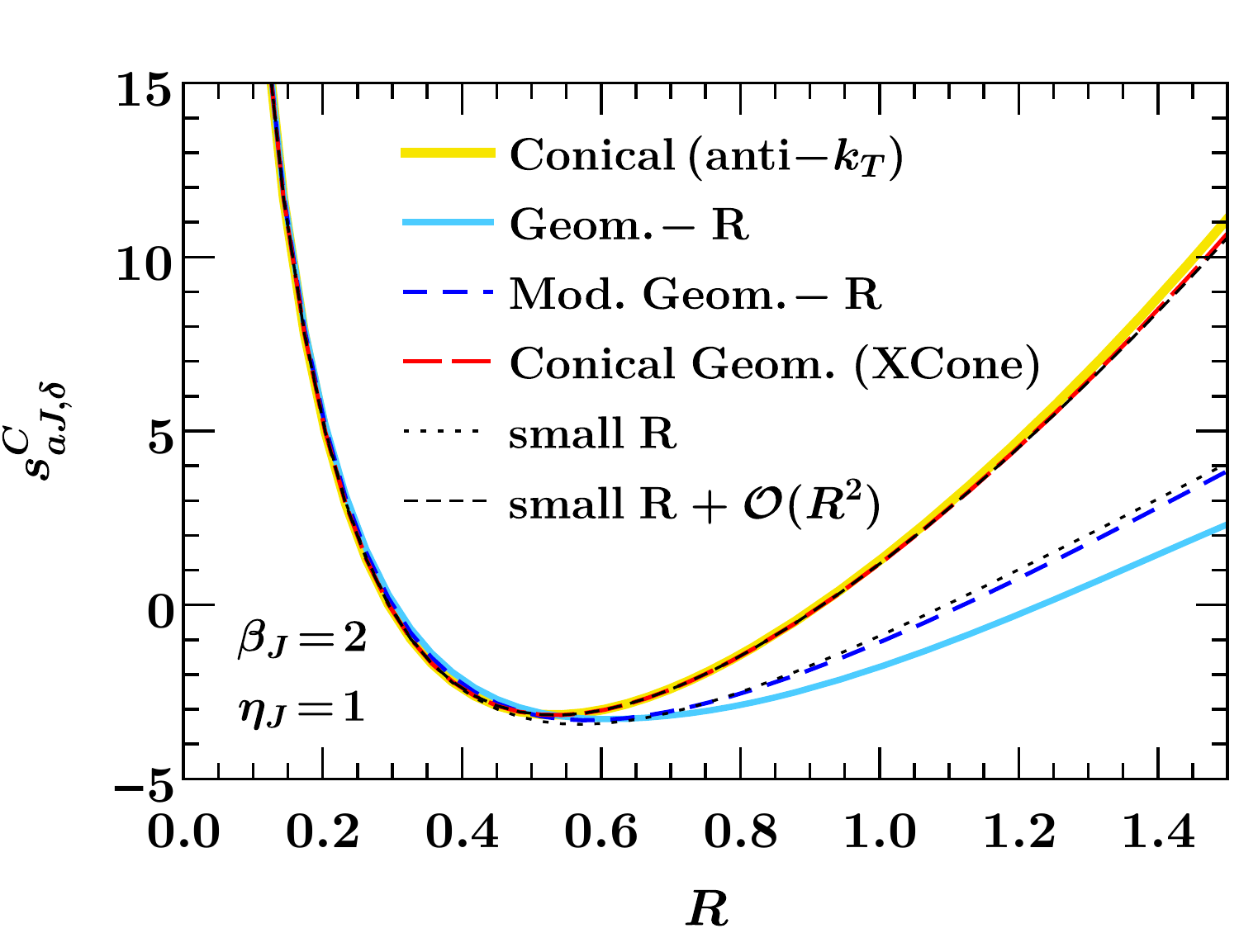}%
\hfill%
\includegraphics[height=3.8cm]{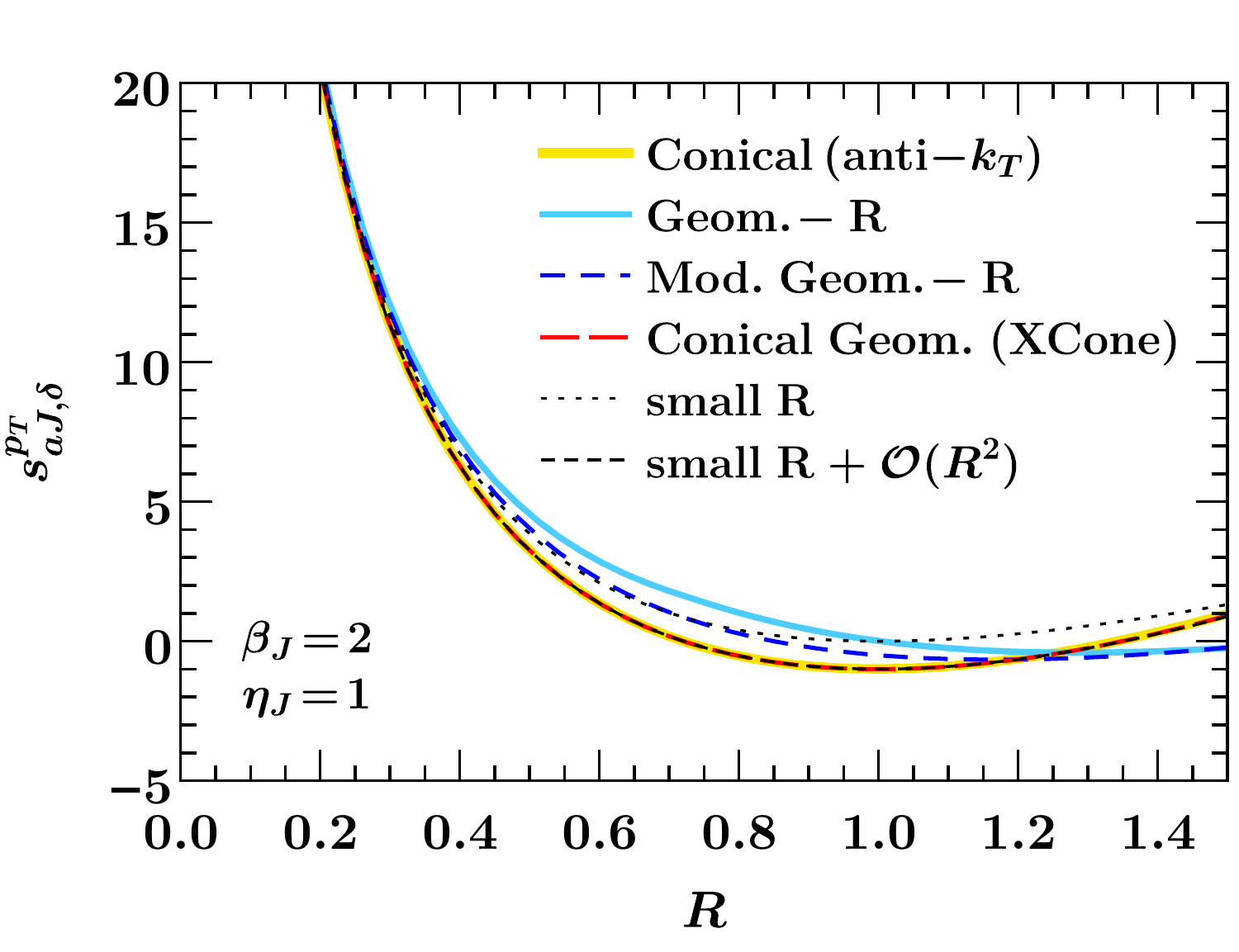}%
\\
\includegraphics[height=3.8cm]{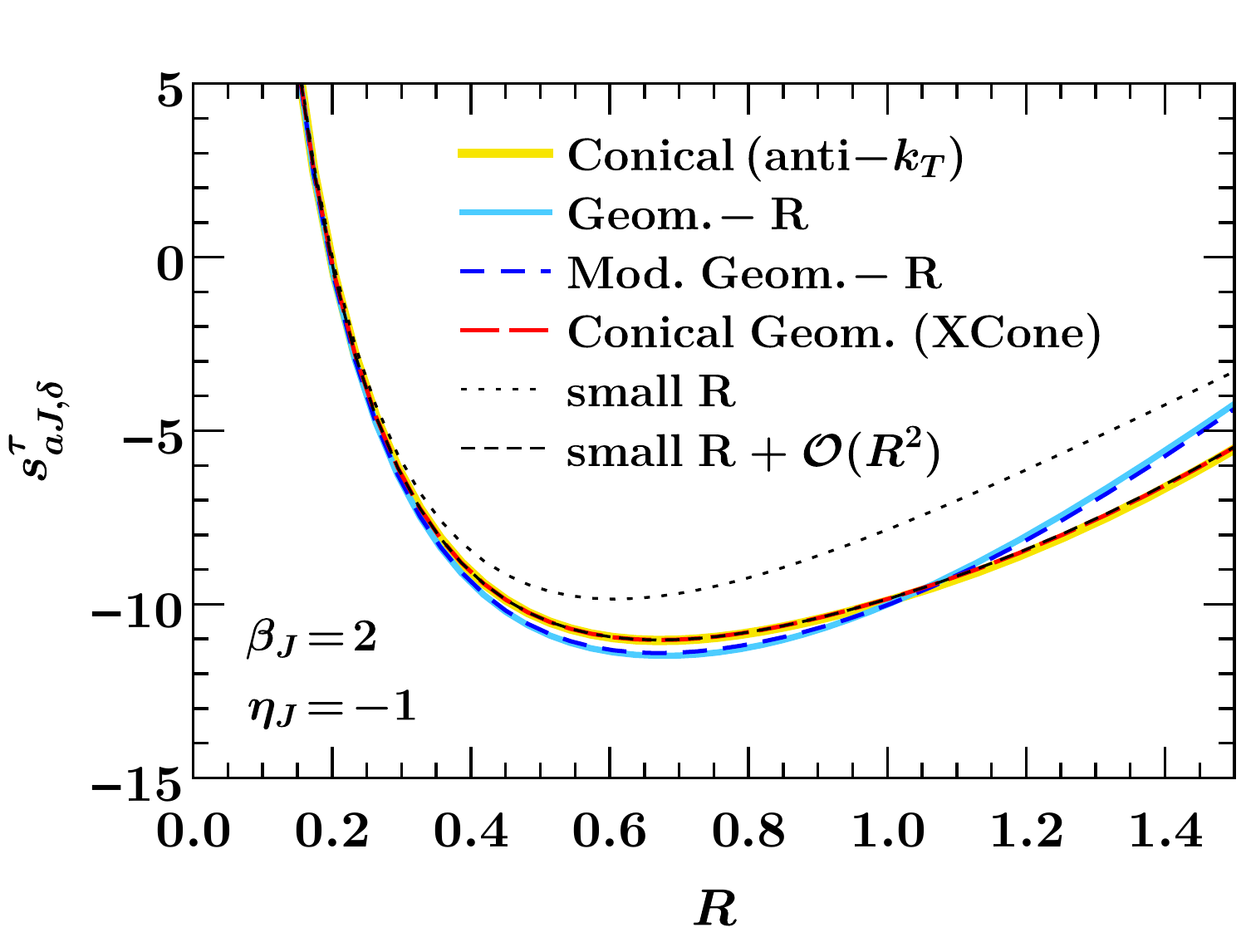}%
\hfill%
\includegraphics[height=3.8cm]{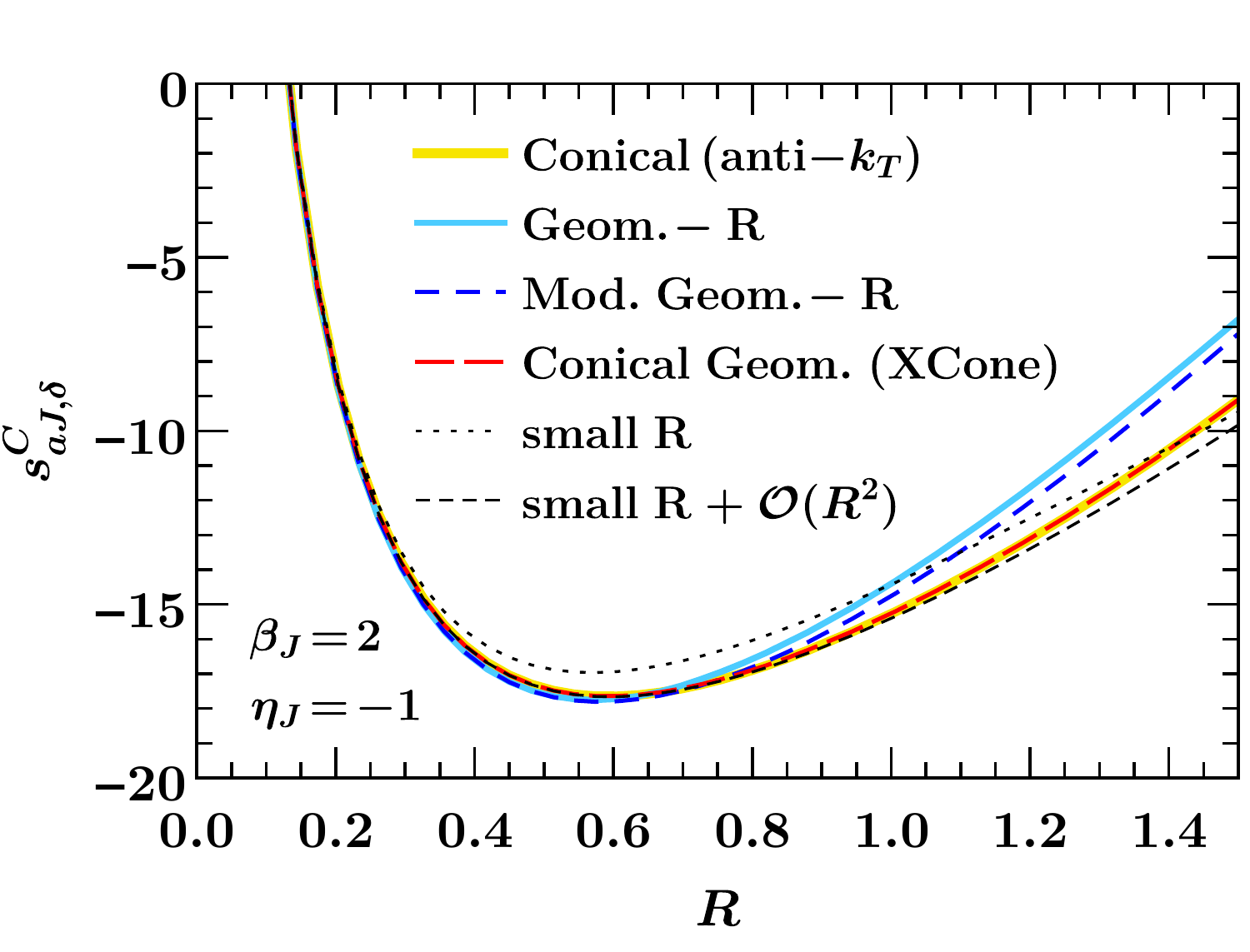}%
\hfill%
\includegraphics[height=3.8cm]{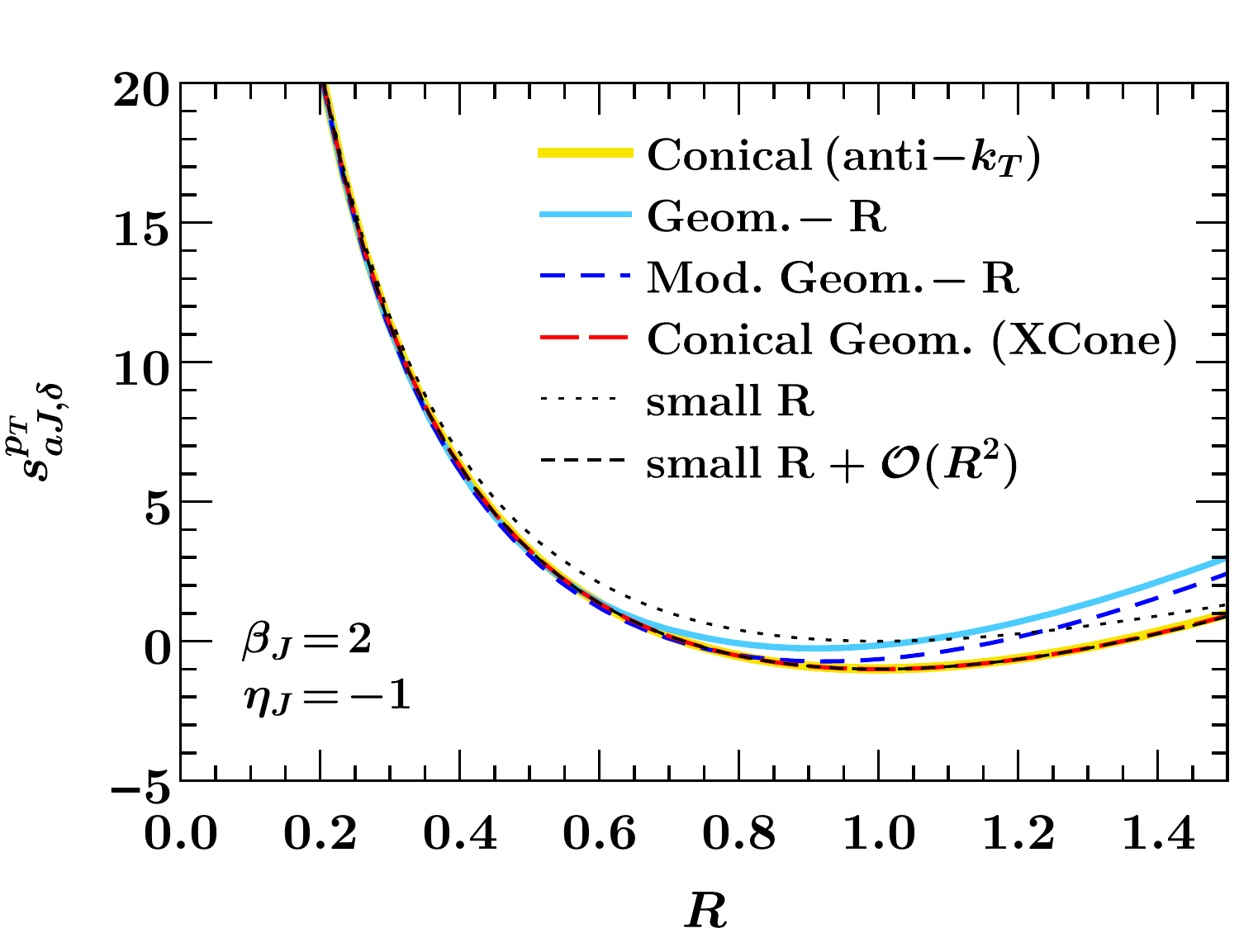}%
\caption{The coefficient $s_{aJ,\delta}$ for the various distance measures and with the small $R$ results for beam thrust (left column), C-parameter (middle column) and $p_T$ (right column) for a jet mass measurement ($\beta=2$) for $\eta_J=0$ (top row), $\eta_J =1$ (middle row) and $\eta_J=-1$ (bottom row) as function of $R$.  \label{fig:SaJ12}} 
\end{figure}

\begin{figure}
\centering
\includegraphics[height=3.8cm]{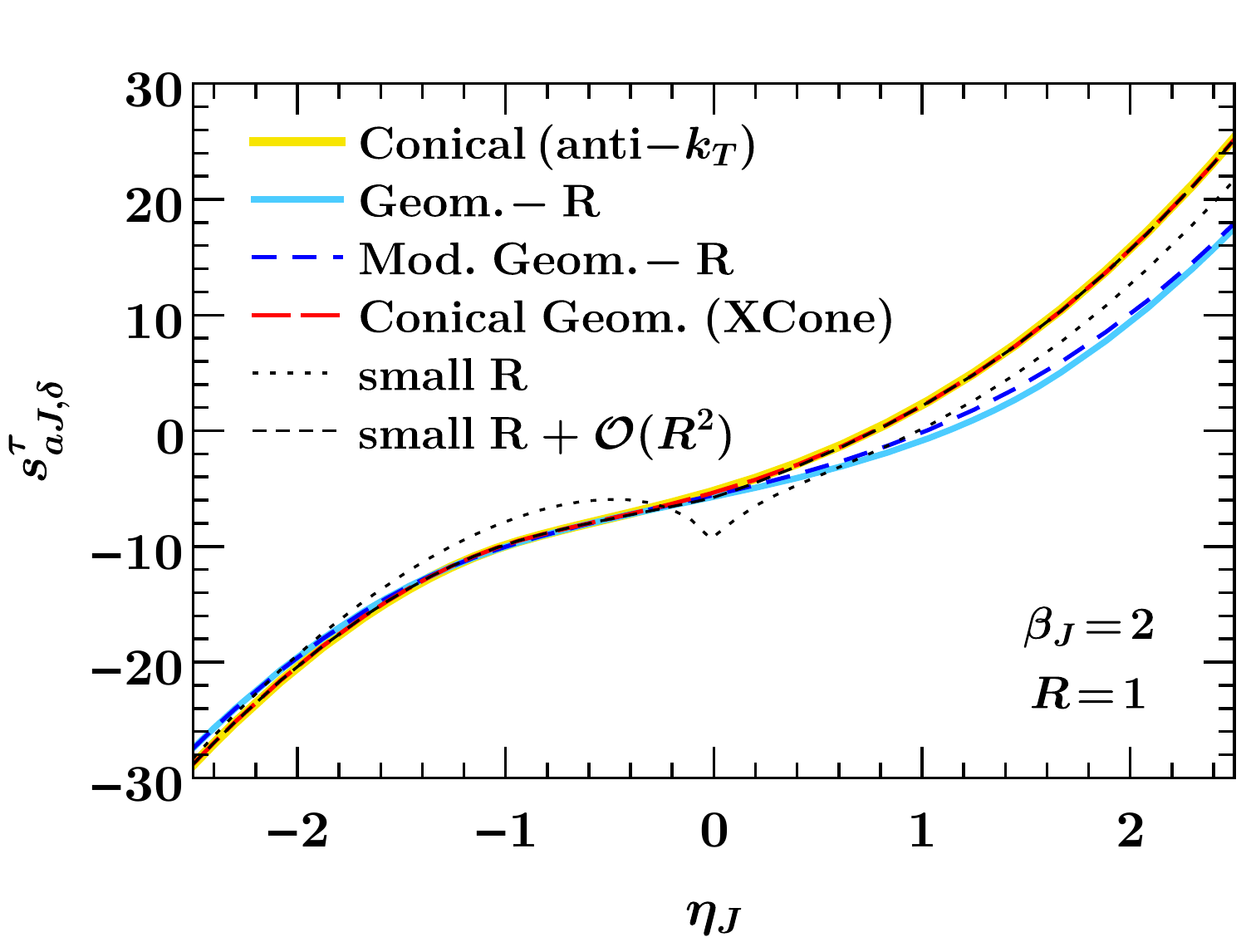}%
\hfill%
\includegraphics[height=3.8cm]{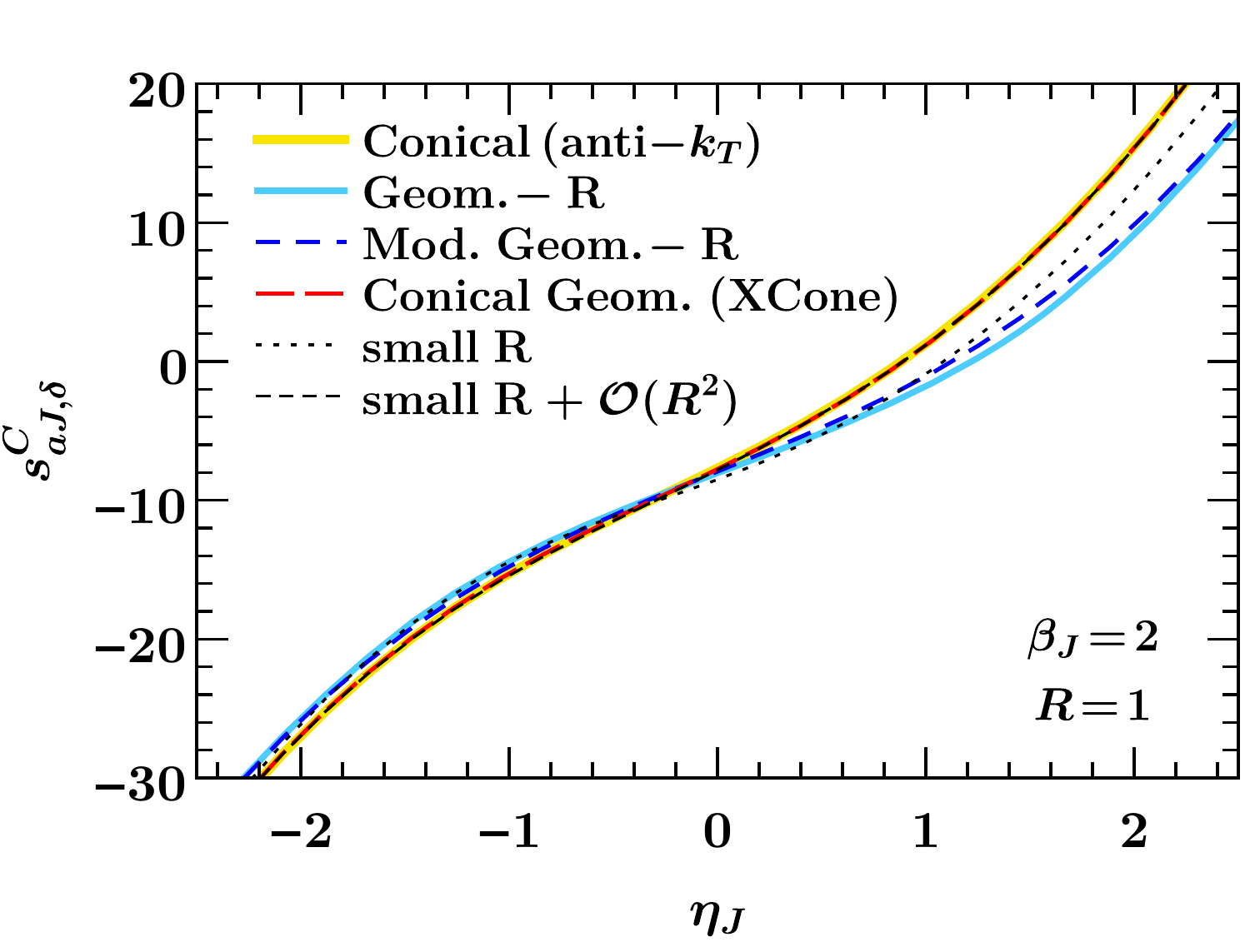}%
\hfill%
\includegraphics[height=3.8cm]{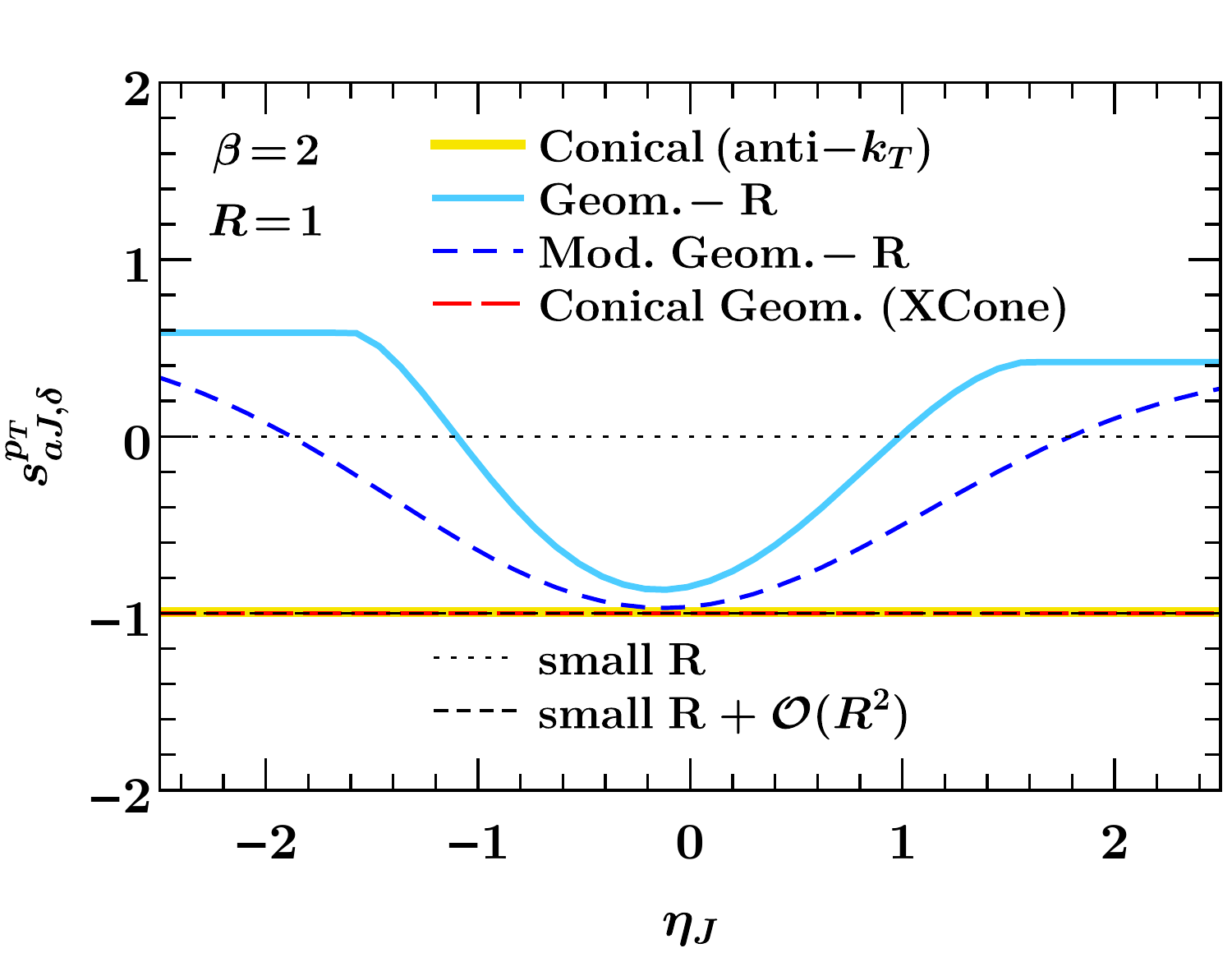}%
\caption{Same as \fig{SaJ12} but for $R=1$ as function of $\eta_J$.
\label{fig:SaJ1_eta}} 
\end{figure}

The contributions from the beam-beam dipole $s_{ab,\delta}$ are shown in \fig{Sab12} for $\eta_J=0$ and $|\eta_J|=1$ as a function of $R$, and in \fig{Sab1_eta} for $R=1$ as function of $\eta_J$. The results deviate from the $\mathcal{O}(R^0)$ result away from $R =0$, in particular also for the phenomenologically relevant values $R \sim 0.5$. However, including the $\mathcal{O}(R^2)$ corrections, the analytic contributions agree very well with the exact results for central rapidities even for values as large as $R \sim 1$. These $\mathcal{O}(R^2)$ corrections are the same for all distance measures, which explains why they behave very similar, and they are enhanced by logarithms of the jet radius, as can be seen from \eqs{sab_delta_R2}{saj_delta_R}. For the transverse momentum beam measurement with a conical anti-$k_T$ jet (red curves in the right panels of \figs{Sab12}{Sab1_eta}), there are in fact no higher order $R$ corrections beyond $\mathcal{O}(R^2)$ for $s_{ab,\delta}$. Otherwise, the next corrections are $\mathcal{O}(R^4)$ except for the beam thrust case with $|\eta_J| \lesssim R$ where they are $\mathcal{O}(R^3)$ due to the kink at $\eta=0$. This explains the larger deviation between the analytic $\mathcal{O}(R^2)$ beam thrust result and the exact result for $\eta_J =0$ as seen in the top-left panel of \fig{Sab12}. At large jet rapidities there are sizable differences between the geometric-R measures and the conical (and conical geometric) measure, which is due to the different jet shapes illustrated in \fig{jetregions}.

Results for the beam-jet dipole coefficients $s_{aJ,B}$ and $s_{aJ,J}$ are shown in \fig{SaJ0} and these coefficients are independent of the measurements in the beam and jet regions. For central rapidities both coefficients differ very little between different distance measures. Away from $\eta_J=0$ there are noticeable differences between the geometric-R, modified geometric-R and conical (anti-$k_T$ and XCone) measures, as can be seen in the right panel of \fig{SaJ0}.
In \fig{SaJ12} we plot $s_{aJ,\delta}$ for $\eta_J= -1,0,1$ as function of $R$ and in \fig{SaJ1_eta} for $R=1$ in terms of $\eta_J$. Once again results are shown for the beam-thrust, C-parameter and $p_T$-measurements and $\beta=2$. Compared to the beam-beam dipole, the coefficients are not any more symmetric in $\eta_J \lra -\eta_J$. Furthermore, the $\mathcal{O}(R^2)$ corrections are not universal for different partitionings, which can lead to sizable deviations for $R \sim 1$, especially for forward jets. This is clearly visible for $s_{aJ,J}$, as shown in the right panel of \fig{SaJ0}, or e.g.~for $s_{aJ,\de}$ with $\eta_J=1$ shown in the middle row of \fig{SaJ12}. The analytic results including $\mathcal{O}(R^2)$ corrections that are shown correspond to the conical partitioning. The difference with respect to the exact result is very small up to values of $R \sim 2$ for all measurements in the beam region, suggesting that the effective expansion parameter is $R/R_0$ with $R_0 \gtrsim 2$. For the geometric-$R$ measures the corresponding $\mathcal{O}(R^2)$ corrections (not shown) are also close to the full results for $R \lesssim 1$, but deviate much stronger for large values of $R$. 

In general, the results for anti-$k_T$ and XCone jets are almost identical for isolated jets and reasonable values of the jet radius, as expected from the very similar shapes displayed in \fig{jetregions}. This will be different when the distance between jets becomes less than $2R$, as illustrated in \fig{ThreeJetAlgorithms}. Furthermore, since the shape of isolated anti-$k_T$ and XCone jets is invariant under boosts along the beam axis, the results for the corresponding soft function coefficients $s_{ab,B}$,  $s_{ab,\delta}$, $s_{aJ,\delta}$, $s_{aJ,J}$ and $s_{aJ,\delta}$ do not depend on the jet rapidity when using the (boost invariant) $p_T$-measurement in the beam region. 

For different values of $\beta$ the qualitative behavior looks similar. To illustrate this, we display the coefficients $s_{ab,\delta}$ and $s_{aJ,\delta}$ for $\beta=1$ and the $p_T$-measurement in \fig{S_beta1}. The most noticeable differences between the distance measures are again between the (modified) Geometric-R and the conical measures away from central rapidity.

%%%%%%%%%%%%%%%%%%%%%%%%%%%%%%%%%%%%%%%%%%%%%%%%%%%%%%%%%%%%%%%%%%%%%%%%%%%%%%%%
\section{Conclusions}
\label{sec:conclusions}
%%%%%%%%%%%%%%%%%%%%%%%%%%%%%%%%%%%%%%%%%%%%%%%%%%%%%%%%%%%%%%%%%%%%%%%%%%%%%%%%

In this paper we worked out a general setup to calculate one-loop soft functions for exclusive $N$-jet processes at hadron colliders. This method applies to any jet algorithm that satisfies soft-collinear factorization, and for generic infrared- and collinear safe jet measurements and jet vetoes, as long as they reduce to an angularity in the limit where they approach the jet/beam axis. The soft function is calculated using a hemisphere decomposition of the phase space, extending the approach that was used in ref.~\cite{Jouttenus:2011wh} to calculate the $N$-jettiness soft function. The divergences are extracted analytically, such that numerical computations only arise for the finite terms.

We also demonstrated how the method works in practice, providing explicit expressions for single jet production $pp\to L+1\text{ jet}$ for several cases: angularities as jet measurements, beam thrust, $C$-parameter, and transverse momentum as jet vetoes, and anti-$k_T$ and XCone as jet algorithms. We optimized our method by expanding the finite corrections in the jet radius $R$, obtaining a fully analytical result in the limit $R\ll 1$. It turns out that the remaining (numerical) contributions are rather small, even for relatively large values of $R$, thus improving the stability.

With the soft functions discussed in this paper, one can calculate resummed cross-section at NNLL or NLL$'$ accuracy for exclusive jet processes at the LHC. This same soft function also enters in jet substructure calculations, see e.g.~the 2-jettiness calculation of ref.~\cite{Feige:2012vc}, and the subtraction techniques could prove useful for other jet substructure calculations as found in ref.~\cite{Larkoski:2015kga}.

\begin{acknowledgments}

P.P. would like to thank Bahman Dehnadi for pointing out some typos in the draft. This work was supported by the German Science Foundation (DFG) through the Emmy-Noether Grant No.~TA 867/1-1, and the Collaborative Research Center (SFB) 676 Particles, Strings and the Early Universe, by the Office of Nuclear Physics of the U.S. Department of Energy under the Grant No.~DE-SC0011090, Grant No.~DE-AC02-05CH11231, Grant No.~DE-AC52-06NA25396, and through the Los Alamos National Lab LDRD Program, by the Simons Foundation through the Investigator grant 327942, by the European Research Council under Grant No.~ERC-STG-2015-677323, by the D-ITP consortium, a program of the Netherlands Organization for Scientific Research (NWO) that is funded by the Dutch Ministry of Education, Culture and Science (OCW), and by a Global MISTI Collaboration Grant from MIT. We also thank the Erwin Schr\"odinger Institute's program ``Challenges and Concepts for Field Theory and Applications in the Era of the
LHC Run-2", where portions of this work were completed.
\end{acknowledgments}

\appendix

%%%%%%%%%%%%%%%%%%%%%%%%%%%%%%%%%%%%%%%%%%%%%%%%%%%%%%%%%%%%%%%%%%%%%%%%%%%%%%%%
\section{Analytic contributions for $pp \to L +1$ jet}
\label{app:anal_soft_pieces}
%%%%%%%%%%%%%%%%%%%%%%%%%%%%%%%%%%%%%%%%%%%%%%%%%%%%%%%%%%%%%%%%%%%%%%%%%%%%%%%%

In this appendix we collect some details about the analytic calculation of several soft function corrections for $pp \to L +1$ jet discussed in \sec{OneJetCase}. We discuss the jet hemisphere correction to the soft function for angularity measurements in \app{hemi_soft}, and compute the analytic results for the $\mathcal{O}(R^2)$ terms of the soft function coefficients in \eq{soft_coeffs} for anti-$k_T$ in \app{R_expansion}.

%===============================================================================
\subsection{Hemisphere soft function correction} \label{app:hemi_soft}
%===============================================================================

We perform the calculation of the jet hemisphere correction for the boost-invariant angularities defined in \eq{fJchoices}, i.e.~$S_{a<J}$ in \eq{SaJ_decomp}. It is given by the integral
\begin{align} \label{eq:Sij}
S_{J<a}(\{k_l\},\rho,\eta_J)\! = -2
\Bigl(\frac{\mu^2 e^{\gamma_E}}{4\pi}\Bigr)^\eps\! g^2\!\!
\int\! \frac{\df^d p}{(2\pi)^d}\,\frac{n_a\cdot n_J}{(n_a\cdot p)(n_J\cdot p)}\,
2\pi \delta(p^2)\theta(p^0)\, F_{J<a}(\{k_m\},\rho,\eta_J,p)  ,
\end{align}
with the size of the hemisphere adjusted by the parameter $\rho$ and the measurement given by
\begin{align}
F_{J<a}(\{k_m\},\rho,\eta_J,p) =\delta\bigl(k_J-p_T\mathcal{R}_{J}^\beta\bigr)\,\theta\Bigl(n_a\cdot p-\frac{n_J\cdot p}{\rho}\Bigr)\,\delta(k_B) \,,
\end{align}   
in analogy to \eq{F_onejet}.
Here $\mathcal{R}_{J} \equiv \mathcal{R}_{sJ}$ denotes the distance of the soft emission with momentum $p^\mu$ with respect to the jet direction in azimuth-rapidity space as defined in \eq{DeltaRdef}.
Let us define the momentum projection $p_k$ along a generic light-like direction $n_k$ and the angular distance between two light-like directions $\hat{s}_{ij}$ as
\begin{equation}
\label{eq:sij}
\begin{split}
p_k&\equiv n_k\cdot p \, , \qquad \hat{s}_{ij} \equiv\frac{n_i\cdot n_j}{2} =\frac{1-\cos\theta_{ij}}{2} \, .
\end{split}
\end{equation}
For any $ij$-dipole, the gluon four-momentum can be decomposed as
\begin{equation}\label{eq:basis}
p^\mu=\frac{p_i}{2\hat{s}_{ij}}n_j^\mu+\frac{p_j}{2\hat{s}_{ij}}n_i^\mu+p_{\perp_{ij}}^\mu \, ,
\end{equation}
with the integration measure given by
\begin{equation}
\df^{4-2\e}p=\frac{p_{\perp_{ij}}^{1-2\e}}{2\hat{s}_{ij}} \,\df p_i\,\df p_j \, \df p_{\perp_{ij}} \,\df\Omega_{2-2\e} \, .
\end{equation}
The boost-invariant jet angularity can be expressed in this basis, by first writing 
\begin{align}
p_T\mathcal{R}_{J}^\beta=(2p_j\cosh\eta_j)^{\beta/2}(p_T)^{1-\b/2},
\end{align}
and then substituting 
\begin{align}
p_T= \frac{p_{\perp_{ij}}\mathcal{G}(q,\phi)}{q}  \quad {\rm with }\,\, q=\frac{p_j}{p_{\perp_{ij}}} \, .
\end{align}
The function $\mathcal{G}(q,\phi)$ is given in general by
\begin{align}
\mathcal{G}(q,\phi)=&\biggl(\hat{s}_{aj}+\frac{\hat{s}_{ai}}{\hat{s}_{ij}}\, q^2-2 \sqrt{\frac{\hat{s}_{aj} \hat{s}_{ai}}{\hat{s}_{ij}}} \,q \cos \phi\, \biggr)^{\frac{1}{2}} \biggl(\hat{s}_{bj}+\frac{\hat{s}_{bi}}{\hat{s}_{ij}}\, q^2-2 \sqrt{\frac{\hat{s}_{bj} \hat{s}_{bi}}{\hat{s}_{ij}}} \,q \cos (\phi-\Delta\phi_{ij})\, \biggr)^{\frac{1}{2}}\, .
\end{align}
Here $\phi$ is the azimuthal angle in the two-dimensional $\perp_{ij}$-space, and $\Delta\phi_{ij}$ is the difference in azimuth (with respect to the beam axis) between the dipole directions $i$ and $j$. Thus the jet angularity can be written as
\begin{equation}
p_T\mathcal{R}_J^\beta=p_{\perp_{ij}}\, q^{\b-1}\,[\mathcal{G}(q,\phi)]^{1-\b/2}\,(2\cosh\eta_J)^{\b/2} \, .
\end{equation}
Let us specialize to the case $i=a$ and $j=J$. The hemisphere phase space is given by
\begin{align}
\text{Hemisphere $J<a$:}\quad \theta(q_0-q),\quad q_0=\sqrt{\rho \,\hat{s}_{aJ}}\, ,
\end{align}
with $\hat{s}_{aJ} = e^{-\eta_J}/(2\cosh \eta_J)$. For the case $\b>1$, dimensional regularization regulates all the divergences. Using the basis of \eq{basis}, after the trivial integrations and changing variable from $p_J$ to $q$, \eq{Sij} reads
\begin{align}\label{eq:Sij1}
S_{J<a} (\{k_l\},\rho,\eta_J)&=-\frac{2g^2}{(2\pi)^{3-2\eps}} \Bigl(\frac{\mu^2 e^{\gamma_E}}{4\pi}\Bigr)^\eps\,\frac{(2\cosh\eta_J)^{\b\eps}}{k_J^{1+2\eps}}\,\delta(k_B) \nn \\
& \quad \times\int\df\Omega_{2-2\eps}\int_{0}^{q_0}\frac{\df q\,}{q^{1-2\eps(\b-1)}}\bigl[\mathcal{G}(q,\phi)\bigr]^{\eps(2-\b)} \, .
\end{align} 
Performing the integrals and expanding in $\eps$,
\begin{align*}
S_{J<a}(\{k_l\},\rho,\eta_J)&= \frac{\alpha_s}{4\pi} \, \frac{\delta(k_B)}{\beta-1}  \biggl[\frac{8}{\mu \,r^{\beta-1}}\,  \mathcal{L}_1\biggl(\frac{k_J}{\mu 
 \,r^{\beta-1}}\biggr) - \frac{4}{\eps} \frac{1}{\mu\,r^{\beta-1}} \,\mathcal{L}_0\biggl(\frac{k_J}{\mu\,r^{\beta-1}}\biggr) \\
  & \quad+\delta(k_J) \biggl(\frac{2}{\eps^2}-\frac{\pi^2}{6}-2(\beta-1)(\beta-2)\,\mathcal{I}\biggr) +\mathcal{O}(\epsilon)\biggr] \, ,\numberthis
  \end{align*}
 where $r=(2\cosh\eta_J \,e^{-\eta_J} \rho)^{1/2}$ and
 \begin{align}
 \mathcal{I}=\frac{1}{\pi}\int_{-\pi}^{\pi}\df\phi\int_0^{q_0}\frac{\df q}{q}\ln\left[2\cosh\eta_J \, \mathcal{G}(q,\phi)\right]=\theta(r-1)\ln^2 r \, .
 \end{align}
 Setting $\rho =\rho_R^J (R,\eta_J)$ as defined in \eq{leadingrhoRJ} yields $r=R$ and thus the result in \eq{S_hemi2}. For the case $\b=1$, one can see from \eq{Sij1} that an additional rapidity regulator is needed as $q\to 0$, which can be chosen to be $(\nu/2p^0)^\eta$, as discussed below \eq{Sijbare_generalN}. Following a similar procedure, one obtains the result of \eq{S_hemi2_b1}.

Alternatively, one can get the hemisphere soft function for boost-invariant angularities by adding the finite correction in \eq{I1_hemi} to \eqs{S_hemi}{S_hemi_b1}, which correspond to the standard angularities in $e^+ e^-$-collisions defined in \eq{Tau_hemi}. Using the same variables defined above, one gets 
\begin{align}
4 I_{1,J<a}&=-\frac{2(\beta-2)}{\pi}\int_{-\pi}^{\pi}\df\phi\int_0^{q_0}\frac{\df q}{q}\ln\left[\frac{2\cosh\eta_J\,\mathcal{G}(q,\phi)}{1+e^{2\eta_J}q^2-2e^{\eta_J}q\cos\phi}\right] \nn \\
&=-2(\beta-2)\biggl[\theta(R-1)\ln^2R-2\,\theta\Bigl(\frac{R \, e^{\eta_J}}{2\cosh \eta_J}-1\Bigr)\ln^2 \Bigl(\frac{R \, e^{\eta_J}}{2\cosh \eta_J}\Bigr)\biggr]\, .\numberthis
\end{align}
By adding this correction to \eqs{S_hemi}{S_hemi_b1}, with $c_J=(2\cosh\eta_J)^{\beta-1}$, one recovers again the results in \eqs{S_hemi2}{S_hemi2_b1}.

%===============================================================================
\subsection{Corrections at $\mathcal{O}(R^2)$}
\label{app:R_expansion}
%===============================================================================

Here we outline the analytic calculation of the soft function corrections in \eq{soft_coeffs} at $\mathcal{O}(R^2)$ in the small jet radius expansion. A similar computation has been performed in ref.~\cite{Liu:2014oog} for a jet mass measurement in dijet processes close to the kinematic threshold. We give the results for a conical (anti-$k_T$) jet with the
measurement of arbitrary jet angularities and general smooth jet vetoes (including in addition the beam thrust case).

First, we consider the contributions from the beam-beam dipole. Here the $\mathcal{O}(R^2)$ corrections are the leading contributions that account for the jet region. Since the deviations between the jet boundaries for different partitionings are in addition power suppressed by the jet radius all sets of distance measures discussed in \sec{genTau} lead to the same result at $\mathcal{O}(R^2)$. 
The term $s_{ab,B}$ in \eq{soft_coeffs} corresponds to the jet area giving $s_{ab,B} =4 R^2$.  The coefficient $s_{ab,\delta}$ is given by the integral in \eq{abnum}, which yields at $\mathcal{O}(R^2)$
\begin{align}\label{eq:sab_delta_R2}
 s_{ab,\delta} &= \frac{4}{\pi} \int_{-\infty}^\infty  \df \Delta \eta \int_{-\pi}^\pi\df \phi \, \Bigl(\frac{\beta}{2}\ln \bigl[(\Delta \eta)^2 + \phi^2\bigr] -\ln f_B(\eta_J)\Bigr)\, \theta\Bigl(R^2- (\Delta \eta)^2 -\phi^2\Bigr) +\mathcal{O}(R^4)\nn \\
& = 2 R^2 \Bigl[\beta\bigl(2\ln R-1\bigr)-2\ln f_B(\eta_J)\Bigr] +\mathcal{O}(R^4)\, .
\end{align}
In fact, for conical jets and a transverse momentum veto, i.e.~$f_B(\eta)=1$, any higher order corrections in $R$ vanish, so that \eq{sab_delta_R2} provides already the exact one-loop result for this case.

Next, we discuss the contributions from the beam-jet dipole, which in general differ for different partitionings. The corrections for real radiation inside the jet region can be written as
\begin{align}\label{eq:SaJ_J}
S_{aJ}^{(J)}= - \frac{\alpha_s}{2\pi} \,e^{\gamma_E \epsilon} \mu^{2\epsilon} \frac{\sqrt{\pi}}{\Gamma(\frac{1}{2}-\epsilon)}\, \frac{1}{k_J^{1+2\epsilon}} \,\Bigl[I_{aJ, R\ll 1}^{(J)} + \underbrace{(I_{aJ}^{(J)} - I_{aJ, R\ll 1}^{(J)} )}_{=\Delta I_{aJ}^{(J)}}\Bigr] \, ,
\end{align}
where $I_{aJ, R\ll 1}^{(J)} $ denotes the leading small-R result at $\mathcal{O}(1)$ and $\Delta I_{aJ}^{(J)}$ contains all corrections which are suppressed by the jet size. The latter term can be expanded in $\epsilon$ and is given up to $\mathcal{O}(\epsilon)$ by
\begin{align}
\Delta I_{aJ}^{(J)} &=\frac{1}{\pi}\int_{-\infty}^\infty  \df \Delta \eta \int_{-\pi}^\pi\df \phi \, \, \theta\bigl[d_B(\eta_J+\Delta\eta) -d_J(\eta_J+\Delta\eta,\phi,R)\bigr]  \\
& \quad \times \biggl[\biggl(\frac{e^{\Delta \eta}}{\cosh \Delta \eta - \cos \phi}-\frac{2}{(\Delta \eta)^2+\phi^2} \biggr) \nn \\
& \qquad  +2\epsilon \biggl(\frac{e^{\Delta \eta}\bigl(\frac{\beta}{2} \ln \bigl[2\cosh \Delta \eta - 2\cos \phi\bigr]-\ln \abs{\sin \phi}\bigr)}{\cosh \Delta \eta - \cos \phi}-\frac{\beta \ln\bigl[(\Delta \eta)^2+ \phi^2\bigr] -2 \ln\abs{\phi}}{(\Delta \eta)^2+ \phi^2} \biggr)\biggr]
\, . \nn \end{align}
Expanding the integrand in $R$ yields for conical jets 
\begin{align}\label{eq:IaJ_J}
\Delta I_{aJ}^{(J, k_T)}\Bigr|_{\mathcal{O}(R^2)} = R^2 \biggl[\frac{1}{2}+\epsilon \biggl(\frac{7}{6}-\frac{\beta}{2}+(\beta-1)\ln R+ \ln 2\biggr)\biggr] \, .
\end{align}
The corrections for real radiation inside the beam region can be similarly written as
\begin{align}\label{eq:SaJ_B}
S_{aJ}^{(B)}= - \frac{\alpha_s}{2\pi} \,e^{\gamma_E \epsilon} \mu^{2\epsilon} \frac{\sqrt{\pi}}{\Gamma(\frac{1}{2}-\epsilon)}\, \frac{1}{k_B^{1+2\epsilon}} \,\Bigl[I_{aJ, R\ll 1}^{(B)} + \underbrace{(I_{aJ}^{(B)} - I_{aJ, R\ll 1}^{(B)} )}_{=\Delta I_{aJ}^{(B)}}\Bigr] \,.
\end{align}
Here $\Delta I_{aJ}^{(B)}$ acts as a subtractive contribution inside the jet region and is given by
\begin{align}
\Delta I_{aJ}^{(B)} &=-\frac{1}{\pi}\int_{-\infty}^\infty  \df \Delta \eta \int_{-\pi}^\pi\df \phi \, \, \theta\bigl[d_B(\eta_J+\Delta\eta) -d_J(\eta_J+\Delta\eta,\phi,R)\bigr]  \nn \\
& \quad \times \biggl[\biggl(\frac{e^{\Delta \eta}}{\cosh \Delta \eta - \cos \phi}-\frac{2}{(\Delta \eta)^2+\phi^2} \biggr) \nn \\
& \qquad  +2\epsilon \biggl(\frac{e^{\Delta \eta}(\ln f_B(\eta)-\ln \abs{\sin \phi})}{\cosh \Delta \eta - \cos \phi}-\frac{2(\ln  f_B(\eta_J) - \ln \abs{\phi})}{(\Delta \eta)^2+ \phi^2} \biggr)\biggr]
\, .\end{align}
Expanding the integrand in $R$ yields for conical jets and a smooth function $f_B(\eta)$
\begin{align}\label{eq:IaJ_B}
\Delta I_{aJ}^{(B,k_T)} =-  R^2\biggl[\frac{1}{2}+\epsilon \biggl(\frac{7}{6}- \ln R + \ln f_B(\eta_J) + \frac{2 f_B'(\eta_J) + f_B''(\eta_J)}{f_B(\eta_J)}-\Bigl(\frac{f_B'(\eta_J) }{f_B(\eta_J)}\Bigr)^2+\ln2\biggr)\biggr] \, .
\end{align}
Using eqs.~(\ref{eq:SaJ_J}),~(\ref{eq:IaJ_J}),~(\ref{eq:SaJ_B}) and~(\ref{eq:IaJ_B}) the soft function coefficients at $\mathcal{O}(R^2)$ for the beam-jet dipole contributions read for anti-$k_T$ jets
\begin{align}
s^{(k_T)}_{aJ,J}\Bigr|_{\mathcal{O}(R^2)}  &=  - s^{(k_T)}_{aJ,B}\Bigr|_{\mathcal{O}(R^2)}  = - R^2 \, , \nn \\
s^{(k_T)}_{aJ,\delta}\Bigr|_{\mathcal{O}(R^2)} & =  R^2 \biggl[ \beta\ln R -\frac{\beta}{2} -\ln f_B(\eta_J)- \frac{2 f_B'(\eta_J) + f_B''(\eta_J)}{f_B(\eta_J)}+\Bigl(\frac{f_B'(\eta_J) }{f_B(\eta_J)}\Bigr)^2 \biggr] \, . \label{eq:saj_delta_R}
\end{align}
Since the beam thrust veto has a kink at $\eta=0$, \eq{saj_delta_R} does not fully determine all power suppressed terms up to $\mathcal{O}(R^2)$ if $|\eta_J|<R$. In this case the next-to leading correction is of $\mathcal{O}(R)$ and the additional contribution with respect to \eq{saj_delta_R} reads
\begin{align}
 \Delta s^{(k_T)}_{aJ,\delta}&= \theta(1-|x|)  \biggl\{ \frac{16R}{\pi}\biggl[\sqrt{1-x^2} + \abs{x} (\ln(2\abs{x})-1) \arccos \abs{x} -\frac{\abs{x}}{2}\, {\rm Cl}_2 \Bigl(\arccos(1 - 2 x^2)\Bigr) \biggr]\nn \\
& \quad + \frac{4R^2}{\pi} \biggl[\frac{\pi}{2}\bigl(\theta(-x) -\theta(x)\bigr) +3x \sqrt{1-x^2}+ \arcsin(x)- 2 x\abs{x} \arccos\abs{x}\Bigr) \biggr]+\mathcal{O}(R^3)\biggr\} \, , 
\end{align}
where $x \equiv \eta_J/R$ and the ${\rm Cl}_2(\theta) \equiv {\rm Im}[\Li_2(e^{i \theta})]$.

Results for jet regions from a different partitioning can be obtained by considering deviations from the circular jet shape in addition. For the conical geometric distance measure in \eq{d_ConicalGeometric} corresponding to a XCone default jet the results at $\mathcal{O}(R^2)$ are the same as for the conical measure (i.e.~for an anti-$k_T$ jet).

%%%%%%%%%%%%%%%%%%%%%%%%%%%%%%%%%%%%%%%%%%%%%%%%%%%%%%%%%%%%%%%%%%%%%%%%%%%%%%%%
\section{Numerical evaluation of soft function integrations}
\label{app:num_evaluation}
%%%%%%%%%%%%%%%%%%%%%%%%%%%%%%%%%%%%%%%%%%%%%%%%%%%%%%%%%%%%%%%%%%%%%%%%%%%%%%%%

We discuss the numerical evaluation of the boundary mismatch integrals $I_{aJ}^{B}$ and $I_{aJ}^{J}$ in \eq{Iedge} for $pp \to L +1$ jet. To compute them efficiently we need to determine the integration bounds. These depend on the relations between the distance measures $d_B(p)$ and $d_J(p)$ and between the projections $n_a \cdot p$ and $n_J \cdot p/\rho_J$ used for the analytic calculation of the hemisphere results. We discuss here the explicit boundaries only for the most important case, the conical (anti-$k_T$) measure. For the geometric measures (including the conical geometric XCone measure) one can follow a strategy similar to \cite{Jouttenus:2011wh} using coordinates based on the lightcone projections $n_a \cdot p$ and $n_J \cdot p$. Furthermore, we also explain why the integrals encoding the corrections to the small $R$ limit give only a moderate numerical impact, even for sizable values of the jet radius.

%===============================================================================
\subsection{Integration bounds for the conical measure}
%===============================================================================

\begin{figure}
\centering
\includegraphics[height=5.1cm]{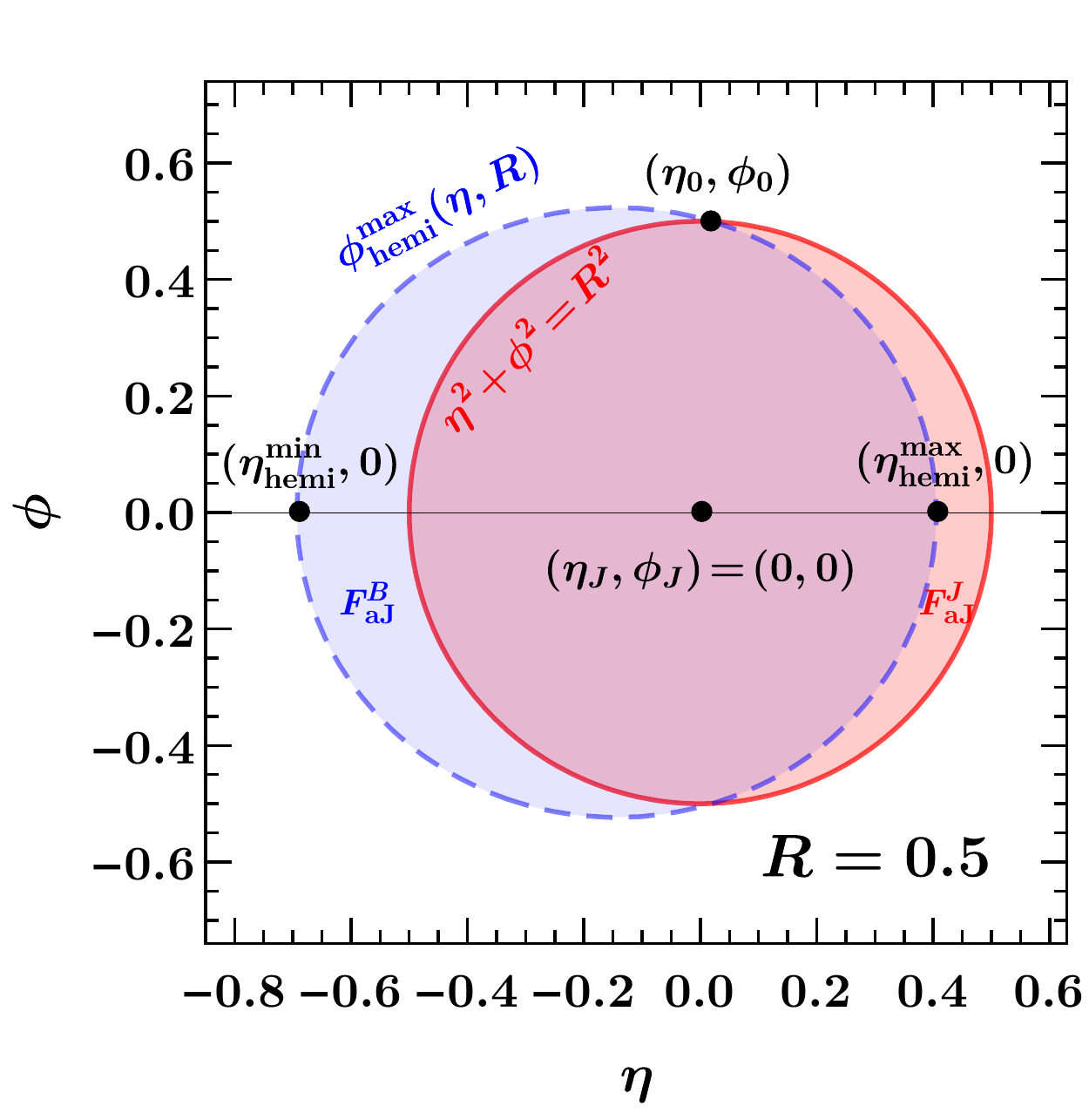} \hfill
\includegraphics[height=5.1cm]{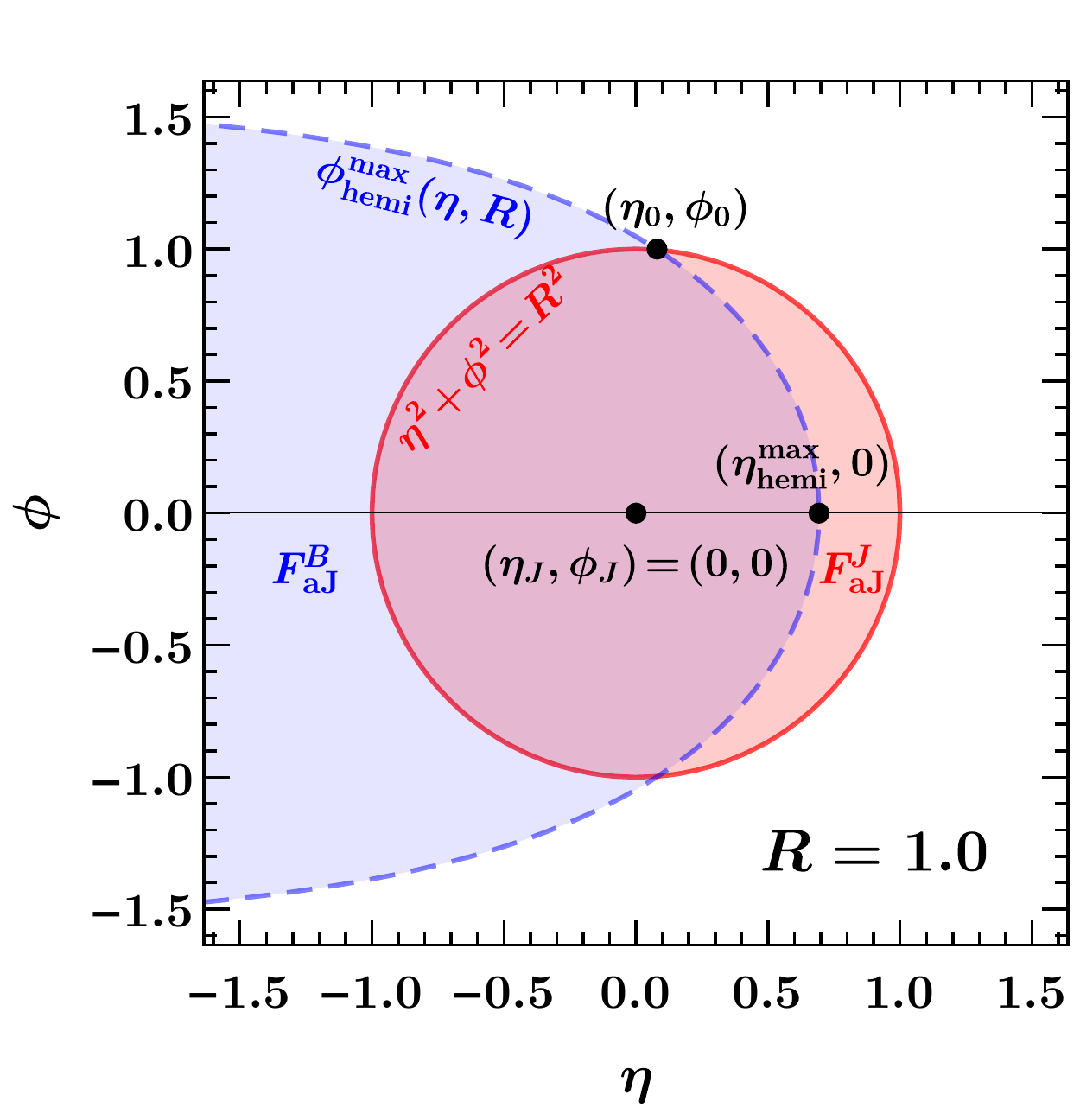} \hfill
\includegraphics[height=5.1cm]{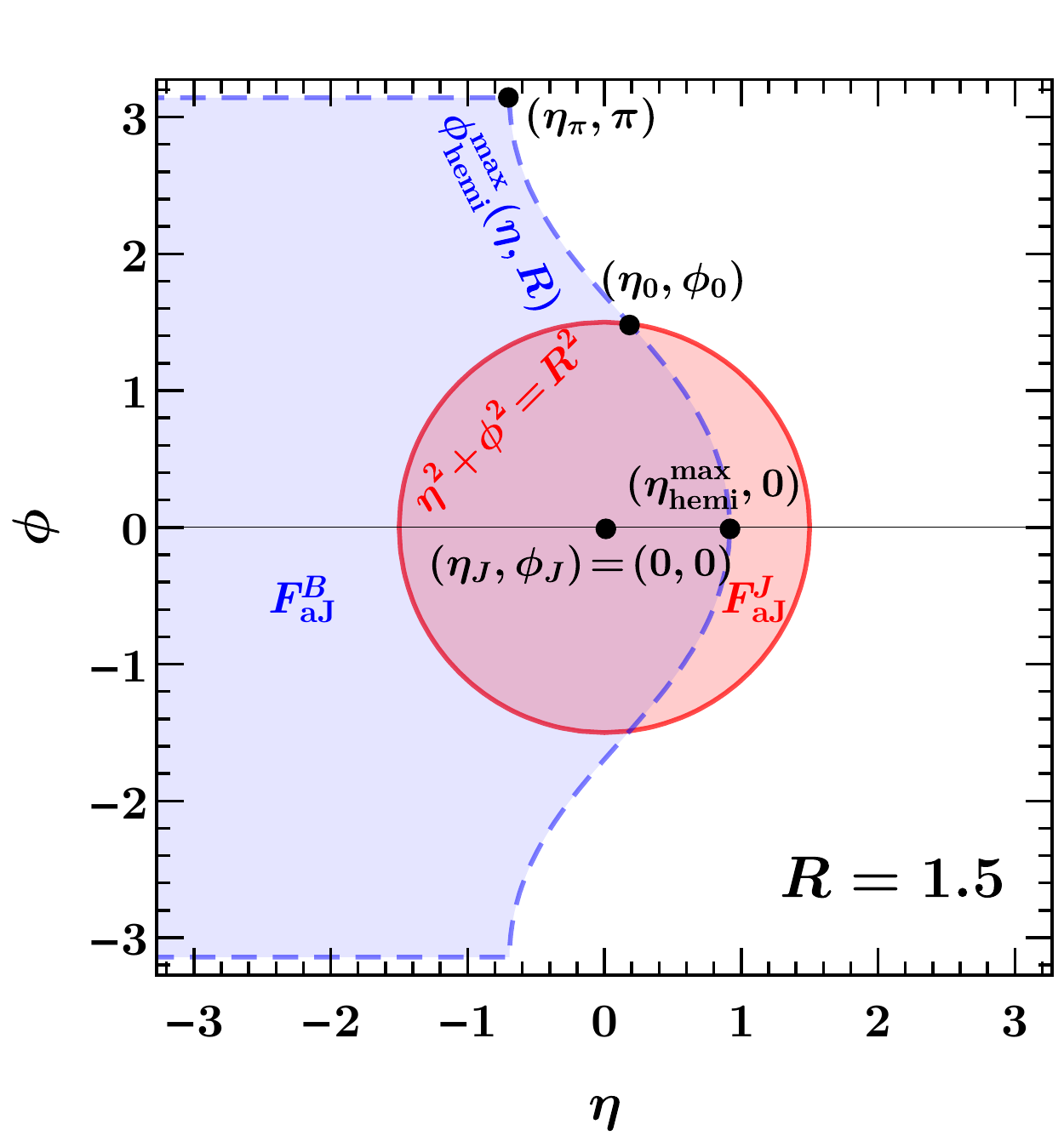}
\caption{Illustration of the phase space misalignment between the hemisphere jet region with $\rho_J=\rho_J^R$ (blue, dashed) and a conical jet area (red, solid) for $\eta_J=0$ and $R=0.5,1.0,1.5$. The areas which do not overlap correspond to the integration domains of the integrals $I_{aJ}^B$ and $I_{aJ}^J$, respectively.
\label{fig:jetregions_R}}
\end{figure}

For the conical measure the integration boundaries can be most easily obtained in beam coordinates $\eta$, $\phi$. The conditions from the measurement functions in \eq{F_onejet} read 
\begin{align}
F^B_{aJ}: \quad & R^2< (\Delta \eta)^2 + \phi^2   \quad  {\rm and} \quad \rho_J e^{-\eta_J} \cosh \eta_J < e^{\Delta \eta}(\cosh \Delta \eta - \cos \phi) \, , \nn \\
F^J_{aJ}: \quad & R^2> (\Delta \eta)^2 + \phi^2   \quad  {\rm and} \quad \rho_J e^{-\eta_J} \cosh \eta_J > e^{\Delta \eta}(\cosh \Delta \eta - \cos \phi) \, .
\end{align}
We use the value $\rho_J= \rho_J^R$ in \eq{leadingrhoRJ}, which eliminates the dependence on the jet rapidity $\eta_J$ (in favor of the jet radius $R$) in the second relation and leads to integrals which are power suppressed in $R$. (The computation for arbitrary $\rho_J$ can be carried out similarly.) The associated hemisphere mismatch regions are displayed in \fig{jetregions_R}. For $F^J_{aJ}$ the integration boundaries read 
\begin{align}
&\int_{-\infty}^\infty \df \Delta \eta \int_{-\pi}^\pi \df \phi \, \theta\bigl(R^2-(\Delta \eta)^2-\phi^2\bigr)\, \theta\Bigl(\frac{n_J \cdot p}{\rho^R_J}-n_a \cdot p\Bigr) \nn \\
&\quad = \int_{\eta_0(R)}^{\eta^{\rm max}_{\rm hemi}(R)} \df \Delta \eta \int_{\phi^{\rm max}_{\rm hemi}(\Delta \eta,R)}^{\sqrt{R^2-(\Delta \eta)^2}} \df \phi + \int_{\eta^{\rm max}_{\rm hemi}(R)}^R \df \Delta \eta \int_0^{\sqrt{R^2-(\Delta \eta)^2}} \df \phi  +(\phi \leftrightarrow - \phi)\, ,
\end{align}
where we have defined
\begin{align}
\phi^{\rm max}_{\rm hemi}(\Delta \eta,R) =\arccos\biggl(\frac{e^{\Delta \eta}+(1-R^2)e^{-\Delta \eta}}{2}\biggr),\quad
\eta^{\rm max}_{\rm hemi}(R) =\ln(1+R),
\end{align}
and $\eta_0(R)$ is the solution of the transcendental equation
\begin{align}
[\eta_0(R)]^2 +\bigl[\phi^{\rm max}_{\rm hemi}(\eta_0(R),R)\bigr]^2 = R^2 \, .
\end{align}
For $F^B_{aJ}$ we get 
\begin{align}
&\int_{-\infty}^\infty \df \Delta \eta \int_{-\pi}^\pi \df \phi \, \theta\bigl((\Delta \eta)^2+\phi^2-R^2\bigr)\, \theta\Bigl(n_a \cdot p -\frac{n_J \cdot p}{\rho^R_J}\Bigr) \nn \\
&= \theta(R \leq 1) \biggl[\int_{\eta^{\rm min}_{\rm hemi}(R)}^{-R} \!\df \Delta \eta \int_0^{\phi^{\rm max}_{\rm hemi}(\Delta \eta,R)} \!\df \phi + \int_{-R}^{\eta_0(R)} \!\df \Delta \eta \int_{\sqrt{R^2-(\Delta \eta)^2}}^{\phi^{\rm max}_{\rm hemi}(\Delta \eta,R)} \df \phi \biggr] \nn \\
& + \theta(R_{\pi} \!> \!R \!>\!1)  \biggl[\int_{-\infty}^{\eta_{\pi}(R)} \!\!\!\!\!\!\!\df \Delta \eta \int_0^{\pi} \!\!\df \phi + \int_{\eta_{\pi}(R)}^{-R} \!\!\df \Delta \eta \int_{0}^{\phi^{\rm max}_{\rm hemi}(\Delta \eta,R)} \!\!\!\!\!\df \phi +\! \int_{-R}^{\eta_0(R)} \!\!\df \Delta \eta \int_{\sqrt{R^2-(\Delta \eta)^2}}^{\phi^{\rm max}_{\rm hemi}(\Delta \eta,R)} \!\df \phi  \biggr] \nn \\
& + \theta(R \geq R_{\pi}) \biggl[\int_{-\infty}^{-R} \!\!\df \Delta \eta \int_0^{\pi} \!\df \phi + \int^{\eta_{\pi}(R)}_{-R} \!\df \Delta \eta \int_{\sqrt{R^2-(\Delta \eta)^2}}^{\pi} \df \phi + \int_{\eta_{\pi}(R)}^{\eta_0(R)} \!\!\!\!\df \Delta \eta \int_{\sqrt{R^2-(\Delta \eta)^2}}^{\phi^{\rm max}_{\rm hemi}(\Delta \eta,R)} \!\df \phi  \biggr]
\nn  \\
& +(\phi \leftrightarrow -\phi)\, ,
\end{align}
where we have defined
\begin{align}
\eta^{\rm min}_{\rm hemi}(R) =\ln(1-R),\quad \eta_{\pi}(R) & = \ln(R-1),
\end{align}
and $R_{\pi} \approx 1.28$ is the solution of the transcendental equation
\begin{align}
\eta_{\pi}(R_\pi) = -R_\pi\, .
\end{align}
With these explicit limits the integrals can be evaluated efficiently.

%===============================================================================
\subsection{Power suppression of boundary integrals}
\label{app:scalingR}
%===============================================================================

\begin{figure}
\centering
\includegraphics[height=5.1cm]{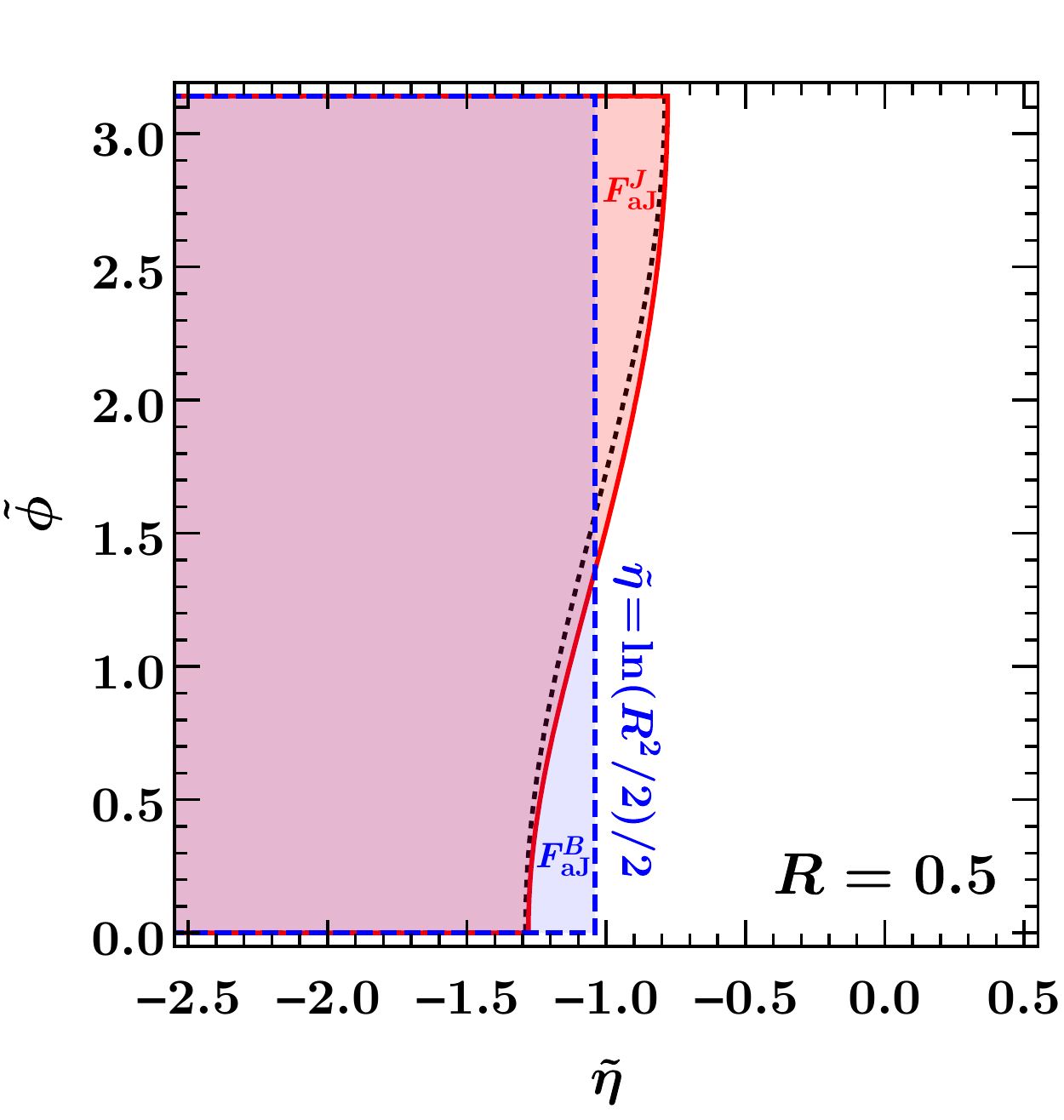}\hfill
\includegraphics[height=5.1cm]{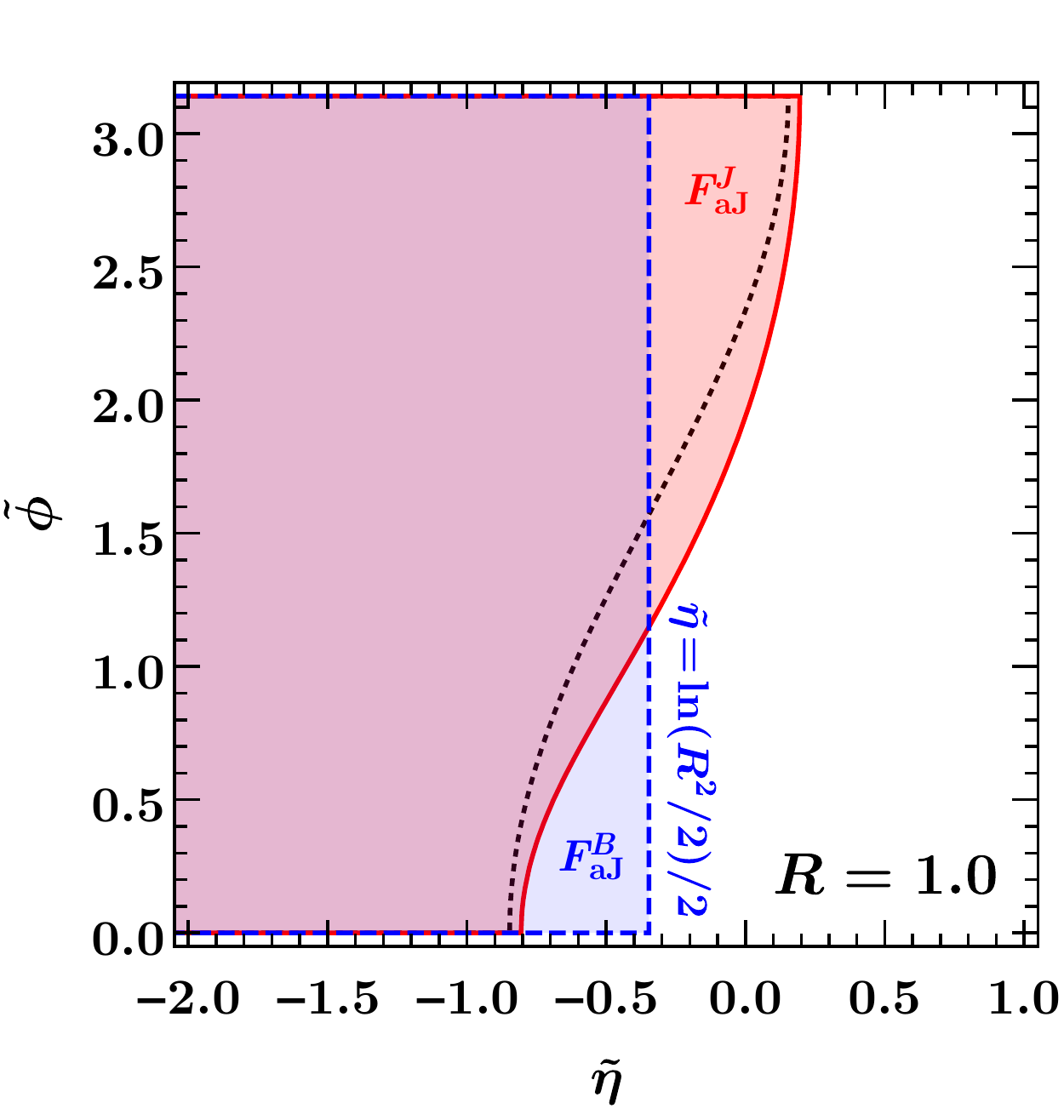} \hfill
\includegraphics[height=5.1cm]{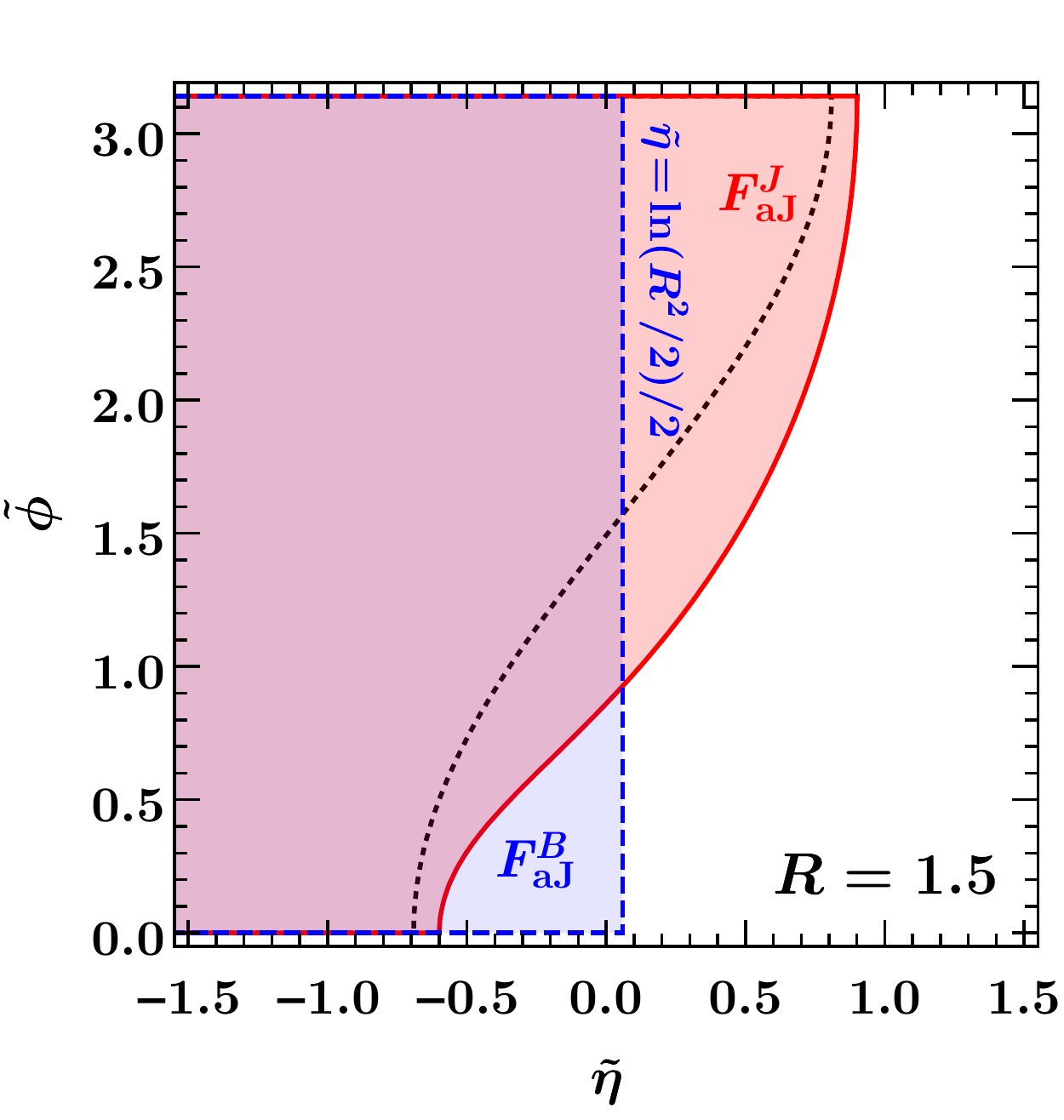}
\caption{Illustration of the phase space misalignment between the hemisphere jet region with $\rho_J=\rho_J^R$ (blue, dashed) and a conical partitioning (red, solid) for $\eta_J=0$ and $R=0.5,1.0,1.5$ in the boosted frame where the jet and beam $a$ are back-to-back. The areas which do not overlap scale as the integrals $I_{aJ}^B$ and $I_{aJ}^J$, respectively. The black dotted lines indicates the analytic result for the conical measure at $\mathcal{O}(R)$. The associated corrections to the coefficients $s_{aJ,J}$, $s_{aJ,B}$ correspond to the areas between the dashed and solid curves.
\label{fig:jetregions_boost_R}}
\end{figure}

We have seen in \fig{jetregions_R} that for a small jet radius the jet region from the hemisphere decomposition with $\rho^R_J$ and the actual conical partitioning largely overlap giving small results for the non-hemisphere corrections. However, for $R \sim 1$ the areas in the $\eta$-$\phi$ plane begin to differ very significantly, which might suggest that the associated corrections become very large in this regime and the results for the small $R$-expansion do not provide a good approximation. As we have seen in \sec{num_softs} this turns out not to be the case since the deviations of the jet areas in the beam coordinates are not representative for the size of the associated corrections. Instead it is more meaningful to compare the jet areas in the boosted frame where the jet and beam direction are back-to-back and soft radiation from the beam-jet dipole $aJ$ is uniform in the respective rapidity-azimuth coordinates $\tilde{\eta}$, $\tilde{\phi}$. The associated transformation rules between the sets of coordinates are explicitly given in ref.~\cite{Kasemets:2015uus}. In \fig{jetregions_boost_R} we display the jet regions in these coordinates for the conical measure (red) and for the hemisphere decomposition with $\rho_J=\rho_J^R$ for different values of $R$. The areas which do not overlap correspond directly to the integrals $I_{0,aJ}^B$ and $I_{0,aJ}^J$, respectively, while $I_{1,aJ}^B$ and $I_{1,aJ}^J$ are (logarithmic) moments in these regions. These are  individually of $\sim \mathcal{O}(R)$, which can be also confirmed by an analytic expansion indicated by the black, dotted line. In total the contributions from $F_{aJ}^B$ and  $F_{aJ}^J$ cancel each other at this order leading to a net contribution to the soft function of $\mathcal{O}(R^2)$.\footnote{For the corrections $s_{aJ,B}$ and  $s_{aJ,J}$ in \eq{soft_coeffs} this is obvious since only the difference between the two mismatch areas in \fig{jetregions_boost_R} enters. For the correction $s_{aJ,\delta}$ this holds for measurements which are continuous functions in $\eta$, $\phi$ due to the fact that at leading order in $R$ the integrands are constant in these areas.}

%%%%%%%%%%%%%%%%%%%%%%%%%%%%%%%%%%%%%%%%%%%%%%%%%%%%%%%%%%%%%%%%%%%%%%%%%%%%%%%%
\section{Analytic corrections for $pp \to $ dijets}
\label{app:dijets}
%%%%%%%%%%%%%%%%%%%%%%%%%%%%%%%%%%%%%%%%%%%%%%%%%%%%%%%%%%%%%%%%%%%%%%%%%%%%%%%%

Beyond single jet production, $pp \to $ dijets is another process of phenomenological relevance for measurements like jet mass. The full computation of the associated soft function corrections for arbitrary jet and beam measurements and partitionings can be carried out following the hemisphere decompositions discussed in secs.~\ref{sec:GenHemiDecomp} and~\ref{sec:OneJetCase}. Here we compute the analytic corrections for $pp \to $ dijets ($j_1,j_2$) in a small $R$ expansion up to terms at $\mathcal{O}(R^2)$, whereas the full $R$ dependence can be determined numerically but now including a jet-jet dipole. For definiteness and simplicity we consider conical jets with a jet mass measurement (i.e.~angularity in defined in \eq{fJchoices} with $\beta=2$) and a $p_T$ jet veto.  For generic $R<\pi/2$ we can write the renormalized one-loop soft function as\footnote{For $R<\pi/2$ (i.e.~as long as the jet regions do not share a common boundary) the measurements and partitioning are invariant under boosts along the beam axis,  such that this correction mainly depends on the relative rapidity of the jets $\Delta \eta_{12}$ and the jet radius. Since the rapidity regularization breaks boost invariance, there is, however, also a residual dependence on the individual jet rapidities appearing in $s_{a1,B}$.} 
 \begin{align} \label{eq:S_dijets}
 &  \widehat{S}^{\kappa(1)}_{2} (\{k_i\},R,\eta_1,\eta_2,\mu,\nu) = \frac{\alpha_s(\mu)}{4\pi}\biggl\{
  \bfT_a \cdot \bfT_b  \biggl[\frac{16}{\mu}\Bigl( \cL_1\Bigl(\frac{k_B}{\mu}\Bigr) - \cL_0\Bigl(\frac{k_B}{\mu}\Bigr) \ln \Bigl(\frac{\nu}{\mu}\Bigr) \Bigr)\,\delta(k_1) \, \delta(k_2) 
  \nn \\ & \quad
  + s_{ab,B}(R) \Bigl(\frac{2}{\mu}\,\cL_0\Bigl(\frac{k_B}{\mu}\Bigr)\, \delta(k_1) \,\delta(k_2)- \frac{1}{\mu}\,\cL_0\Bigl(\frac{k_1}{\mu}\Bigr)\,\delta(k_B)\,\delta(k_2) - \frac{1}{\mu}\,\cL_0\Bigl(\frac{k_2}{\mu}\Bigr)\,\delta(k_B)\,\delta(k_1)  \Bigr) \nn \\
 & \quad  +  s_{ab,\delta}(R)\,\delta(k_B) \,\de(k_1)\,\de(k_2)\biggr] + \bfT_1 \cdot \bfT_2 \biggl[\frac{8}{\mu}\,\cL_1\Bigl(\frac{k_1}{\mu}\Bigr) \,\delta(k_2)\,\delta(k_B) +\frac{8}{\mu}\,\cL_1\Bigl(\frac{k_2}{\mu}\Bigr) \,\delta(k_1)\,\delta(k_B)  \nn  \\
   & \quad + s_{12,J}(R,\Delta \eta_{12}) \Bigl(\frac{1}{\mu}\, \cL_0\Bigl(\frac{k_1}{\mu}\Bigr) \,\delta(k_2) +\frac{1}{\mu}\, \cL_0\Bigl(\frac{k_2}{\mu}\Bigr) \,\delta(k_1) \Bigr)\delta(k_B) \nn \\
   & \quad + s_{12,B}(R,\Delta \eta_{12})\,\frac{1}{\mu}\, \cL_0\Bigl(\frac{k_B}{\mu}\Bigr) \,\delta(k_1)\,\delta(k_2)
      + s_{12,\delta}(R,\Delta \eta_{12}) \, \delta(k_1) \, \delta(k_2) \,\delta(k_B)  \biggr] \nn \\
  & \quad + \bfT_a \cdot \bfT_1
  \biggl[\frac{8}{\mu}\,\cL_1\Bigl(\frac{k_1}{\mu}\Bigr) \,\delta(k_B) \, \delta(k_2) + \frac{8}{\mu}\,\cL_1\Bigl(\frac{k_B}{\mu}\Bigr) \,\delta(k_1) \,\delta(k_2)-\frac{8}{\mu}\,\mathcal{L}_0\Bigl(\frac{k_B}{\mu}\Bigr) \ln \Bigl(\frac{\nu}{\mu}\Bigr) \,\delta(k_1) \,\delta(k_2) \nn \\
  & \quad  +s_{a1,1}(R)\, \frac{1}{\mu}\, \cL_0\Bigl(\frac{k_1}{\mu}\Bigr) \,\delta(k_B) \, \delta(k_2) + s_{a1,2}(R,\Delta \eta_{12})\, \frac{1}{\mu}\, \cL_0\Bigl(\frac{k_2}{\mu}\Bigr) \,\delta(k_B) \, \delta(k_1) \nn \\
  & \quad + s_{a1,B}(R,\eta_1,\Delta \eta_{12})\,\frac{1}{\mu}\, \cL_0\Bigl(\frac{k_B}{\mu}\Bigr) \,\delta(k_1) \, \delta(k_2) 
     + s_{a1,\delta}(R,\Delta \eta_{12}) \, \delta(k_B) \, \delta(k_1) \, \delta(k_2) \biggr] \nn \\
  & \quad + \bfT_b \cdot \bfT_1\biggl[\Delta \eta_{12} \to -\Delta\eta_{12}\biggr] + \bfT_b \cdot \bfT_2\biggl[(k_1,k_2,\eta_1)\to (k_2,k_1,\eta_2)\biggr] \nn \\
  & \quad + \bfT_a \cdot \bfT_2\biggl[(k_1,k_2,\eta_1,\Delta \eta_{12})\to (k_2,k_1,\eta_2,-\Delta \eta_{12})\biggr] \biggr\} \, ,
  \end{align}
  where $\Delta \eta_{12} \equiv \eta_1-\eta_2$ is the difference between the rapidities of the two jets and $R_1=R_2 \equiv R <\pi/2$. The replacements in the last line are always with respect to the terms with the color factor $\bfT_a \cdot\bfT_1$.
  
The contributions from the beam-beam dipole are equivalent to the case of single production given in \eq{soft_coeffs} and \app{R_expansion}, i.e.
\begin{align}
s_{ab,B}(R)=4R^2 +\mathcal{O}(R^4) \, , \quad s_{ab,\delta}(R)= -\frac{\pi^2}{3} +4 R^2 (2 \ln R-1) + \mathcal{O}(R^4) \, .
\end{align}

\begin{figure}
\centering
\includegraphics[height=3.8cm]{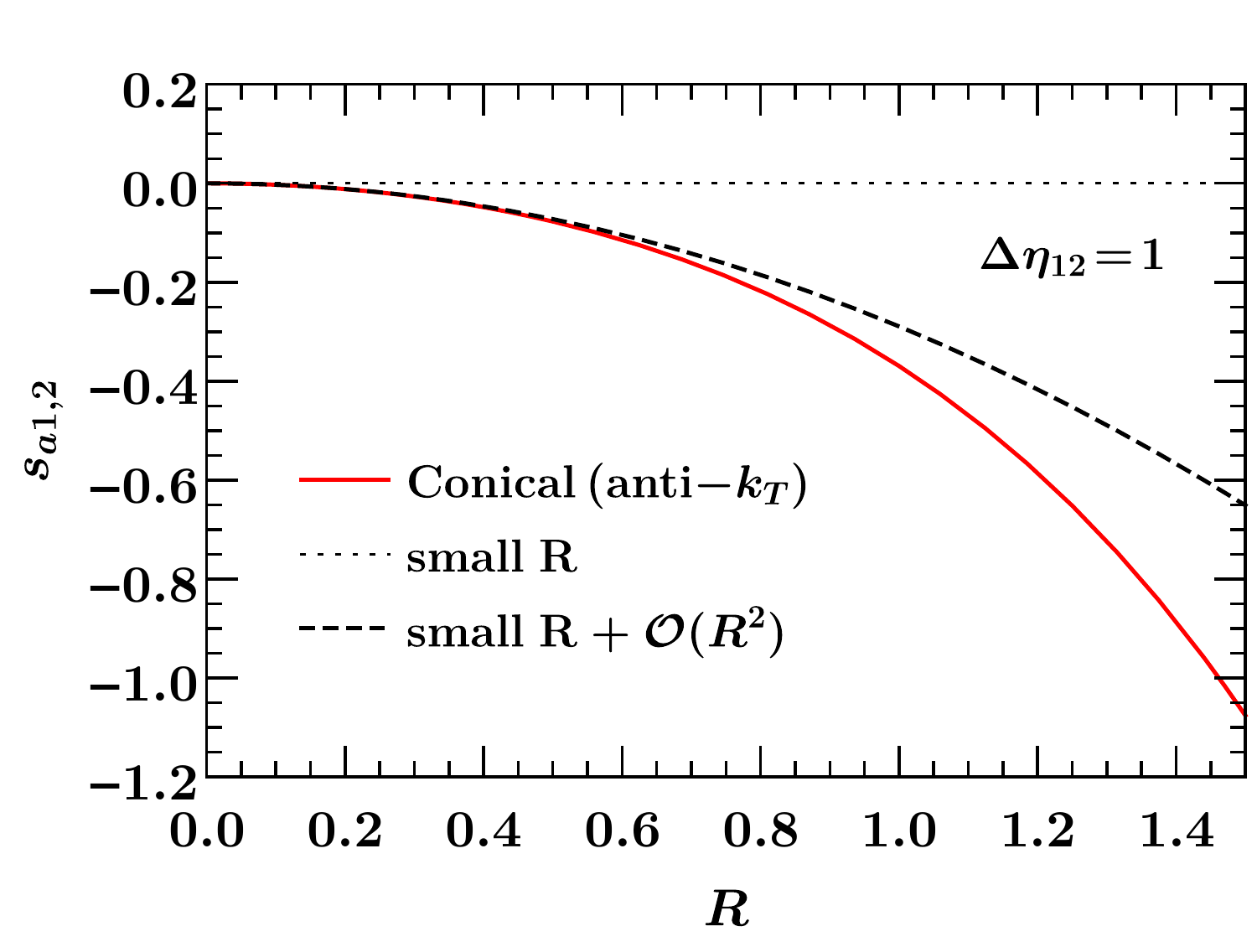}\hspace{0.2cm}
\includegraphics[height=3.8cm]{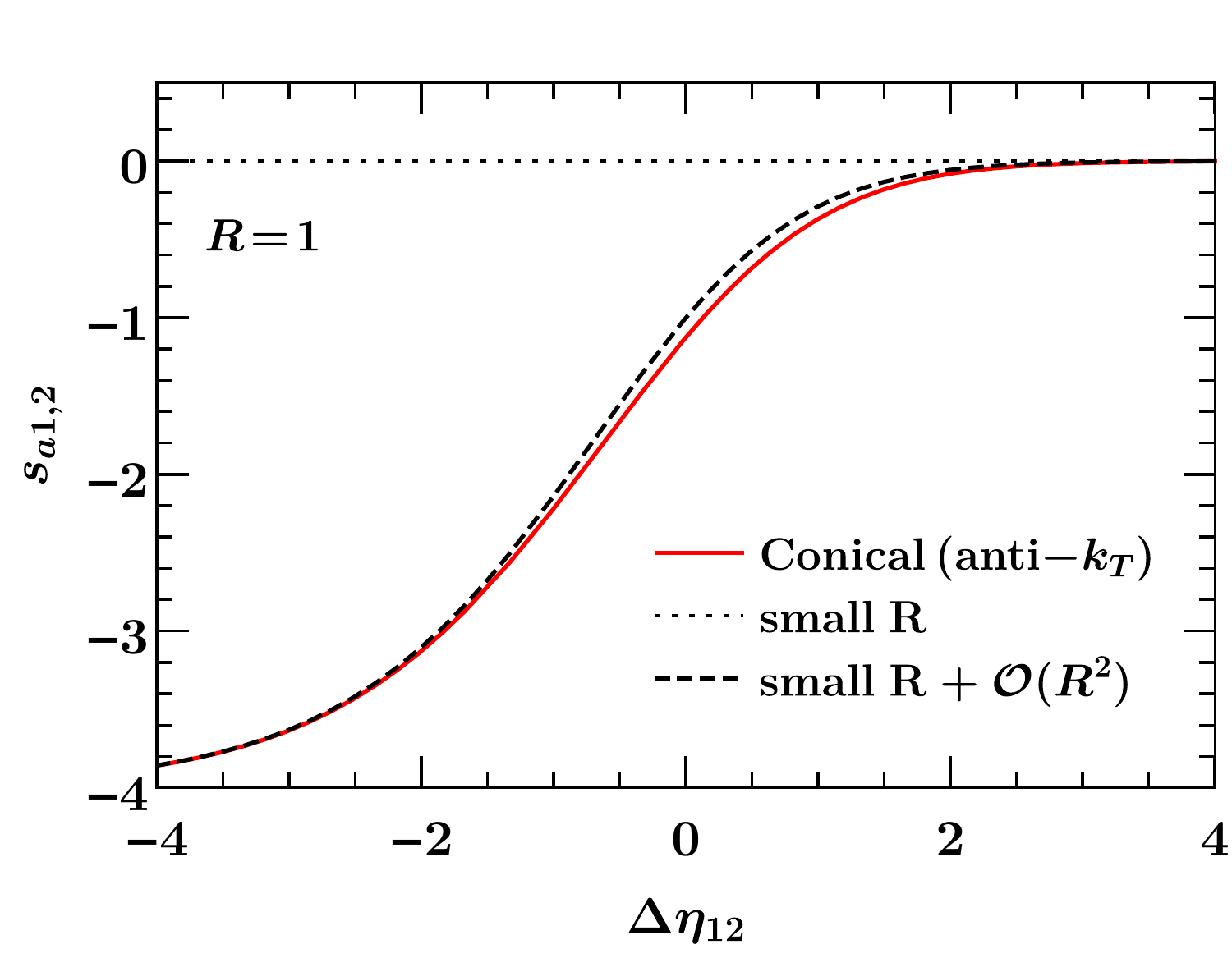}
\\
\includegraphics[height=3.8cm]{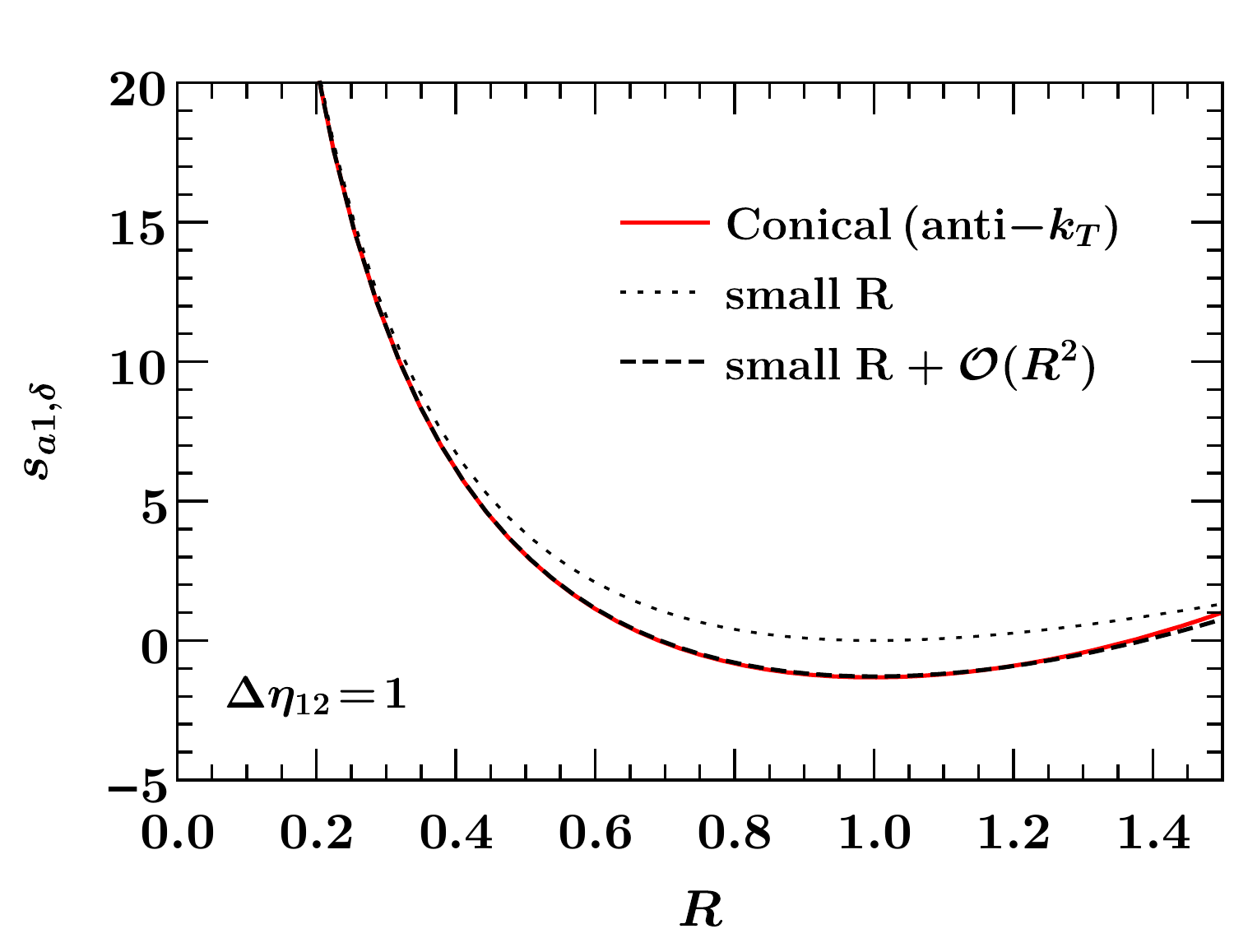}\hspace{0.2cm}
\includegraphics[height=3.8cm]{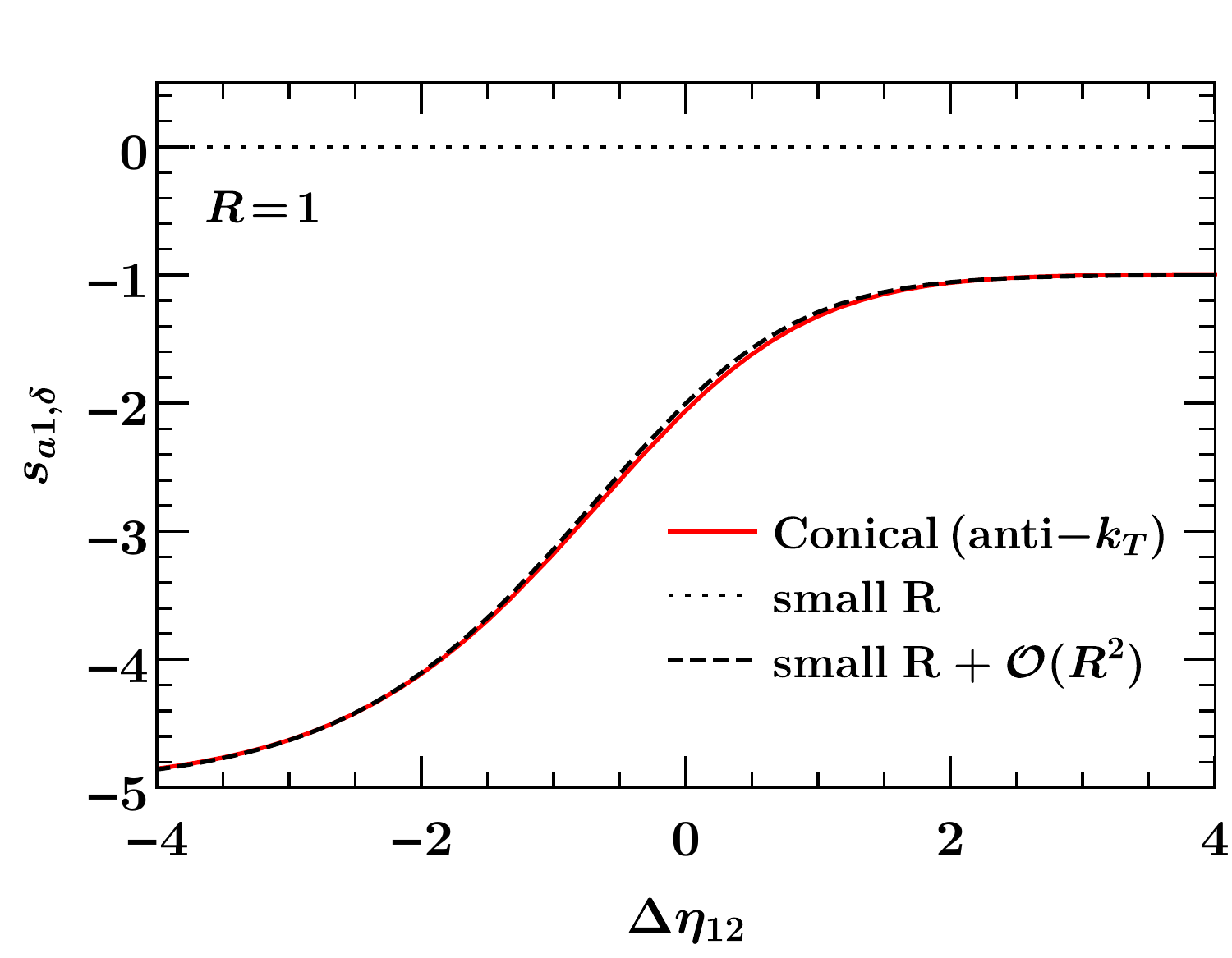}
\caption{The full coefficients $s_{a1,2}$ (top) and $s_{a1,\delta}$ (bottom) together with the small $R$ for conical (anti-$k_T$) jets for a jet mass measurement  ($\beta=2$) and a $p_T$ veto, for $\Delta \eta_{12} =1$ in terms of $R$ (left) and for $R=1$ in terms of $\Delta \eta_{12}$ (right).\label{fig:Sa12}}
\end{figure}

The contributions from the beam-jet dipoles are also closely related to the ones for single production given in \eq{soft_coeffs} and \app{R_expansion} with the difference that starting at $\mathcal{O}(R^2)$ there is now also a correction due to emissions into the phase space region of the second jet, which concerns the coefficients $s_{a1,2}$, $s_{a1,B}$ and $s_{a1,\delta}$ and can be easily computed analytically in analogy to \app{R_expansion}. We get
\begin{align}
s_{a1,1}(R) &= -8 \ln R -R^2+ \mathcal{O}(R^4) \, , \\
s_{a1,2}(R,\Delta\eta_{12}) &=  - R^2  \frac{e^{-\Delta\eta_{12}}}{\cosh^2\Bigl(\frac{\Delta\eta_{12}}{2}\Bigr)}+ \mathcal{O}(R^4) \, , \nn \\
s_{a1,B}(R,\eta,\Delta\eta_{12}) &= 8\ln R+ 8\eta +R^2 \biggl[1+ \frac{e^{-\Delta\eta_{12}}}{\cosh^2\Bigl(\frac{\Delta\eta_{12}}{2}\Bigr)} \biggr]+\mathcal{O}(R^4) \, , \nn \\
s_{a1,\delta}(R,\Delta\eta_{12}) &=4 \ln^2 R (2-\theta(R-1))
+R^2 (2\ln R-1)\biggl[1+ \frac{ e^{-\Delta\eta_{12}}}{\cosh^2\Bigl(\frac{\Delta\eta_{12}}{2}\Bigr)} \biggr]+\mathcal{O}(R^4) \, . \nn 
\end{align}
We demonstrate in \fig{Sa12} that including the terms up to $\mathcal{O}(R^2)$ gives a very good approximation of the full results, even for $R\sim 1$.

\begin{figure}
\centering
\includegraphics[height=3.8cm]{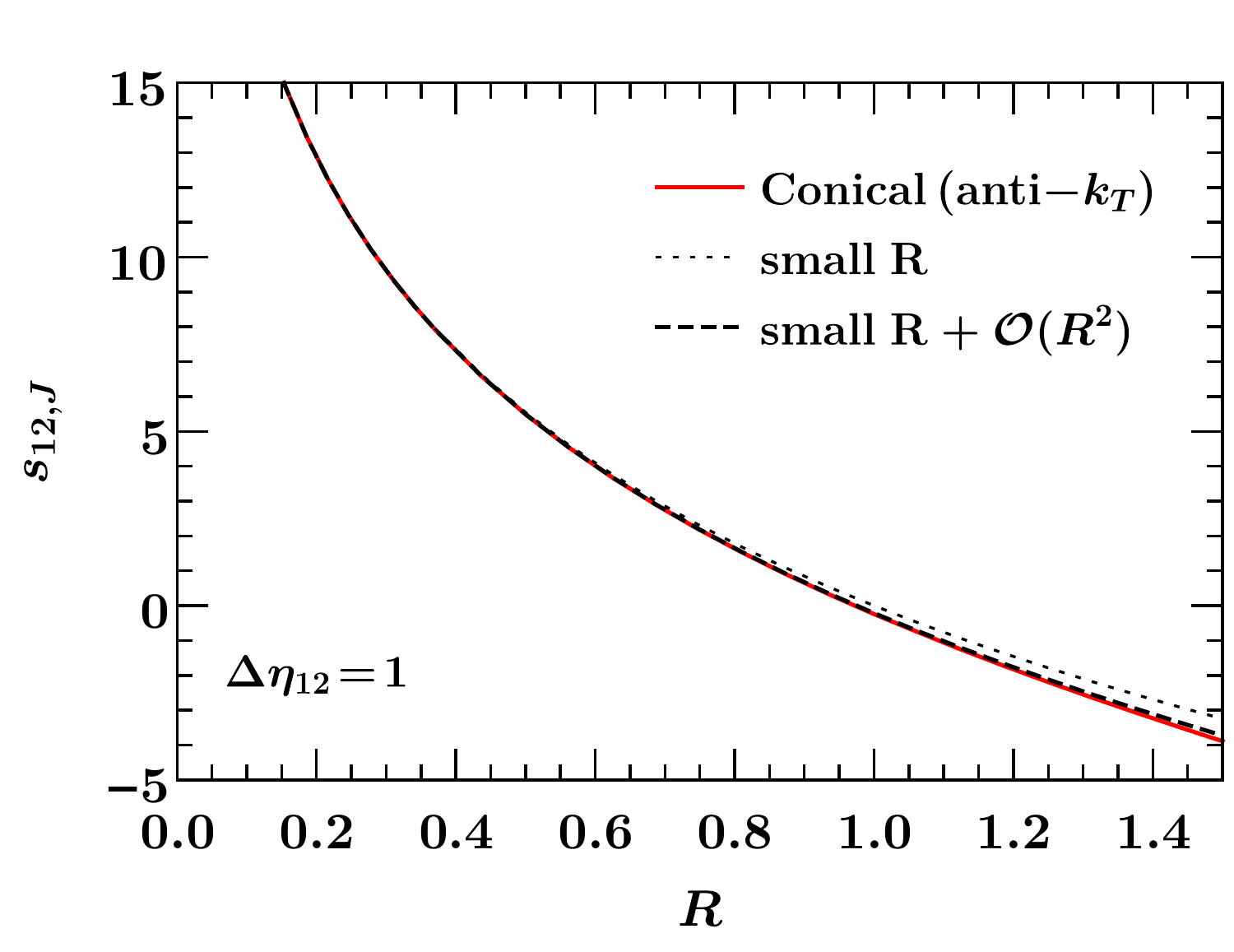}\hspace{0.2cm}
\includegraphics[height=3.8cm]{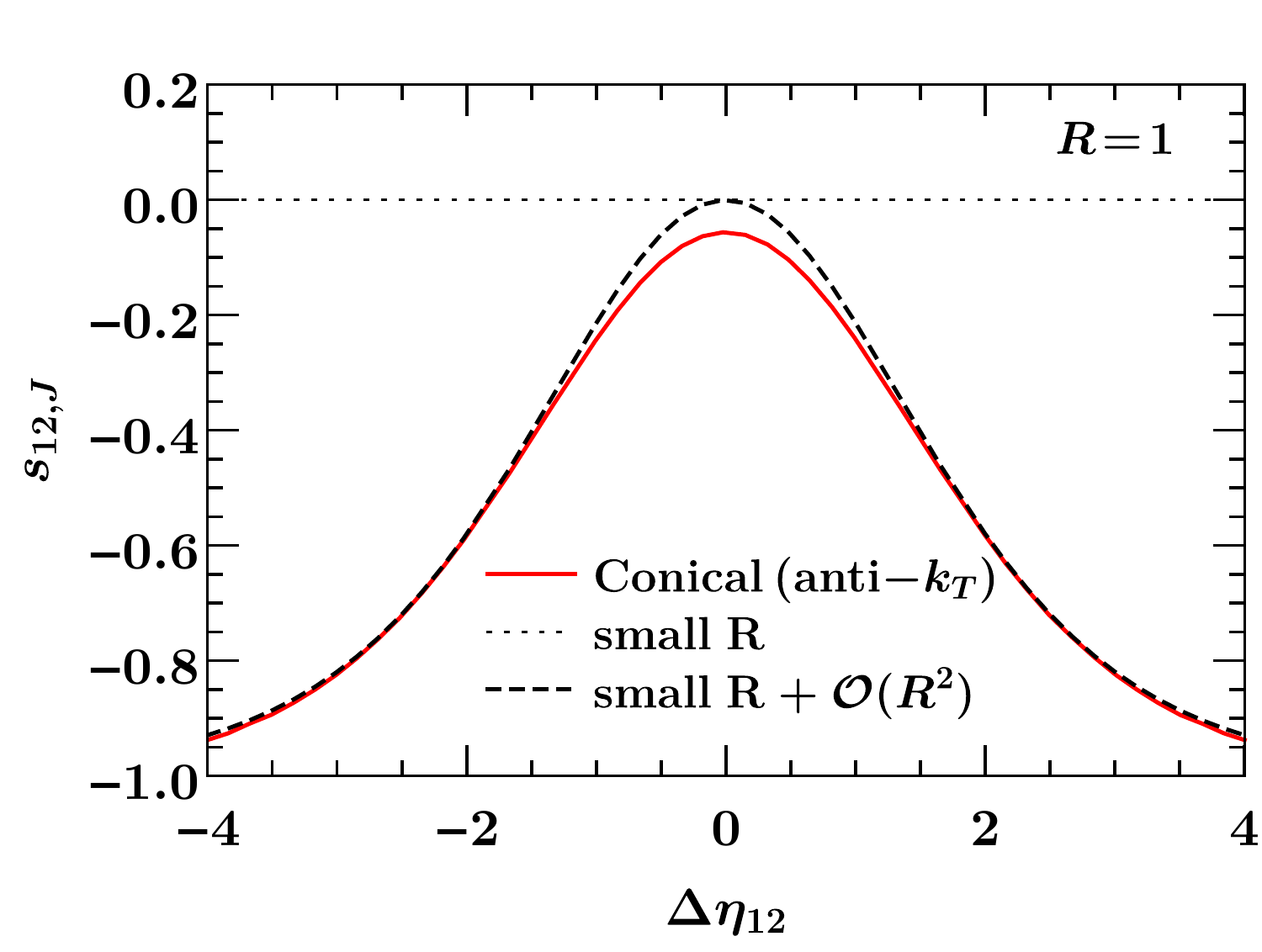}
\\
\includegraphics[height=3.8cm]{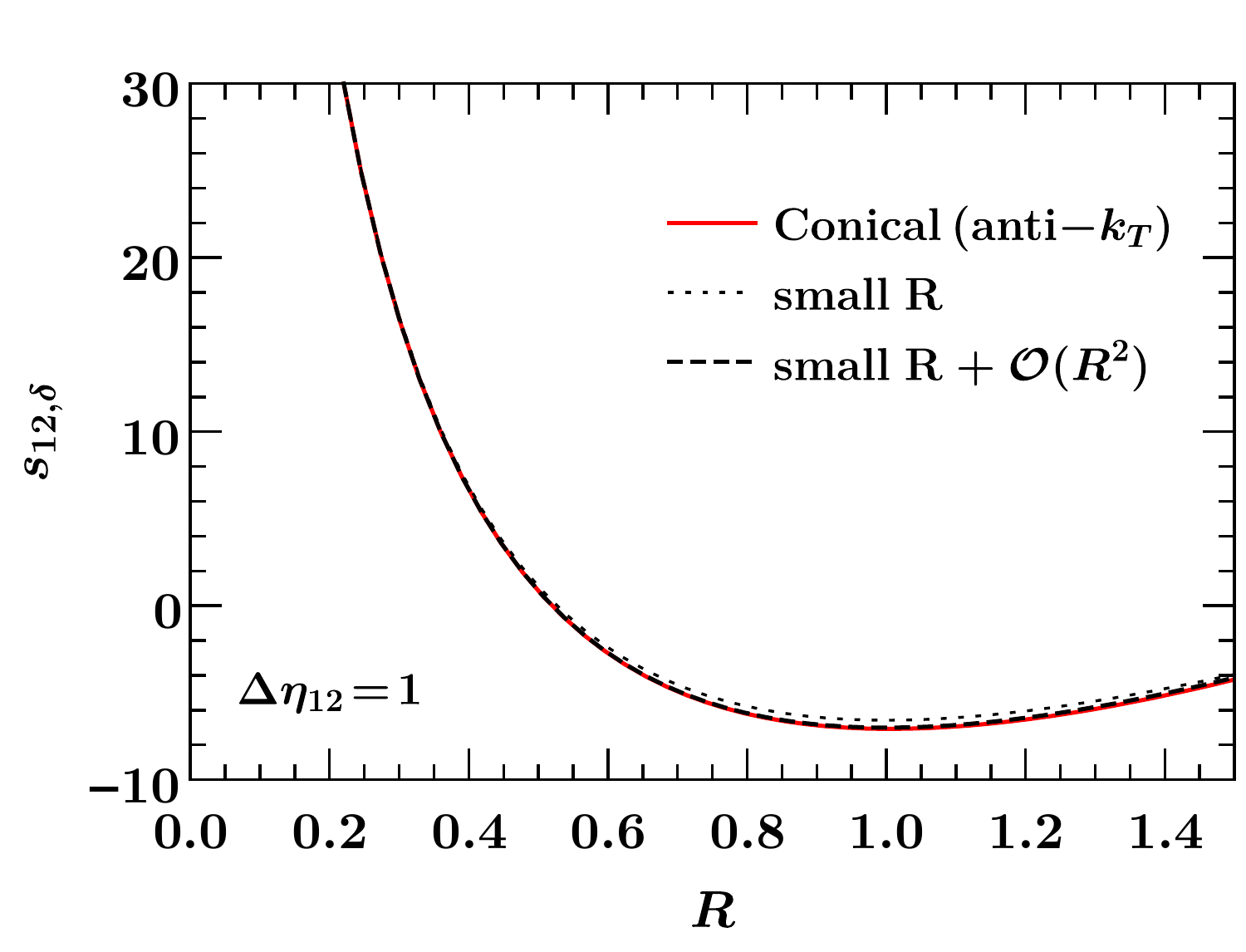}\hspace{0.2cm}
\includegraphics[height=3.8cm]{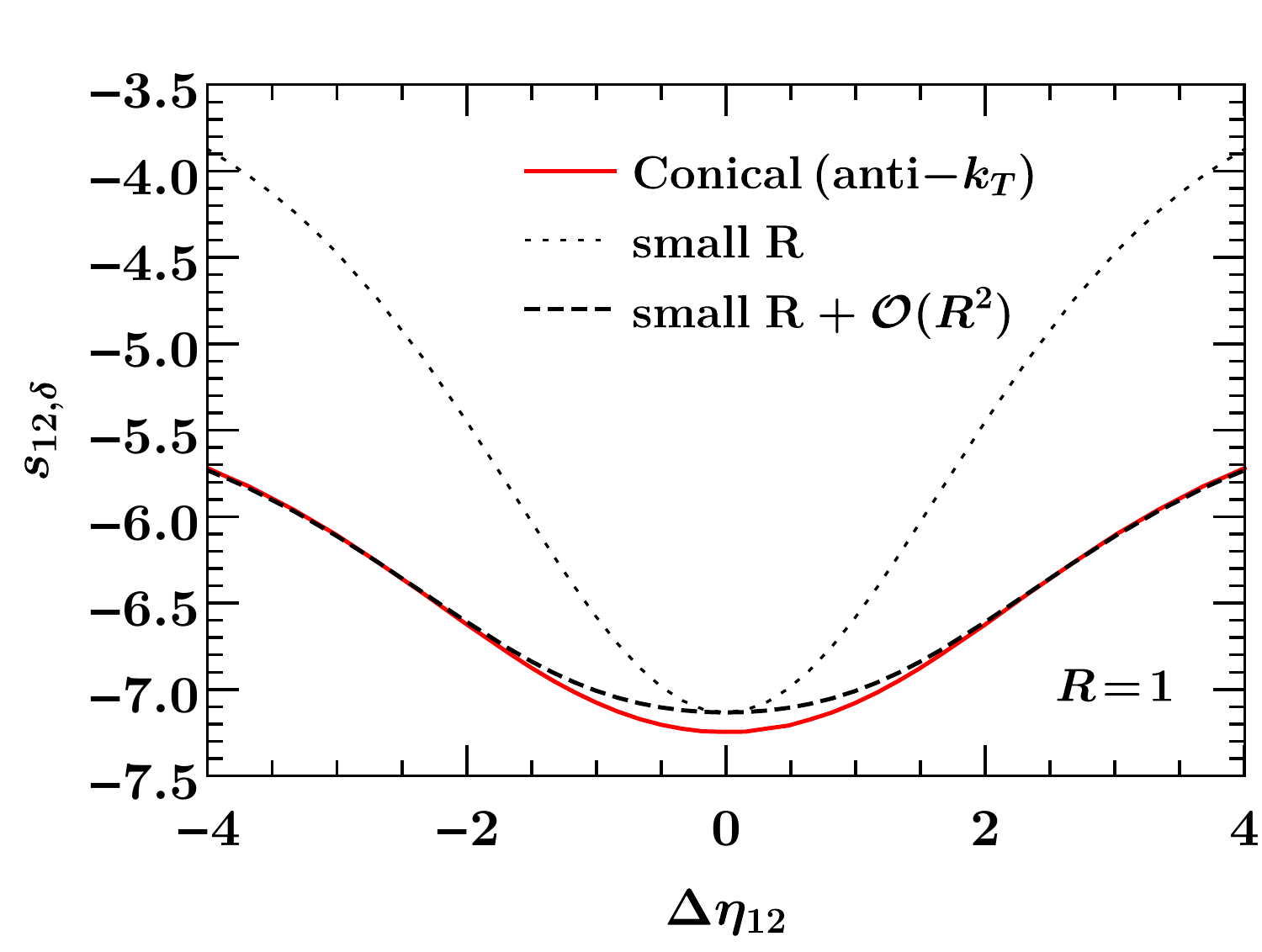}
\caption{The full coefficients $s_{12,B}$ (top) and $s_{12,\delta}$ (bottom) together with the small $R$ results for conical (anti-$k_T$) jets for a jet mass measurement  ($\beta=2$) and a $p_T$ veto, for $\Delta \eta_{12} =1$ in terms of $R$ (left) and for $R=1$ in terms of $\Delta \eta_{12}$ (right). \label{fig:S12}}
\end{figure}

The only remaining ingredient is the correction from the jet-jet dipole.
 The leading small-$R$ results have been computed in ref.~\cite{Hornig:2016ahz}, which we have reproduced.\footnote{Reference~\cite{Hornig:2016ahz} considers a $p_T$-veto with a rapidity cutoff $\eta_{\rm cut}$. For the jet-jet dipole the effect due to $\eta_{\rm cut}$ is power suppressed in $1/e^{\eta_{\rm cut}}$, while for the other dipole contributions it leads to different results than those given above.} The $\mathcal{O}(R^2)$ corrections can be computed following \app{R_expansion}. This gives
\begin{align}
s_{12,J}(R,\Delta \eta_{12}) & =-8 \ln R - R^2  \tanh^2 \frac{\Delta \eta_{12}}{2} +\mathcal{O}(R^4)\, ,\\
s_{12,B}(R,\Delta \eta_{12}) & = -16 \ln \Bigl( 2\cosh \frac{\Delta \eta_{12}}{2}\Bigr) -2 s_{12,J}(\Delta \eta_{12},R)\, ,\nn \\
s_{12,\delta}(R,\Delta \eta_{12}) & =16 \ln^2 R -8\ln^2\Bigl( 2\cosh\frac{ \Delta \eta_{12}}{2}\Bigr) + 2 (\Delta \eta_{12})^2 -\frac{\pi^2}{3}\nn\\
&\quad+ R^2 \Bigl[2 (2\ln R-1) \tanh^2 \frac{\Delta \eta_{12}}{2}\biggr] +\mathcal{O}(R^4)
\nn \, .\end{align}
In \fig{S12} we compare the full numeric results for these coefficients to the analytic expressions. Again the small $R$ expansion provides an excellent approximation of the full result for the jet-jet dipole contribution. Together with the findings for the beam-beam and beam-jet dipole corrections this indicates that keeping terms up to $\mathcal{O}(R^2)$ is likely sufficient for phenomenological purposes.

We remark that for jet vetoes which are not boost invariant, all of the dipoles, in particular also the jet-jet dipole, depend  on the individual jet rapidities. For multijet processes or an additional recoiling color singlet state the soft function depends in addition on the separation of the jets in azimuth. The analytic computation for these cases is significantly more involved.

%% Bibliography
\phantomsection
\addcontentsline{toc}{section}{References}
\bibliographystyle{jhep}
\bibliography{softfunc}

\end{document}